\documentclass[11pt]{article}
\usepackage{makeidx}
\usepackage{amsfonts}
\usepackage{amssymb}
\usepackage{graphicx}
\usepackage{amsmath}
\usepackage{geometry}
\usepackage{appendix}

\DeclareMathSizes{10.95}{10}{7}{6}
\input{tcilatex}
\begin{document}

\title{A Statistical Field Perspective on Capital Allocation and
Accumulation: Individual dynamics}
\author{Pierre Gosselin\thanks{%
Pierre Gosselin : Institut Fourier, UMR 5582 CNRS-UGA, Universit\'{e}
Grenoble Alpes, BP 74, 38402 St Martin d'H\`{e}res, France.\ E-Mail:
Pierre.Gosselin@univ-grenoble-alpes.fr} \and A\"{\i}leen Lotz\thanks{%
A\"{\i}leen Lotz: Cerca Trova, BP 114, 38001 Grenoble Cedex 1, France.\
E-mail: a.lotz@cercatrova.eu}}
\date{May 2023}
\maketitle

\begin{abstract}
We have shown, in a series of articles, that a classical description of a
large number of economic agents can be replaced by a statistical fields
formalism.

To better understand the accumulation and allocation of capital among
different sectors, the present paper applies this statistical fields
description to a large number of heterogeneous agents divided into two
groups. The first group is composed of a large number of firms in different
sectors that collectively own the entire physical capital. The second group,
investors, holds the entire financial capital and allocates it between firms
across sectors according to investment preferences, expected returns, and
stock prices variations on financial markets. In return, firms pay dividends
to their investors. Financial capital is thus a function of dividends and
stock valuations, whereas physical capital is a function of the total
capital allocated by the financial sector. Whereas our previous work focused
on the background fields that describe potential long-term equilibria, here
we compute the transition functions of individual agents and study their
probabilistic dynamics in the background field, as a function of their
initial state. We show that capital accumulation depends on various factors.
The probability associated with each firm's trajectories is the result of
several contradictory effects: the firm tends to shift towards sectors with
the greatest long-term return, but must take into account the impact of its
shift on its attractiveness for investors throughout its trajectory. Since
this trajectory depends largely on the average capital of transition
sectors, a firm's attractiveness during its relocation depends on the
relative level of capital in those sectors. Thus, an under-capitalized firm
reaching a high-capital sector will experience a loss of attractiveness, and
subsequently, in investors. Moreover, the firm must also consider the
effects of competition in the intermediate sectors. An under-capitalized
firm will tend to be ousted out towards sectors with lower average capital,
while an over-capitalized firm will tend to shift towards higher
average-capital sectors. For investors, capital allocation depends on their
short and long-term returns. These returns are not independent: in the
short-term, returns are composed of both the firm's dividends and the
increase in its stock prices. In the long-term, returns are based on the
firm's growth expectations, but also, indirectly, on expectations of higher
stock prices. Investors' capital allocation directly depends on the
volatility of stock prices and firms' dividends. Investors will tend to
reallocate their capital to maximize their short and long-term returns. The
higher their level of capital, the stronger the re-allocation will be.


\end{abstract}

\pagebreak

\section{Introduction}

We have shown in a series of articles that a classic description of a large
number of economic agents can be replaced by a description in terms of
statistical fields. This formalism is used to study both the collective
behavior of the system, by finding the so called background field of the
system, and the probabilistic behavior of agents in this collective state,
through the transition functions of an arbitrary number of agents.

As in (Gosselin Lotz Wambst 2022), we apply this formalism to the
accumulation and allocation of capital between different sectors.
Specifically, we consider two groups of agents: producers representing the
real economy and investors representing the financial markets. Producers
collectively own the entire physical capital, whereas investors hold and
allocate the entire financial capital between firms across sectors according
to investment preferences, expected returns, and stock prices variations on
financial markets. Firms pay back their investors through dividends.
Financial capital is thus a function of dividends and stock valuations, and
physical capital is a function of the overall capital allocated by the
financial sector.

These two groups of agents are described by two different interacting fields
and by the system action functional. The solutions to the minimization
equations of the action functional are the background fields of the system.
From a sector perspective, the background fields determine the average
distribution of capital and the firm density within sectors, given external
parameters, such as expected returns, technological advances, and their
dynamics.

Whereas (Gosselin Lotz Wambst 2022) presented the background fields modeling
the potential collective states, the present paper will study the transition
functions of individual agents and their probabilistic dynamics in the
background field, as a function of their initial state.\textbf{\ }We
demonstrate that several factors influence the probability of each firm's
relocation path.

First, firms tend to relocate in sectors with highest long-term returns.\
However, the path followed by the firm to realocate depends on the
characteristics of the transition sectors, that are themselves determined by
the collective state of the system. The attractiveness of the firm during
its relocation process depends on the average capital of the transition
sectors it stumbles into. Depending on the sector, investors may over or
underinvest in the firm.\ An under-capitalized firm may fail to attract
investors in and either end up being stuck in this sector or be repelled to
a less attractive one.

Second, competition along the transition sectors, depends on the background
state of the system and impact differently the firm's level or capital and
attractiveness. An overcapitalized firm facing many less-endowed
competitors, will oust them out of the sector. On the contrary, an
under-capitalized firm will be ousted out from its own sector and move
towards less-capitalized and denser sectors in average. A capital gain - or
loss - may follow. Under-capitalized firms tend to move towards lower than
average capitalized sectors, while over-capitalized firms tend to move
towards higher than average-capitalized sectors.

Third, investors' capital allocation depends on short and long-term
returns.\ Yet these returns are not independent: short-term returns,
dividends and stock prices variations are correlated to the long-term that
depend on growth expectations and stock prices expectations.

Changes in investors' capital allocation are therefore directly dependent on
stock prices' volatility and firms' dividends. Changes in growth
expectations impact stock prices and incite investors to reallocate capital
to maximize their returns. The higher their level of capital, the stronger
the reallocation will be.

The paper is organized as follows. The second section is a literature
review. The first part of the paper presents the use of field theory in
economic modeling. Section three presents the principles of this approach.
In section four, we explain the translation of a classical framework into a
field model and section five derives the transition functions of the model.
These functions compute the probabilities of the model to evolve from an
initial to a final state. In section 6, we introduce the notions of
background field, effective action and their use in computing and
interpreting the transition functions.

The second part of the paper applies the field approach to a model of
capital accumulation. Section 7 presents the classical framework of the
model and section 8 translates this framework into a field model. In section
9, we derive the transition functions for one and two agents in this
particular model. The results are presented and discussed in section 10
while section 11 concludes.

\section{Literature review}

Several branches of the economic literature seek to replace the
representative agent with a collection of heterogeneous ones. However, among
other things, they differ in the way they model this collection of agents.
The comparison between these approaches and our formalism has been detailed
in (paper), where we have also compared the notion of collective states in
these approaches to our own. In the present paper, we instead detail the
differences concerning individual dynamics.

Several branches share a common approach to finding the probability density
of agent behavior. This approach is followed by mean-field theory,
heterogeneous agents, new Keynesian (HANK) models, and the
information-theoretic approach to economics. In mean field theory (see
(Bensoussan et al. 2018; Lasry et al. 2010a, b), (Gomes et al., 2015)),
agents interact with the population as a whole, described by the mean field.
However, this literature does not consider the direct interactions between
agents.

Heterogeneous agents new Keynesian (HANK) models (see Kaplan and Violante
2018 for an account) consider probabilistic techniques close to mean field
theory: an equilibrium probability distribution for each group of agents is
derived from a set of optimizing heterogeneous agents. Information theory
(see Yang 2018) considers probabilistic states around the equilibrium
derived from an entropy maximization program. Lastly, the rational
inattention theory (Sims 2006) derives non-gaussian density laws from
limited information and constraints.

Our approach concerning differs from this mean field-like technique. We do
not seek individual transition at equilibrium. We first instead consider the
whole set of agents and their interactions. Our statistical weight is also a
probabilistic approach. Nevertheless, it is directly built from the
microeconomic dynamic equations of N-interacting agents. This corresponds to
working with a much larger space of dynamic variables than the ones
considered in the other papers. At a collective level, this allows us to
describe the model with fields and find the background field that arises
from the interactions at the individual level. Once the collective states
have been found, we can recover both the types of individual dynamics
depending on the initial conditions and the "effective" form of interactions
between two or more agents.

At the individual level, which is at stake here, agents are distributed
along some probability law. However, this probability law is directly
conditioned by the collective state of the system and the effective
interactions. Different collective states, given different parameters, yield
different individual dynamics. This approach allows for coming back and
forth between collective and individual aspects of the system.

Different categories of agents appear in the emerging collective state.
Dynamics may present very different patterns, given the collective state's
form and the agents' initial conditions.

A second branch of the literature is closest to our approach since it
considers the interacting system of agents. Finally, it is the multi-agent
systems literature, notably agent-based models (see Gaffard Napoletano 2012,
Mandel et al. 2010 2012) and economic networks (Jackson 2010).

Agent-based models use microfounded general macroeconomics models, whereas
network models lower-scale models such as contract theory, behaviour
diffusion, information sharing, or learning. In both settings, agents are
typically defined by and follow various rules, leading to equilibria and
dynamics otherwise inaccessible to the representative agent setup. Both
approaches are, however, highly numerical and model-dependent. As a
consequence, ABM models tend to reveal collective states numerically. Unlike
our approach, they do not seek to reveal these collective states' relation
to individual dynamics. In other words, individual dynamics are considered
together to produce a collective state, as in our approach. However, the
non-analytical approach to this collective state prevents a precise
inspection of the emergence of several phases and the impact of the
collective state on individual trajectories. Statistical fields theory, on
the contrary, accounts for transitions between scales. Macroeconomic
patterns are grounded in behaviours and interaction structures. Describing
these structures in terms of field theory allows for the emergence of
collective states at the macro scale and, in turn, the study of their impact
at the individual level.

A third branch of the literature, Econophysics, is also related to ours
since it often considers the set of agents as a statistical system (for a
review, see Abergel et al. 2011a,b and references therein; or Lux 2008,
2016). However, it tends to focus on empirical laws rather than apply the
full potential of field theory to economic systems. Kleinert (2009) uses
path integrals to model stock price dynamics, but our approach differs in
that it keeps track of usual microeconomic concepts, such as utility
functions and expectations, and includes them into the analytical treatment
of multi-agent systems by translating the main characteristics of optimizing
agents in terms of statistical systems.

The literature on interactions between finance and real economy or capital
accumulation mainly occurs in the context of DGSE models. A review of the
literature is provided by Cochrane (2006), while further developments can be
found in Grassetti et al. (2022), Grosshans and Zeisberger (2018), B\"{o}hm
et al. (2008), Caggese and Orive (2012), Bernanke et al. (1999), Campello et
al. (2010), Holmstrom and Tirole (1997), Jermann and Quadrini (2012), Khan
and Thomas (2013), and Monacelli et al. (2011). Theoretical models in this
literature typically include several types of agents at the aggregated
level, such as producers for possibly several sectors, consumers, and
financial intermediaries. They aim to describe the interactions between a
few representative agents and determine interest rates, levels of
production, and asset pricing, all in the context of ad-hoc anticipations.

Our formalism differs from this literature in three ways. First, we consider
several groups of a large number of agents to describe the emergence of
collective states and study the continuous space of sectors. Second, we
consider expected returns and the longer-term horizon somewhat exogenous or
structural. Expected returns combine elements such as technology, returns,
productivity, sectoral capital stock, expectations, and beliefs. These
returns are also a function defined over the sectors' space: the system's
background fields are functionals of these expected returns. Taken together,
the background fields of a field model describe an economic environment for
a given configuration of expected returns. As such, expected returns are at
first seen as exogenous. In a second step, when we consider the dynamics
between capital accumulation and expectations, expectations may be seen as
endogenous. Even then, the relations between actual and expected variables
specified are general enough to derive some possible dynamics.

The literature on interactions between finance and real economy or capital
accumulation occurs mainly in the context of DGSE models. (for a review of
the literature, see Cochrane 2006; for further developments, see Grassetti
et al. 2022, Grosshans and Zeisberger 2018, B\"{o}hm et al. 2008, Caggese
and Orive, Bernanke e al. 1999, Campello et al. 2010, Holmstrom and Tirole
1997, Jermann, and Quadrini 2012, Khan Thomas 2013, Monacelli et al. 2011).
Theoretical models include several types of agents at the aggregated level.
They describe the interactions between a few representative agents, such as
producers for possibly several sectors, consumers, and financial
intermediaries, to determine interest rates, levels of production, and asset
pricing, in a context of ad-hoc anticipations.

Our formalism differs from this literature in three ways. First, we model
several groups of agents across a continuous space of economic sectors to
capture the emergence of collective states. Second, we consider expected
returns and the long-term horizon as somewhat exogenous or structural,
incorporating elements such as technology, returns, productivity, sectoral
capital stock, expectations, and beliefs. These returns are a function
defined over the sectors' space, and the system's background fields are
themselves functionals of these expected returns. Thus, the background
fields of the system describe an economic environment for a given
configuration of expected returns, which are first seen as exogenous.
However, we later consider the dynamics between capital accumulation and
expectations, allowing expectations to become endogenous.

Last, we do not seek aggregated dynamics, and the individual dynamics depend
on a particular collective state. In such a background, agents' typical
dynamics are computed through transition functions. These functions compute
the transition probabilities from one capital-sector point to another. Since
backgrounds may be considered dynamic quantities, structural or long-term
variations in the returns' landscape may modify the background and, in turn,
the individual dynamics. Expected returns themselves depend on and interact
with capital accumulation.

\part*{Field theory and economic modeling}

The field formalism combines the micro and macro aspects of an economic
system by considering the interactions among the entire set of agents and
their environment. This allows for a back-and-forth analysis between micro
and macro scales: once the collective state of a system is determined, it
enables a return to individual dynamics within that collective state. We
will first translate a microeconomic model its field equivalent, then
explain how this translation allows to retrieve collective or macro states
and the agents transition functions.

\section{Principles}

The field-formalism used in this paper is rooted in a probabilistic
description of economic systems with large number of agents. Classically,
each agent's dynamics is described by an optimal path\ for\textbf{\ }some
vector variable, say $A_{i}\left( t\right) $, from an initial to a final
point, up to some fluctuations. However, the same system of agents can be
viewed probabilistically. An agent can be described by a \emph{probability
density }that is, due to idiosyncratic uncertainties, centered around the
classical optimal path\footnote{%
Due to the infinite number of possible paths, each individual path has a
null probability to exist.\ We, therefore, use the word "probability
density" rather than "probability".} (see Gosselin, Lotz and Wambst 2017,
2020, 2021). In this probabilistic approach, each possible dynamics for the
set of $N$ agents must be taken into account and weighted by its
probability. The system is then described by a \emph{statistical weight},
the probability density for any configuration of $N$\ arbitrary individual
paths. Once this statistical weight is found, we can compute the transition
probabilities of the system which are the probabilities for any number of
agents to evolve from an initial to a final state in $A_{i}$, $B_{i}$ in a
given time.

Because this probabilistic approach implies keeping track of the $N$\
agents' probability transitions, it is practically untractable for a large
number of agents. However, this step is necessary since it can be translated
into a more compact \emph{field formalism}\textbf{\ }(see Gosselin, Lotz,
and Wambst 2017, 2020, 2021), which preserves the essential information
encoded in the model but implements a change in perspective.\ It does not
keep track of the $N$-indexed agents but describes their dynamics and
interactions as a collective thread of all possible unlabeled paths. This
collective thread can be seen as an environment that conditions the dynamics
of individual agents from one state to another. The field formalism eases
the computation of transition functions. More importantly, it detects the
collective states or phases encompassed in the field, that would otherwise
remain indetectable using the probabilistic formulation.

To translate the probabilistic approach into a field model, the $N$\ agents'
trajectories $\mathbf{A}_{i}\left( t\right) $ is replaced by a field $\Psi $%
, a complex-valued function that solely depends on a single set of
variables, $\mathbf{A}$. The statistical weight of the probabilistic
approach is translated into a probability density on the space of
complex-valued functions of the variables $\mathbf{A}$.\textbf{\ }For the
configuration $\Psi \left( \mathbf{A}\right) $, this probability density has
the form $\exp \left( -S\left( \Psi \right) \right) $. The functional $%
S\left( \Psi \right) $\ is called the \emph{field action functional}. It
captures the microscopic features of individual agents' dynamics and
interactions. The idea is that of a dictionary that would translate the
various terms of the classical description in terms of their field
equivalent.\ The integral of $\exp \left( -S\left( \Psi \right) \right) $
over the configurations $\Psi $ is the \emph{partition function} of the
system. The fields that minimize the action functional are the \emph{%
classical background fields, }or more simply the\emph{\ background fields}.\
They encapsulate the collective states of the system.\ 

For several types of agents, the generalisation is straightforward. Each
type $\alpha $ is described by a field $\Psi _{\alpha }\left( \mathbf{A}%
_{\alpha }\right) $. The field action depends on the whole set of fields $%
\left\{ \Psi _{\alpha }\right\} $. It accounts for all types of agents and
their interactions, and writes $S\left( \left\{ \Psi _{\alpha }\right\}
\right) $. The form of $S\left( \left\{ \Psi _{\alpha }\right\} \right) $\
is obtained directly from the classical description of our model.

In the following, we will detail a shortcut of this translation method and
apply it to the microeconomic framework below.

\section{From a classical framework to its field expression}

To translate a classical economic framework into a field model, we must
first consider the various types of agents in the model.\ We will rewrite
their dynamics as the minimization equations of some initial functions, in
the same way as, for instance, consumption dynamics could be derived from an
utility function. These minimization functions\textbf{\ }will then be
translated into field functionals of several independent fields\footnote{%
The term functional refers to a function of a function, i.e. a function
whose argument is itself a function.}, one for each type of agent. The sum
of these functionals, the "field action functional", describes the whole
system in terms of fields\footnote{%
Details about the probabilistic step will be given as a reminder along the
text and in appendix 1.}. We will detail the process for a classical and
relatively general model.

\subsection{A general type of classical framework}

We consider two types of agents, characterized by vector-variables $\left\{ 
\mathbf{A}_{i}\left( t\right) \right\} _{i=1,...N},$ and $\left\{ \mathbf{%
\hat{A}}_{l}\left( t\right) \right\} _{l=1,...\hat{N}}$... respectively,
where $N$ and $\hat{N}$ are the number of agents of each type.

In a classical model, the optimization by agents of their characteristic
function, such as utility, consumption, production, ..., leads to a system
of dynamic equations.

Our approach is the opposite: using dynamic equations that could result from
the optimization of agents, we reconstruct what we will call a "minimization
function," $s\left( \left\{ \mathbf{A}_{l}\right\} ,\left\{ \mathbf{\hat{A}}%
_{l}\right\} ,t\right) $, that depends on the entire system and whose
minimization would yield the dynamic equations of the system. We will use in
the following the general form of $s\left( \left\{ \mathbf{A}_{l}\right\}
,\left\{ \mathbf{\hat{A}}_{l}\right\} ,t\right) $ provided in (Gosselin Lotz
Wambst 2022) \footnote{%
A generalisation of equation (\ref{dnn}) in which agents interact at
different times is presented in Appendix 1 of (Gosselin Lotz Wambst 2022),
along with its translation in term of field.}:

\begin{eqnarray}
s\left( \left\{ \mathbf{A}_{i}\right\} ,\left\{ \mathbf{\hat{A}}_{l}\right\}
,t\right) &=&\sum_{i}\left( \frac{d\mathbf{A}_{i}\left( t\right) }{dt}%
-\sum_{j,k,l...}f\left( \mathbf{A}_{i}\left( t\right) ,\mathbf{A}_{j}\left(
t\right) ,\mathbf{A}_{k}\left( t\right) ,\mathbf{\hat{A}}_{l}\left( t\right)
,\mathbf{\hat{A}}_{m}\left( t\right) ...\right) \right) ^{2}  \label{dnn} \\
&&+\sum_{i}\left( \frac{d\mathbf{\hat{A}}_{i}\left( t\right) }{dt}%
-\sum_{j,k,l...}\hat{f}\left( \mathbf{A}_{i}\left( t\right) ,\mathbf{A}%
_{j}\left( t\right) ,\mathbf{A}_{k}\left( t\right) ,\mathbf{\hat{A}}%
_{l}\left( t\right) ,\mathbf{\hat{A}}_{m}\left( t\right) ...\right) \right)
^{2}  \notag \\
&&+\sum_{i}\sum_{j,k,l...}g\left( \mathbf{A}_{i}\left( t\right) ,\mathbf{A}%
_{j}\left( t\right) ,\mathbf{A}_{k}\left( t\right) ,\mathbf{\hat{A}}%
_{l}\left( t\right) ,\mathbf{\hat{A}}_{m}\left( t\right) ...\right)  \notag
\end{eqnarray}%
Note incidentally that, setting $g=0$, the minimization of $s\left( \left\{ 
\mathbf{A}_{l}\right\} ,\left\{ \mathbf{\hat{A}}_{l}\right\} ,t\right) $
would indeed yield:%
\begin{equation}
\frac{d\mathbf{A}_{i}\left( t\right) }{dt}-\sum_{j,k,l...}f\left( \mathbf{A}%
_{i}\left( t\right) ,\mathbf{A}_{j}\left( t\right) ,\mathbf{A}_{k}\left(
t\right) ,\mathbf{\hat{A}}_{l}\left( t\right) ,\mathbf{\hat{A}}_{m}\left(
t\right) ...\right) =0  \label{dnt}
\end{equation}%
and:

\begin{equation}
\frac{d\mathbf{\hat{A}}_{i}\left( t\right) }{dt}-\sum_{j,k,l...}\hat{f}%
\left( \mathbf{A}_{i}\left( t\right) ,\mathbf{A}_{j}\left( t\right) ,\mathbf{%
A}_{k}\left( t\right) ,\mathbf{\hat{A}}_{l}\left( t\right) ,\mathbf{\hat{A}}%
_{m}\left( t\right) ...\right) =0  \label{dnw}
\end{equation}%
that would be the usual type of dynamic equations for agents $\mathbf{A}_{i}$
and $\mathbf{\hat{A}}_{l}$.

It is specifically the last term in formula (\ref{dnn})\ that is a specific
feature of our large number of agents approach. It represents the whole set
of interactions both among and between two groups of agents, modifying the
standard equations (\ref{dnt}) and (\ref{dnw}).

In a classical approach, the model is solved by numerically solving equation
(\ref{dnt}) for the n agents. Minimizing the function $s\left( \left\{ 
\mathbf{A}_{l}\right\} ,\left\{ \mathbf{\hat{A}}_{l}\right\} ,t\right) $
would not yield any new insight compared to the dynamic equations we started
with if we consider that they are satisfied exactly. The key feature of
introducing the minimization function $s\left( \left\{ \mathbf{A}%
_{l}\right\} ,\left\{ \mathbf{\hat{A}}_{l}\right\} ,t\right) $ only arises
when the standard dynamic equations such as (\ref{dnt}) and (\ref{dnw}) are
satisfied only up to error terms. These errors represent the fluctuations of
the agents around an optimal dynamics.%
\begin{equation}
\frac{d\mathbf{A}_{i}\left( t\right) }{dt}-\sum_{j,k,l...}f\left( \mathbf{A}%
_{i}\left( t\right) ,\mathbf{A}_{j}\left( t\right) ,\mathbf{A}_{k}\left(
t\right) ,\mathbf{\hat{A}}_{l}\left( t\right) ,\mathbf{\hat{A}}_{m}\left(
t\right) ...\right) =\varepsilon _{i}
\end{equation}%
and:

\begin{equation}
\frac{d\mathbf{\hat{A}}_{i}\left( t\right) }{dt}-\sum_{j,k,l...}\hat{f}%
\left( \mathbf{A}_{i}\left( t\right) ,\mathbf{A}_{j}\left( t\right) ,\mathbf{%
A}_{k}\left( t\right) ,\mathbf{\hat{A}}_{l}\left( t\right) ,\mathbf{\hat{A}}%
_{m}\left( t\right) ...\right) =\hat{\varepsilon}_{l}
\end{equation}%
These errors confer a probabilistic nature to the system, for they imply a
change in perspective: we do no longer consider solely optimal trajectories.
Rather, we deem any trajectory as possible, with a probability decreasing
with the distance of the trajectory from the optimum.\ This allows to
compute the probabilities for the entire system to evolve from any initial
state to any final state.

This is the main purpose of the function\ $s\left( \left\{ \mathbf{A}%
_{l}\right\} ,\left\{ \mathbf{\hat{A}}_{l}\right\} ,t\right) $ defined in (%
\ref{dnn}). By construction, it is minimal for the optimal path and in
general, it increases with the distance to this optimal path. To gain a
better insight, consider a case without interactions, i.e. $g=0$. In this
particular case, the first two terms of equation (\ref{dnn}) represent the
squared deviation from the dynamics of an agent of each type. The higher
this deviation, the higher $s$. We will associate to each trajectory a
probability, its \emph{statistical weight}, that decreases with the
deviation, written:%
\begin{equation}
\exp \left( -W\left( \left\{ \mathbf{A}_{l}\right\} ,\left\{ \mathbf{\hat{A}}%
_{l}\right\} \right) \right) =\exp \left( -\int s\left( \left\{ \mathbf{A}%
_{l}\right\} ,\left\{ \mathbf{\hat{A}}_{l}\right\} ,t\right) dt\right)
\label{ST}
\end{equation}%
The integral over time accounts for summing over all deviations along the
trajectory.

By assigning a probability to each trajectory of the system, the statistical
weight allows us to describe all evolutions of the entire agent ensemble
from one state to another by summing the probabilities of the trajectories
connecting these two states. However, because this calculation requires
considering the entire agent ensemble, it is generally intractable.

On the other hand, the statistical weight can be replaced by an equivalent
in terms of fields that contains the same information and also allows the
detection of collective states that are undetectable by the statistical
weight.

\subsection{Field Translation of the framework}

We will translate the classical model described above into field terms. A
field is an object that takes into account all possible states of the system
for all variables, independently of individual agents. For example, if a set
of agents is described by their individual consumptions, the field
associated with this number of agents will overlook the details of the
agents but describe all possible states of consumption within the system.

Mathematically, a field is a random variable whose realisations run over a
space of complex-valued functions that depend on the variables of the
system. The realizations of the field represent the various states of the
system, and the associated probability for each state is given by a
statistical weight. This weight is a function of the field and is
constructed by translating $\exp \left( -W\left( \left\{ \mathbf{A}%
_{l}\right\} ,\left\{ \mathbf{\hat{A}}_{l}\right\} \right) \right) $ given
by (\ref{ST}) into field terms. To obtain the equivalent of (\ref{ST}) in
terms of fields, we proceed as follows:

For each function that characterize agents in a classical system, we write a
function whose variables are the fields. This function translates the
properties of the initial function but incorporates the change in
perspective associated with the translation. It describes the range of
possibilities for the function for each system state.

To translate formula (\ref{dnn}) and (\ref{ST}) into field terms, recall
that formula (\ref{dnn}) includes three terms that correspond to three
functions of the field. The last term in formula (\ref{dnn})\textbf{\ }%
describes the whole set of interactions both among and between two groups of
agents, and is of the form:%
\begin{equation}
\sum_{i}\sum_{j,k,l,m...}g\left( \mathbf{A}_{i}\left( t\right) ,\mathbf{A}%
_{j}\left( t\right) ,\mathbf{A}_{k}\left( t\right) ,\mathbf{\hat{A}}%
_{l}\left( t\right) ,\mathbf{\hat{A}}_{m}\left( t\right) ...\right)
\label{ptl}
\end{equation}%
It involves terms with indexed variables but no temporal derivative terms.

These terms are the easiest to translate.\ Here, agents are characterized by
their variables $\mathbf{A}_{i}\left( t\right) ,\mathbf{A}_{j}\left(
t\right) ,\mathbf{A}_{k}\left( t\right) $... and $\mathbf{\hat{A}}_{l}\left(
t\right) ,\mathbf{\hat{A}}_{m}\left( t\right) $... respectively, for
instance in our model firms and investors.

In the field translation, agents of type $\mathbf{A}_{i}\left( t\right) $
and $\mathbf{\hat{A}}_{l}\left( t\right) $ are described by a field $\Psi
\left( \mathbf{A}\right) $ and $\hat{\Psi}\left( \mathbf{\hat{A}}\right) $,
respectively. The variables indexed $j$, $k$, $l$, $m$..., such as $\mathbf{A%
}_{j}\left( t\right) $, $\mathbf{A}_{k}\left( t\right) $, $\mathbf{\hat{A}}%
_{l}\left( t\right) ,\mathbf{\hat{A}}_{m}\left( t\right) $... are replaced
by $\mathbf{A}^{\prime },\mathbf{A}^{\prime \prime }$, $\mathbf{\hat{A}}$, $%
\mathbf{\hat{A}}^{\prime }$ , and so on for all the indices in the function.
The translation of (\ref{ptl}) in terms of fields is:%
\begin{eqnarray}
&&\int g\left( \mathbf{A},\mathbf{A}^{\prime },\mathbf{A}^{\prime \prime },%
\mathbf{\hat{A},\hat{A}}^{\prime }...\right) \left\vert \Psi \left( \mathbf{A%
}\right) \right\vert ^{2}\left\vert \Psi \left( \mathbf{A}^{\prime }\right)
\right\vert ^{2}\left\vert \Psi \left( \mathbf{A}^{\prime \prime }\right)
\right\vert ^{2}\times ...d\mathbf{A}d\mathbf{A}^{\prime }d\mathbf{A}%
^{\prime \prime }...  \label{truc} \\
&&\times \left\vert \hat{\Psi}\left( \mathbf{\hat{A}}\right) \right\vert
^{2}\left\vert \hat{\Psi}\left( \mathbf{\hat{A}}^{\prime }\right)
\right\vert ^{2}\times ...d\mathbf{\hat{A}}d\mathbf{\hat{A}}^{\prime }... 
\notag
\end{eqnarray}%
where the dots stand for the products of square fields and the necessary
integration symbols.

In formula (\ref{dnn}), the terms that imply a variable temporal derivative
are of the form:%
\begin{equation}
\sum_{i}\left( \frac{d\mathbf{A}_{i}^{\left( \alpha \right) }\left( t\right) 
}{dt}-\sum_{j,k,l,m...}f^{\left( \alpha \right) }\left( \mathbf{A}_{i}\left(
t\right) ,\mathbf{A}_{j}\left( t\right) ,\mathbf{A}_{k}\left( t\right) ,%
\mathbf{\hat{A}}_{l}\left( t\right) ,\mathbf{\hat{A}}_{m}\left( t\right)
...\right) \right) ^{2}  \label{edr}
\end{equation}%
This particular form represents the dynamics of the $\alpha $-th coordinate
of a variable $\mathbf{A}_{i}\left( t\right) $ as a function of the other
agents.

Their translation is of the form:%
\begin{equation}
\int \Psi ^{\dag }\left( \mathbf{A}\right) \left( -\nabla _{\mathbf{A}%
^{\left( \alpha \right) }}\left( \frac{\sigma _{\mathbf{A}^{\left( \alpha
\right) }}^{2}}{2}\nabla _{\mathbf{A}^{\left( \alpha \right) }}+\Lambda (%
\mathbf{A})\right) \right) \Psi \left( \mathbf{A}\right) d\mathbf{A}
\label{Trl}
\end{equation}%
with:

\begin{equation}
\Lambda (\mathbf{A})=\int f^{\left( \alpha \right) }\left( \mathbf{A},%
\mathbf{A}^{\prime },\mathbf{A}^{\prime \prime },\mathbf{\hat{A},\hat{A}}%
^{\prime }...\right) \left\vert \Psi \left( \mathbf{A}^{\prime }\right)
\right\vert ^{2}\left\vert \Psi \left( \mathbf{A}^{\prime \prime }\right)
\right\vert ^{2}d\mathbf{A}^{\prime }d\mathbf{A}^{\prime \prime }\left\vert 
\hat{\Psi}\left( \mathbf{\hat{A}}\right) \right\vert ^{2}\left\vert \hat{\Psi%
}\left( \mathbf{\hat{A}}^{\prime }\right) \right\vert ^{2}d\mathbf{\hat{A}}d%
\mathbf{\hat{A}}^{\prime }  \label{bdt}
\end{equation}%
The variance $\sigma _{\mathbf{A}^{\left( \alpha \right) }}^{2}$ depicts the
probabilistic nature of the model hidden behind the field formalism. This
variance represents the characteristic level of uncertainty of the system's
dynamics. It is a parameter of the model.

Ultimately, the field description is obtained by summing all the terms
translated above and introducing a time dependency. This sum is called the
action functional, denoted $S\left( \Psi ,\Psi ^{\dag }\right) $.\ It is the
sum of terms of the form (\ref{truc}) and (\ref{Trl}). We obtain:%
\begin{eqnarray}
S\left( \Psi ,\hat{\Psi}\right) &=&\int \Psi ^{\dag }\left( \mathbf{A}%
\right) \left( -\nabla _{\mathbf{A}^{\left( \alpha \right) }}\left( \frac{%
\sigma _{\mathbf{A}^{\left( \alpha \right) }}^{2}}{2}\nabla _{\mathbf{A}%
^{\left( \alpha \right) }}+\Lambda _{1}(\mathbf{A})\right) \right) \Psi
\left( \mathbf{A}\right) d\mathbf{A} \\
&&\mathbf{+}\int \hat{\Psi}^{\dag }\left( \mathbf{\hat{A}}\right) \left(
-\nabla _{\mathbf{\hat{A}}^{\left( \alpha \right) }}\left( \frac{\sigma _{%
\mathbf{\hat{A}}^{\left( \alpha \right) }}^{2}}{2}\nabla _{\mathbf{\hat{A}}%
^{\left( \alpha \right) }}+\Lambda _{2}(\mathbf{\hat{A}})\right) \right) 
\hat{\Psi}\left( \mathbf{\hat{A}}\right) d\mathbf{\hat{A}}  \notag \\
&&+\sum_{m}\int g_{m}\left( \mathbf{A},\mathbf{A}^{\prime },\mathbf{A}%
^{\prime \prime },\mathbf{\hat{A},\hat{A}}^{\prime }...\right) \left\vert
\Psi \left( \mathbf{A}\right) \right\vert ^{2}\left\vert \Psi \left( \mathbf{%
A}^{\prime }\right) \right\vert ^{2}\left\vert \Psi \left( \mathbf{A}%
^{\prime \prime }\right) \right\vert ^{2}\times ...d\mathbf{A}d\mathbf{A}%
^{\prime }d\mathbf{A}^{\prime \prime }...  \notag \\
&&\times \left\vert \hat{\Psi}\left( \mathbf{\hat{A}}\right) \right\vert
^{2}\left\vert \hat{\Psi}\left( \mathbf{\hat{A}}^{\prime }\right)
\right\vert ^{2}\times ...d\mathbf{\hat{A}}d\mathbf{\hat{A}}^{\prime }... 
\notag
\end{eqnarray}%
where the sequence of functions $g_{m}$\ describes the various types of
interactions in the system.

The field formalism allows to compute the transition functions of the
system, i.e. the probability for any set of agent to evolve from a state to
an other in a given time.

Several results can be derived from the field action, $S\left( \Psi \right) $%
,\ and its statistical weight, $\exp \left( -S\left( \Psi \right) \right) $.
The background states that minimize $S\left( \Psi \right) $\ can be found,
which in turn allows to compute the average quantities of the system, its
associated effective action, and the transition functions. We will focus in
the following specifically on the transition functions\footnote{%
The concept of background state and its corresponding average quantities
have been presented in (Gosselin Lotz Wambst 2022).}.

\section{Transition functions}

The transition functions describe the probabilities of agents to evolve
within the system. They calculate the probability of any group of agents to
transition from any initial state to any final state. Consequently, there
exists an infinite number of transition functions, one for each possible
group of agents. For instance there can be a transition function for a
single agent of a certain type, two agents of the same type, an agent of one
type and an agent of another type, multiple agents of the same type, and a
single agent of another type, etc.

The transition functions depend on the initial and final positions of each
agent, the agent's type, and their interactions with others in the group.
Furthermore, we will see that the transition functions also depend on the
collective state of the system defined by the background field, which
reflects the dependence of the group of agents on the entire macroeconomic
system.

In principle, the transition functions could be computed using the classical
system and the classical weight $W$ by directly summing the probabilities of
some trajectories. However, this approach is intractable\textbf{\ }and lacks
interpretability. We will therefore present an alternative method to compute
these transition functions using the field model.

\subsection{Transition functions in a classical framework}

In a classical perspective, the statistical weight (\ref{ST}) can be used to
compute the transition probabilities of the system, i.e. the probabilities
for any number of agents of both types to evolve from an initial state $%
\left\{ \mathbf{A}_{l}\right\} _{l=1,...},\left\{ \mathbf{\hat{A}}%
_{l}\right\} _{l=1,...}$\textbf{\ }to a final state in a given timespan.
These transition functions describe the dynamic of the agents of the system.

To do so, we first compute the integral of equation (\ref{ST}) over all
paths between the initial and the final points considered. Defining $\left\{ 
\mathbf{A}_{l}\left( s\right) \right\} _{l=1,...,N}$ and $\left\{ \mathbf{%
\hat{A}}_{l}\left( s\right) \right\} _{l=1,...,\hat{N}}$ the sets of paths
for agents of each type, where $N$\ and $\hat{N}$\ are the numbers of agents
of each type, we consider the set of $N+\hat{N}$\ independent paths written:

\begin{equation*}
\mathbf{Z}\left( s\right) =\left( \left\{ \mathbf{A}_{l}\left( s\right)
\right\} _{l=1,...,N},\left\{ \mathbf{\hat{A}}_{l}\left( s\right) \right\}
_{l=1,...,\hat{N}}\right)
\end{equation*}%
\ The weight (\ref{ST}) can now be written $\exp \left( -W\left( \mathbf{Z}%
\left( s\right) \right) \right) $.

The transition functions $T_{t}\left( \underline{\left( \mathbf{Z}\right) },%
\overline{\left( \mathbf{Z}\right) }\right) $\ compute the probability for
the $(N,\hat{N}$\ $)$\ agents to evolve from\ the initial points $Z\left(
0\right) \equiv \underline{\mathbf{Z}}$\ to the\ final points $Z\left(
t\right) \equiv \overline{\left( \mathbf{Z}\right) }$\ during a time span $t$%
. This probability is defined by:%
\begin{equation}
T_{t}\left( \underline{\mathbf{Z}},\overline{\left( \mathbf{Z}\right) }%
\right) =\frac{1}{\mathcal{N}}\int_{\substack{ \mathbf{Z}\left( 0\right)
\equiv \underline{\mathbf{Z}}  \\ \mathbf{Z}\left( t\right) \equiv \overline{%
\left( \mathbf{Z}\right) }}}\exp \left( -W\left( \mathbf{Z}\left( s\right)
\right) \right) \mathcal{D}\left( \mathbf{Z}\left( s\right) \right)
\label{tsn}
\end{equation}%
The integration symbol $D\mathbf{Z}\left( s\right) $\ covers all sets of $%
N\times \hat{N}$\ paths constrained by $\mathbf{Z}\left( 0\right) \equiv 
\underline{\mathbf{Z}}$\ and $\mathbf{Z}\left( t\right) \equiv \overline{%
\left( \mathbf{Z}\right) }$. The normalisation factor sets the total
probability defined by the weight (\ref{ST}) to $1$ and is equal to:%
\begin{equation*}
\mathcal{N=}\int \exp \left( -W\left( \mathbf{Z}\left( s\right) \right)
\right) \mathcal{D}\mathbf{Z}\left( s\right)
\end{equation*}%
The interpretation of (\ref{tsn}) is straightforward. Instead of studying
the full trajectory of one or several agents, we compute their probability
to evolve from one configuration to another, and in average, the usual
trajectory approach remains valid.

Equation (\ref{tsn}) can be generalized to define the transition functions
for $k\leqslant N$\ and $\hat{k}\leqslant \hat{N}$\ agents of each type. The
initial and final points respectively for this set of $k+\hat{k}$\ agents
are written:%
\begin{equation*}
\mathbf{Z}\left( 0\right) ^{\left[ k,\hat{k}\right] }\equiv \underline{%
\mathbf{Z}}^{\left[ k,\hat{k}\right] }
\end{equation*}%
and:%
\begin{equation*}
\mathbf{Z}\left( t\right) ^{\left[ k,\hat{k}\right] }\equiv \overline{\left( 
\mathbf{Z}\right) }^{\left[ k,\hat{k}\right] }
\end{equation*}%
The transition function for these agents is written:%
\begin{equation*}
T_{t}\left( \underline{\left( \mathbf{Z}\right) }^{\left[ k,\hat{k}\right] },%
\overline{\left( \mathbf{Z}\right) }^{\left[ k,\hat{k}\right] }\right)
\end{equation*}%
and the generalization of equation (\ref{tsn}) is: \ 
\begin{equation}
T_{t}\left( \underline{\left( \mathbf{Z}\right) }^{\left[ k,\hat{k}\right] },%
\overline{\left( \mathbf{Z}\right) }^{\left[ k,\hat{k}\right] }\right) =%
\frac{1}{\mathcal{N}}\int_{\substack{ \mathbf{Z}\left( 0\right) ^{\left[ k,%
\hat{k}\right] }=\underline{\left( \mathbf{Z}\right) }^{\left[ k,\hat{k}%
\right] }  \\ \mathbf{Z}\left( t\right) ^{\left[ k,\hat{k}\right] }=%
\overline{\left( \mathbf{Z}\right) }^{\left[ k,\hat{k}\right] }}}\exp \left(
-W\left( \left( \mathbf{Z}\left( s\right) \right) \right) \right) \mathcal{D}%
\left( \left( \mathbf{Z}\left( s\right) \right) \right)  \label{krt}
\end{equation}%
The difference with (\ref{tsn}) is that only $k$\ paths are constrained by
their initial and final points.

Ultimately, the Laplace transform of $T_{t}\left( \underline{\left( Z\right) 
}^{\left[ k,\hat{k}\right] },\overline{\left( Z\right) }^{\left[ k,\hat{k}%
\right] }\right) $ computes the - time averaged - transition function for
agents with random lifespan of mean $\frac{1}{\alpha }$, up to a factor $%
\frac{1}{\alpha }$, and is given by:%
\begin{equation}
G_{\alpha }\left( \underline{\left( \mathbf{Z}\right) }^{\left[ k,\hat{k}%
\right] },\overline{\left( \mathbf{Z}\right) }^{\left[ k,\hat{k}\right]
}\right) =\int_{0}^{\infty }\exp \left( -\alpha t\right) T_{t}\left( 
\underline{\left( \mathbf{Z}\right) }^{\left[ k,\hat{k}\right] },\overline{%
\left( \mathbf{Z}\right) }^{\left[ k,\hat{k}\right] }\right) dt  \label{krv}
\end{equation}%
This formulation of the transition functions is relatively intractable.
Therefore, we will now propose an alternative method based on the field
model.

\subsection{Field-theoretic expression}

The transition functions (\ref{krt}) and (\ref{krv}) can be retrieved using
the$\ $field theory transition functions - or\ Green functions, which
compute the probability for a variable number $\left( k,\hat{k}\right) $\ of
agents to transition from an initial state $\underline{\left( \mathbf{%
Z,\theta }\right) }^{\left[ k,\hat{k}\right] }$\ to a final state $\overline{%
\left( \mathbf{Z,\theta }\right) }^{\left[ k,\hat{k}\right] }$, where $%
\underline{\left( \mathbf{\theta }\right) }^{\left[ k,\hat{k}\right] }$ and
\ $\overline{\left( \mathbf{\theta }\right) }^{\left[ k,\hat{k}\right] }$ are%
\textbf{\ }vectors of initial and final times for $k+\hat{k}$\textbf{\ }%
agents respectively.

We will write: 
\begin{equation*}
T_{t}\left( \underline{\left( \mathbf{Z,\theta }\right) }^{\left[ k,\hat{k}%
\right] },\overline{\left( \mathbf{Z,\theta }\right) }^{\left[ k,\hat{k}%
\right] }\right)
\end{equation*}%
\ the transition function between $\underline{\left( \mathbf{Z},\mathbf{%
\theta }\right) }^{\left[ k,\hat{k}\right] }$\ and $\overline{\left( \mathbf{%
Z},\mathbf{\theta }\right) }^{\left[ k,\hat{k}\right] }$ with $\overline{%
\left( \mathbf{\theta }\right) }_{i}<t$, $\forall i$,\ \ and: 
\begin{equation*}
G_{\alpha }\left( \underline{\left( \mathbf{Z,\theta }\right) }^{\left[ k,%
\hat{k}\right] },\overline{\left( \mathbf{Z,\theta }\right) }^{\left[ k,\hat{%
k}\right] }\right)
\end{equation*}%
\ its Laplace transform. Setting $\underline{\left( \mathbf{\theta }\right) }%
_{i}=0$\ and $\overline{\left( \mathbf{\theta }\right) }_{i}=t$\ for $%
i=1,...,k+\hat{k}$, these functions\ reduce to (\ref{krt}) or (\ref{krv}):
the probabilistic formalism of the transition functions is thus a particular
case of the field formalism definition. In the sequel we therefore will use
the term transition function indiscriminately.

The computation of the transition functions relies on the fact that $\exp
\left( -S\left( \Psi \right) \right) $\ itself represents a statistical
weight for the system. Gosselin, Lotz, Wambst (2020) showed that $S\left(
\Psi \right) $\ can be modified in a straightforward manner to include
source terms:%
\begin{equation}
S\left( \Psi ,J\right) =S\left( \Psi \right) +\int \left( J\left( Z,\theta
\right) \Psi ^{\dag }\left( Z,\theta \right) +J^{\dag }\left( Z,\theta
\right) \Psi \left( Z,\theta \right) \right) d\left( Z,\theta \right)
\label{SwtS}
\end{equation}%
where $J\left( Z,\theta \right) $ is an arbitrary complex function, or
auxiliary field.

Introducing $J\left( Z,\theta \right) $ in $S\left( \Psi ,J\right) $ allows
to compute the transition functions by successive derivatives. Actually, we
can show that:%
\begin{equation}
G_{\alpha }\left( \underline{\left( \mathbf{Z},\theta \right) }^{\left[ k,%
\hat{k}\right] },\overline{\left( \mathbf{Z},\theta \right) }^{\left[ k,\hat{%
k}\right] }\right) =\left[ \prod\limits_{l=1}^{k}\left( \frac{\delta }{%
\delta J\left( \underline{\left( \mathbf{Z},\theta \right) }_{i_{l}}\right) }%
\frac{\delta }{\delta J^{\dag }\left( \overline{\left( \mathbf{Z},\theta
\right) }_{i_{l}}\right) }\right) \int \exp \left( -S\left( \Psi ,J\right)
\right) \mathcal{D}\Psi \mathcal{D}\Psi ^{\dag }\right] _{J=J^{\dag }=0}
\label{trnsgrtx}
\end{equation}%
where the notation $\mathcal{D}\Psi \mathcal{D}\Psi ^{\dag }$ denotes an
integration over the space of functions $\Psi \left( Z,\theta \right) $ and $%
\Psi ^{\dag }\left( Z,\theta \right) $, i.e. an integral in an infinite
dimensional space. Even though these integrals can only be computed in
simple cases, a series expansion of $G_{\alpha }\left( \underline{\left( 
\mathbf{Z},\theta \right) }^{\left[ k,\hat{k}\right] },\overline{\left( 
\mathbf{Z},\theta \right) }^{\left[ k,\hat{k}\right] }\right) $ can be found
using Feynman graphs techniques.

Once $G_{\alpha }\left( \underline{\left( \mathbf{Z},\theta \right) }^{\left[
k,\hat{k}\right] },\overline{\left( \mathbf{Z},\theta \right) }^{\left[ k,%
\hat{k}\right] }\right) $ is computed, the expression of $T_{t}\left( 
\underline{\left( \mathbf{Z},\theta \right) }^{\left[ k,\hat{k}\right] },%
\overline{\left( \mathbf{Z},\theta \right) }^{\left[ k,\hat{k}\right]
}\right) $ can be\ retrieved in principle by an inverse Laplace transform.
In field theory, formula (\ref{trnsgrtx}) shows that the transition functions%
\textbf{\ }(\ref{krv}) are correlation functions of the field theory with
action $S\left( \Psi \right) $.

\section{Field-theoretic computations of transition functions}

The formula (\ref{trnsgrtx}) provides a precise and compact definition of
the transition functions for multiple agents in the system. However, in
practice, this formula is not directly applicable and does not shed much
light on the connection between the collective and microeconomic aspects of
the considered system. To calculate the dynamics of the agents, we will
proceed in three steps.\textbf{\ }

Firstly, we will minimize the system's action functional and determine the
background field, which represents the collective state of the system. Once
the background field is found, we will perform a series expansion of the
action functional around this background field, referred to as the effective
action of the system. It is with this effective action that we can compute
the transition functions for the state defined by the background field. We
will discover that each term in this expansion has an interpretation in
terms of a transition function.

Instead of directly computing the transition functions, we can consider a
series expansion of the action functional around a specific background field
of the system.

\subsection{Step 1: finding the background field}

For a particular type of agent, background fields are defined as the fields $%
\Psi _{0}\left( Z,\theta \right) $\ that maximize the statistical weight $%
\exp \left( -S\left( \Psi \right) \right) $ or, alternatively, minimize $%
S\left( \Psi \right) $: 
\begin{equation*}
\frac{\delta S\left( \Psi \right) }{\delta \Psi }\mid _{\Psi _{0}\left(
Z,\theta \right) }=0\text{, }\frac{\delta S\left( \Psi ^{\dag }\right) }{%
\delta \Psi ^{\dag }}\mid _{\Psi _{0}^{\dag }\left( Z,\theta \right) }=0
\end{equation*}%
The field $\Psi _{0}\left( Z,\theta \right) $\ represents the most probable
configuration, a specfic state of the entire system that influences the
dynamics of agents. It serves as the background state from which probability
transitions and average values can be computed. As we will see, the agents'
transitions explicitely depend on the chosen background field $\Psi
_{0}\left( Z,\theta \right) $, which represents the macroeconomic state in
which the agents evolve.

When considering two or more types of agents, the background field satisfies
the following condition:%
\begin{eqnarray*}
\frac{\delta S\left( \Psi ,\hat{\Psi}\right) }{\delta \Psi } &\mid &_{\Psi
_{0}\left( Z,\theta \right) }=0\text{, }\frac{\delta S\left( \Psi ,\hat{\Psi}%
\right) }{\delta \Psi ^{\dag }}\mid _{\Psi _{0}^{\dag }\left( Z,\theta
\right) }=0 \\
\frac{\delta S\left( \Psi ,\hat{\Psi}\right) }{\delta \hat{\Psi}} &\mid &_{%
\hat{\Psi}_{0}\left( Z,\theta \right) }=0\text{, }\frac{\delta S\left( \Psi ,%
\hat{\Psi}\right) }{\delta \hat{\Psi}^{\dag }}\mid _{\hat{\Psi}_{0}^{\dag
}\left( Z,\theta \right) }=0
\end{eqnarray*}

\subsection{Step 2: Series expansion around the background field}

In a given background state, the \emph{effective action}\footnote{%
Actually, this paper focuses on the \emph{classical} \emph{effective action}%
, which is an approximation sufficient for the computations at hand.} is the
series expansion of the field functional $S\left( \Psi \right) $\ around $%
\Psi _{0}\left( Z,\theta \right) $. We will present the expansion for one
type of agent, but generalizing it to two or several agents is
straightforward.

The series expansion around the background field simplifies the computations 
\textbf{of transition functions} and provides an interpretation of these
functions in terms of individual interactions within the collective state.\
To perform this series expansion, we decompose $\Psi $\ as: 
\begin{equation*}
\Psi =\Psi _{0}+\Delta \Psi
\end{equation*}%
and write the series expansion of the action functional: 
\begin{eqnarray}
S\left( \Psi \right) &=&S\left( \Psi _{0}\right) +\int \Delta \Psi ^{\dag
}\left( Z,\theta \right) O\left( \Psi _{0}\left( Z,\theta \right) \right)
\Delta \Psi \left( Z,\theta \right)  \label{SRP} \\
&&+\sum_{k>2}\int \prod\limits_{i=1}^{k}\Delta \Psi ^{\dag }\left(
Z_{i},\theta \right) O_{k}\left( \Psi _{0}\left( Z,\theta \right) ,\left(
Z_{i}\right) \right) \prod\limits_{i=1}^{k}\Delta \Psi \left( Z_{i},\theta
\right)  \notag
\end{eqnarray}%
The series expansion can be interpreted economically as follows. The first
term,\textbf{\ }$S\left( \Psi _{0}\right) $, describes the system of all
agents in a given macroeconomic state, $\Psi _{0}$. The other terms
potentially describe all the fluctuations or movements of the agents around
this macroeconomic state considered as given. Therefore, the expansion
around the background field represents the microeconomic reality of a
historical macroeconomic state. More precisely, it describes the range of
microeconomic possibilities allowed by a macroeconomic state.

The quadratic approximation is the first term of the expansion and can be
written as:%
\begin{equation}
S\left( \Psi \right) =S\left( \Psi _{0}\right) +\int \Delta \Psi ^{\dag
}\left( Z,\theta \right) O\left( \Psi _{0}\left( Z,\theta \right) \right)
\Delta \Psi \left( Z,\theta \right)  \label{kr}
\end{equation}%
\textbf{\ }It will allow us to find the transition functions of agents in
the historical macro state, where all interactions are averaged. The other
terms of the expansion allow us to detail the interactions within the
nebula, and are written as follows:%
\begin{equation*}
\sum_{k>2}\int \prod\limits_{i=1}^{k}\Delta \Psi ^{\dag }\left( Z_{i},\theta
\right) O_{k}\left( \Psi _{0}\left( Z,\theta \right) ,\left( Z_{i}\right)
\right) \prod\limits_{i=1}^{k}\Delta \Psi \left( Z_{i},\theta \right)
\end{equation*}%
\textbf{\ }They detail, given the historical macroeconomic state, how the
interactions of two or more agents can impact the dynamics of these agents.
Mathematically, this corresponds to correcting the transition probabilities
calculated in the quadratic approximation.

Here, we provide an interpretation of the third and fourth-order terms.

The third order introduces possibilities for an agent, during its
trajectory, to split into two, or conversely, for two agents to merge into
one. In other words, the third-order terms take into account or reveal, in
the historical macroeconomic environment, the possibilities for any agent to
undergo modifications along its trajectory. However, this assumption has
been excluded from our model.

The fourth order reveals that in the macroeconomic environment, due to the
presence of other agents and their tendency to occupy the same space, all
points in space will no longer have the same probabilities for an agent. In
fact, the fourth-order terms reveal the notion of geographical or sectoral
competition and potentially intertemporal competition. These terms describe
the interaction between two agents crossing paths, which leads to a
deviation of their trajectories due to the interaction.

We do not interpret higher-order terms, but the idea is similar. The
even-order terms (2n) describe interactions among n agents that modify their
trajectories. The odd-order terms modify the trajectories but also include
the possibility of agents reuniting or splitting into multiple agents. We
will see in more detail how these terms come into play in the transition
functions.

\subsection{Step 3: Computation of the transition functions}

\subsubsection{Quadratic approximation}

In the first approximation, transition functions in a given background field%
\textbf{\ }$\Psi _{0}\left( Z,\theta \right) $\textbf{\ }can be computed by
replacing $S\left( \Psi \right) $\ in (\ref{trnsgrtx}), with its quadratic
approximation (\ref{kr}). In formula (\ref{kr}), $O\left( \Psi _{0}\left(
Z,\theta \right) \right) $\ is a differential operator of second order. This
operator depends explicitly on $\Psi _{0}\left( Z,\theta \right) $. As a
consequence, transition functions and agent dynamics explicitly depend on
the collective state of the system.\ In this approximation, the formula for
the transition functions (\ref{trnsgrtx}) becomes:\ 
\begin{eqnarray}
G_{\alpha }\left( \underline{\left( Z,\theta \right) }^{\left[ k\right] },%
\overline{\left( Z,\theta \right) }^{\left[ k\right] }\right) &=&\left[
\prod\limits_{l=1}^{k}\left( \frac{\delta }{\delta J\left( \underline{\left(
Z,\theta \right) }_{i_{l}}\right) }\frac{\delta }{\delta J^{\dag }\left( 
\overline{\left( Z,\theta \right) }_{i_{l}}\right) }\right) \right. \\
&&\times \left. \int \exp \left( -\int \Delta \Psi ^{\dag }\left( Z,\theta
\right) O\left( \Psi _{0}\left( Z,\theta \right) \right) \Delta \Psi \left(
Z,\theta \right) \right) \mathcal{D}\Psi \mathcal{D}\Psi ^{\dag }\right]
_{J=J^{\dag }=0}  \notag
\end{eqnarray}%
Using this formula, we can show that the one-agent transition function is
given by:%
\begin{equation}
G_{\alpha }\left( \underline{\left( Z,\theta \right) }^{\left[ 1\right] },%
\overline{\left( Z,\theta \right) }^{\left[ 1\right] }\right) =O^{-1}\left(
\Psi _{0}\left( Z,\theta \right) \right) \left( \underline{\left( Z,\theta
\right) }^{\left[ 1\right] },\overline{\left( Z,\theta \right) }^{\left[ 1%
\right] }\right)  \label{rk}
\end{equation}%
where: 
\begin{equation*}
O^{-1}\left( \Psi _{0}\left( Z,\theta \right) \right) \left( \underline{%
\left( Z,\theta \right) }^{\left[ 1\right] },\overline{\left( Z,\theta
\right) }^{\left[ 1\right] }\right)
\end{equation*}%
\ is the kernel of the inverse operator $O^{-1}\left( \Psi _{0}\left(
Z,\theta \right) \right) $. It can be seen as the $\left( \underline{\left(
Z,\theta \right) }^{\left[ 1\right] },\overline{\left( Z,\theta \right) }^{%
\left[ 1\right] }\right) $\ matrix element of $O^{-1}\left( \Psi _{0}\left(
Z,\theta \right) \right) $\footnote{%
The differential operator $O\left( \Psi _{0}\left( Z,\theta \right) \right) $
can be seen as an infinite dimensional matrix indexed by the double
(infinite) entries $\left( \underline{\left( Z,\theta \right) }^{\left[ 1%
\right] },\overline{\left( Z,\theta \right) }^{\left[ 1\right] }\right) $.
With this description, the kernel $O^{-1}\left( \Psi _{0}\left( Z,\theta
\right) \right) \left( \underline{\left( Z,\theta \right) }^{\left[ 1\right]
},\overline{\left( Z,\theta \right) }^{\left[ 1\right] }\right) $ is the $%
\left( \underline{\left( Z,\theta \right) }^{\left[ 1\right] },\overline{%
\left( Z,\theta \right) }^{\left[ 1\right] }\right) $ element of the inverse
matrix.}.

More generally, the $k$-agents transition functions are the product of
individual transition functions:%
\begin{equation}
G_{\alpha }\left( \underline{\left( Z,\theta \right) }^{\left[ k\right] },%
\overline{\left( Z,\theta \right) }^{\left[ k\right] }\right)
=\prod\limits_{i=1}^{k}G_{\alpha }\left( \underline{\left( Z,\theta \right)
_{i}}^{\left[ 1\right] },\overline{\left( Z,\theta \right) _{i}}^{\left[ 1%
\right] }\right)  \label{gnr}
\end{equation}%
The above formula shows that in the quadratic approximation, the transition
probability from one state to another for a group is the product of
individual transition probabilities. In this approximation, the trajectories
of these agents are therefore independent. The agents do not interact with
each other and only interact with the environment described by the
background field.

The quadratic approximation must be corrected to account for individual
interactions within the group, by including higher-order terms in the
expansion of the action.

\subsubsection{Higher-order corrections}

To compute the effects of interactions between agents of a given group, we
consider terms of order greater than $2$ in the effective action. These
terms\ modify the transition functions. Writing the expansion:%
\begin{equation*}
\exp \left( -S\left( \Psi \right) \right) =\exp \left( -\left( S\left( \Psi
_{0}\right) +\int \Delta \Psi ^{\dag }\left( Z,\theta \right) O\left( \Psi
_{0}\left( Z,\theta \right) \right) \right) \right) \left(
1+\sum_{n\geqslant 1}\frac{A^{n}}{n!}\right)
\end{equation*}%
where:%
\begin{equation*}
A=\sum_{k>2}\int \prod\limits_{i=1}^{k}\Delta \Psi ^{\dag }\left(
Z_{i},\theta \right) O\left( \Psi _{0}\left( Z,\theta \right) ,\left(
Z_{i}\right) \right) \prod\limits_{i=1}^{k}\Delta \Psi \left( Z_{i},\theta
\right)
\end{equation*}%
is the sum of all possible interaction terms, leads to the series expansion
of (\ref{trnsgrtx}):%
\begin{eqnarray}
G_{\alpha }\left( \underline{\left( Z,\theta \right) }^{\left[ k\right] },%
\overline{\left( Z,\theta \right) }^{\left[ k\right] }\right) &=&\left[
\prod\limits_{l=1}^{k}\left( \frac{\delta }{\delta J\left( \underline{\left(
Z,\theta \right) }_{i_{l}}\right) }\frac{\delta }{\delta J^{\dag }\left( 
\overline{\left( Z,\theta \right) }_{i_{l}}\right) }\right) \right.
\label{trg} \\
&&\left. \int \exp \left( -\int \Delta \Psi ^{\dag }\left( Z,\theta \right)
O\left( \Psi _{0}\left( Z,\theta \right) \right) \Delta \Psi \left( Z,\theta
\right) \right) \left( 1+\sum_{n\geqslant 1}\frac{A^{n}}{n!}\right) \mathcal{%
D}\Psi \mathcal{D}\Psi ^{\dag }\right] _{J=J^{\dag }=0}  \notag
\end{eqnarray}%
These corrections can be computed using graphs' expansion. \ 

More precisely, the first term of the series:%
\begin{equation}
\left[ \prod\limits_{l=1}^{k}\left( \frac{\delta }{\delta J\left( \underline{%
\left( Z,\theta \right) }_{i_{l}}\right) }\frac{\delta }{\delta J^{\dag
}\left( \overline{\left( Z,\theta \right) }_{i_{l}}\right) }\right) \int
\exp \left( -\int \Delta \Psi ^{\dag }\left( Z,\theta \right) O\left( \Psi
_{0}\left( Z,\theta \right) \right) \Delta \Psi \left( Z,\theta \right)
\right) \mathcal{D}\Psi \mathcal{D}\Psi ^{\dag }\right] _{J=J^{\dag }=0}
\end{equation}%
is a transition function in the quadratic approximation. The other
contributions of the series expansion correct the approximated $n$ agents
transtns functions (\ref{gnr}).

Typically a contribution:%
\begin{eqnarray}
G_{\alpha }\left( \underline{\left( Z,\theta \right) }^{\left[ k\right] },%
\overline{\left( Z,\theta \right) }^{\left[ k\right] }\right) &=&\left[
\prod\limits_{l=1}^{k}\left( \frac{\delta }{\delta J\left( \underline{\left(
Z,\theta \right) }_{i_{l}}\right) }\frac{\delta }{\delta J^{\dag }\left( 
\overline{\left( Z,\theta \right) }_{i_{l}}\right) }\right) \right. \\
&&\left. \int \exp \left( -\int \Delta \Psi ^{\dag }\left( Z,\theta \right)
O\left( \Psi _{0}\left( Z,\theta \right) \right) \Delta \Psi \left( Z,\theta
\right) \right) \frac{A^{n}}{n!}\mathcal{D}\Psi \mathcal{D}\Psi ^{\dag }%
\right] _{J=J^{\dag }=0}  \notag
\end{eqnarray}%
can be depicted by a graph. The power $\frac{A^{n}}{n!}$ translates that
agents interact $n$ times along their path. The trajectories of each agent
of the group is broken $n$ times between its initial and final points. At
each time of interaction the trajectories of agents are deviated. To such a
graph is associated a probility that modifies the quadratic approximation
transition functions.

In the sequel we will only focus on the first order corrections to the
two-agents transition functions.

\part*{Application to a model of capital accumulation}

\section{The microeconomic framework}

The microeconomic framework presented in this section will be turned into a
field model. Since our goal is to picture the interactions between the real
and the financial economy, we consider two groups of agents, producers, and
investors. We will refer to producers or firms $i$ indistinctively in the
following sections, and use the upper script $\symbol{94}$ for variables
describing investors. This microframework is similar to the one developed in
(Gosselin, Lotz, Wambst 2022), with the exception of two minor adjustments
that do not affect the collective level, but that will provide further
insight into individual dynamics.

\subsection{Producers}

Producers are modeled by firms evolving in a sector space composed of an
infinite number of sectors. In this framework, the notions of firm and
sector are flexible. A single firm with subsidiaries in different countries
or offering differentiated products can be modeled as independent firms.
Similarly, a sector refers to a group of firms with similar activities, but
this criterion is loose: sectors can be decomposed into sectors per country,
to account for local specificities, or in several sectors for that matter.

Producers move across sectors described by a vector space of arbitrary
dimension. The position of producer $i$ in this space is denoted $X_{i}$ and
his physical capital, $K_{i}$. Producers are defined by these two variables,
which are both subject to dynamic changes. Producers may change their
capital stocks over time or altogether shift sectors.

Each firm produces a single differentiated good.\ However, in the following,
we will merely consider the return each producer may provide to its
investors.

The return of producer $i$ at time $t$, denoted $r_{i}$, depends on $K_{i}$, 
$X_{i}$ and on the level of competition in the sector.\ It is written: 
\begin{equation}
r_{i}=r\left( K_{i},X_{i}\right) -\gamma \sum_{j}\delta \left(
X_{i}-X_{j}\right) \frac{K_{j}}{K_{i}}  \label{dvd}
\end{equation}%
The first term is an arbitrary function that depends on the sector and the
level of capital per firm in this sector.\ It represents the return of
capital in a specific sector $X_{i}$ under no competition. We deliberately
keep the form of $r\left( K_{i},X_{i}\right) $\ unspecified, since most of
the results of the model rely on the general properties of the functions
involved. When needed, we will give a standard Cobb-Douglas form to\ the
returns $r\left( K_{i},X_{i}\right) $.\ The second term in (\ref{dvd}) is
the decreasing return of capital. In any given sector, it is proportional to
both the number of competitors and the specific level of capital per firm
used.

We also assume that, for all $i$, firm $i$ has a market valuation defined by
both its price, $P_{i}$, and the variation of this price on financial
markets, $\dot{P}_{i}$.\ This variation is itself assumed to be a function
of an expected long-term return denoted $R\left( K_{i},X_{i}\right) $, and
more precisely the relative return $\bar{R}\left( K_{i},X_{i}\right) $\ of
firm $i$\ against the whole set of firms in its sector: 
\begin{equation}
\frac{\dot{P}_{i}}{P_{i}}=F_{1}\left( \bar{R}\left( K_{i},X_{i}\right)
\right)  \label{pr}
\end{equation}%
with:%
\begin{equation}
\bar{R}\left( K_{i},X_{i}\right) =\frac{R\left( K_{i},X_{i}\right) }{%
\sum_{l}R\left( K_{l},X_{l}\right) }  \label{RBR}
\end{equation}%
The function $F_{1}$ is arbitrary and reflects the preferences of the market
relatively to the firm's relative returns.

We assume that firms shift their production in the sector space according to
returns, in the direction of the gradient of the expected long-term return $%
R\left( K_{i},X_{i}\right) $.\ Yet, the accumulation of agents at any point
of the space creates a repulsive force, so that the evolution of $X_{i}$
minimizes, up to some shocks, the following quantity:%
\begin{equation}
L_{i}\left( X_{i},\frac{dX_{i}}{dt}\right) =\left( \frac{dX_{i}}{dt}-\nabla
_{X}R\left( K_{i},X_{i}\right) H\left( K_{i}\right) \right) ^{2}+\tau \frac{%
K_{X_{i}}}{K_{i}}\sum_{j}\delta \left( X_{i}-X_{j}\right)  \label{dnp}
\end{equation}%
where $K_{X_{i}}$ is the average capital per firm in sector $X_{i}$.

When $\tau =0$, there are no repulsive forces and the move towards the
gradient of $R$ is given by the expression:

\begin{equation*}
\frac{dX_{i}}{dt}=\nabla _{X}R\left( K_{i},X_{i}\right) H\left( K_{i}\right)
\end{equation*}%
When $\tau \neq 0$, repulsive forces deviate the trajectory.\ The dynamic
equation associated to the minimization of (\ref{dnp}) is given by the
general formula of the dynamic optimization:%
\begin{equation}
\frac{d}{dt}\frac{\partial }{\partial \frac{dX_{i}}{dt}}L_{i}\left( X_{i},%
\frac{dX_{i}}{dt}\right) =\frac{\partial }{\partial X_{i}}L_{i}\left( X_{i},%
\frac{dX_{i}}{dt}\right)  \label{lgd}
\end{equation}%
This last equation does not need to be developed further, since formula (\ref%
{dnp}) is sufficient to switch to the field description of the system. Note
for later purpose that the expression $\frac{dX_{i}}{dt}$\ stands for the
continuous version of a discrete variation, $X_{i}\left( t+1\right)
-X_{i}\left( t\right) $.

Finally, it should be noted that we have introduced a difference from our
previous work. In (Gosselin, Lotz, Wambst, 2022), we assumed a constant
competition term\textbf{\ }$\tau $ between firms for simplicity. However, in
this work, we focus on individual dynamics within the background field, and
depart slightly from this assumption.\textbf{\ }We will consider that the
strength of competition depends on the value of the firm's capital, and
replace:%
\begin{equation}
\tau \rightarrow \tau \frac{K_{X_{i}}}{K_{i}}  \label{tfc}
\end{equation}%
\textbf{\ }This modification does not affect the system on average, i.e. at
the collective level, but it allows for more precise results about the
effect of interactions at the individual level. Formula (\ref{tfc}) models
that the\textbf{\ }lower a firm's capital is compared to the sector average,
the stronger the effect of competition.

\subsection{Investors}

Each investor $j$ is defined by his level of capital $\hat{K}_{j}$ and his
position $\hat{X}_{j}$ in the sector space. Investors can invest in the
entire sector space, but tend to invest in sectors close to their position.

Besides, investors tend to diversify their capital: each investor $j$ chose
to allocate parts of his entire capital $\hat{K}_{j}$ between various firms $%
i$. The capital allocated by investor $j$ to firm $i$ is denoted $\hat{K}%
_{j}^{\left( i\right) }$, and given by:

\begin{equation}
\hat{K}_{j}^{\left( i\right) }\left( t\right) =\left( \hat{F}_{2}\left(
s_{i},R\left( K_{i},X_{i}\right) \right) \hat{K}_{j}\right) \left( t\right)
\label{grandf2}
\end{equation}%
where:%
\begin{equation}
\hat{F}_{2}\left( s_{i},R\left( K_{i},X_{i}\right) ,\hat{X}_{j}\right) =%
\frac{F_{2}\left( s_{i},R\left( K_{i},X_{i}\right) \right) G\left( X_{i}-%
\hat{X}_{j}\right) }{\sum_{l}F_{2}\left( s_{l},R\left( K_{l},X_{l}\right)
\right) G\left( X_{l}-\hat{X}_{j}\right) }  \label{FRLV}
\end{equation}%
The function $F_{2}$ is arbitrary. It depends on the expected return of firm 
$i$ and on the distance between sectors $X_{i}$ and $\hat{X}_{j}$. The
function $\hat{F}_{2}\left( s_{i},R\left( K_{i},X_{i}\right) ,\hat{X}%
_{j}\right) $\ is the relative version of $F_{2}$ \textbf{t}hat reflects the
dependence of investments on the relative attractiveness of firms.

Compared to our previous work, we introduced a variable $s_{i}$ that
individualize the shape of $\hat{F}_{2}$ for the firm, whereas in (Gosselin,
Lotz, Wambst 2022) $\hat{F}_{2}$ was only capital and sector-dependent. On
average, this modification does not change our previous results regarding
the background field. However, it enables us to understand the role of
investors' perception of a particular firm in individual transitions.
Mathematically\textbf{, }$s$\ is a variable that determines the shape of the
attractiveness function $\hat{F}_{2}$\ for the firm.\textbf{\ }While a
dynamics equation for this variable could be introduced, treating $s$\ as
static is sufficient for studying the influence of this shape parameter on
capital dynamics, without specifying its interaction with thecapital level
and sector shifts.

We now define $\varepsilon $ as the time scale for capital accumulation.\
The variation in capital of investor $j$\textbf{\ }between $t$\ and $%
t+\varepsilon $\ is the sum of two terms: the short-term returns $r_{i}$\ of
the firms\ in which $j$\ invested, and the stock price variations of these
same firms:

\begin{equation}
\hat{K}_{j}\left( t+\varepsilon \right) -\hat{K}_{j}\left( t\right)
=\sum_{i}\left( r_{i}+\frac{\dot{P}_{i}}{P_{i}}\right) \hat{K}_{j}^{\left(
i\right) }=\sum_{i}\left( r_{i}+F_{1}\left( \bar{R}\left( K_{i},X_{i}\right)
,\frac{\dot{K}_{i}\left( t\right) }{K_{i}\left( t\right) }\right) \right) 
\hat{K}_{j}^{\left( i\right) }  \label{fsn}
\end{equation}%
Note that in equation (\ref{dnp}), the time scale of firms' displacement
within the sectors space is normalized to one. On the contrary, in equation (%
\ref{fsn}) we assume that the time scale for investors shifts is $%
\varepsilon $, where $\varepsilon <<1$, meaning that capital dynamics is
faster than firms' mobility within the sectors space. To rewrite (\ref{fsn})
on the same time-span as $\frac{dX_{i}}{dt}$, we write:

\begin{eqnarray*}
\hat{K}_{j}\left( t+1\right) -\hat{K}_{j}\left( t\right) &=&\sum_{k=1}^{%
\frac{1}{\varepsilon }}\hat{K}_{j}\left( t+k\varepsilon \right) -\hat{K}%
_{j}\left( t\right) \\
&=&\sum_{k=1}^{\frac{1}{\varepsilon }}\sum_{i}\left( r_{i}+\frac{\dot{P}_{i}%
}{P_{i}}\right) \hat{K}_{j}^{\left( i\right) }\left( t+k\varepsilon \right)
\\
&\simeq &\frac{1}{\varepsilon }\sum_{i}\left( r_{i}+F_{1}\left( \bar{R}%
\left( K_{i},X_{i}\right) ,\frac{\dot{K}_{i}\left( t\right) }{K_{i}\left(
t\right) }\right) \right) \hat{K}_{j}^{\left( i\right) }
\end{eqnarray*}%
where the quantities in the sum have to be understood as averages over the
time span\textbf{\ }$\left[ t,t+1\right] $. Using equation (\ref{pr}),
equation (\ref{fsn}) becomes in the continuous approximation: 
\begin{equation}
\frac{d}{dt}\hat{K}_{j}\left( t\right) =\frac{1}{\varepsilon }\sum_{i}\left(
r_{i}+F_{1}\left( \frac{R\left( K_{i},X_{i}\right) }{\sum_{l}\delta \left(
X_{l}-X_{i}\right) R\left( K_{l},X_{l}\right) },\frac{\dot{K}_{i}\left(
t\right) }{K_{i}\left( t\right) }\right) \right) \hat{F}_{2}\left(
s_{i},R\left( K_{i},X_{i}\right) ,\hat{X}_{j}\right) \hat{K}_{j}  \label{nfs}
\end{equation}%
where $\frac{d}{dt}\hat{K}_{j}\left( t\right) =\hat{K}_{j}\left( t+1\right) -%
\hat{K}_{j}\left( t\right) $\ is now normalized to the time scale of $\frac{%
dX_{i}}{dt}$, i.e. 1.

\subsection{Interactions between financial and physical capital}

The entire financial capital is, at any time, completely allocated by
investors between firms. For producers, there is no alternative source of
financing: self-financing is discarded, since it amounts to consider two
agents, a producer and an investor, as one. The physical capital of a any
given firm is thus the sum of all capital allocated to this firm by all its
investors. Physical capital entirely depends on the arbitrage and
allocations of the financial sector. Firms do not possess their capital:
they return it fully at the end of each period with a dividend, though
possibly negative. Investors then entirely reallocate their capital between
firms at the beginning of the next period.

This set up may not be fully accurate in the short-run but, since physical
capital cannot subsist without investment, it holds in the long-run.\ When
investors choose not to finance a firm, this firm is bound to disappear in
the long run. Under these assumptions, the following identity holds\textbf{:}%
\begin{equation}
K_{i}\left( t+\varepsilon \right) =\sum_{j}\hat{K}_{j}^{\left( i\right)
}=\sum_{j}\hat{F}_{2}\left( s_{i},R\left( K_{i}\left( t\right) ,X_{i}\left(
t\right) \right) ,\hat{X}_{j}\left( t\right) \right) \hat{K}_{j}\left(
t\right)  \label{phs}
\end{equation}%
where $K_{i}$ stands for the physical capital of firm $i$ at time $t$, and $%
\sum_{j}\hat{K}_{j}^{\left( i\right) }$ for the sum of capital invested in
firm $i$ by investors $j$. Recall that the parameter $\varepsilon $ accounts
for the specific time scale of capital accumulation.\ It differs from that
of mobility within the sector space (\ref{dnp}), which is normalized to one.

The dynamics (\ref{phs}) rewrites:%
\begin{equation}
\frac{K_{i}\left( t+\varepsilon \right) -K_{i}\left( t\right) }{\varepsilon }%
=\frac{1}{\varepsilon }\left( \sum_{j}\hat{F}_{2}\left( s_{i},R\left(
K_{i}\left( t\right) ,X_{i}\left( t\right) \right) ,\hat{X}_{j}\left(
t\right) \right) \hat{K}_{j}\left( t\right) -K_{i}\left( t\right) \right)
\label{dnK}
\end{equation}%
Using the same token as in the derivation of (\ref{nfs}), we obtain in the
continuous approximation:%
\begin{equation}
\frac{d}{dt}K_{i}\left( t\right) +\frac{1}{\varepsilon }\left( K_{i}\left(
t\right) -\sum_{j}\hat{F}_{2}\left( s_{i},R\left( K_{i}\left( t\right)
,X_{i}\left( t\right) \right) ,\hat{X}_{j}\left( t\right) \right) \hat{K}%
_{j}\left( t\right) \right) =0  \label{dnk}
\end{equation}%
where $\frac{d}{dt}K_{i}\left( t\right) $\ stands for $K_{i}\left(
t+1\right) -K_{i}\left( t\right) $.

\subsection{Capital allocation dynamics}

Investors choose to allocate their capital within sectors, and may modify
their portfolio according to the returns of the sector or firms they invest
in. This is modelled by a move along the sectors' space in the direction of
the gradient of $R\left( K_{i},X_{i}\right) $ Investors located within a
particular sector shift to neighbouring sectors that offer higher expected
long-term returns. The shift of $\hat{X}_{j}$ is described by a dynamic
equation that models its movement over time.

\begin{equation}
\frac{d}{dt}\hat{X}_{j}-\frac{1}{\sum_{i}\delta \left( X_{i}-\hat{X}%
_{j}\right) }\sum_{i}\left( \nabla _{\hat{X}}F_{0}\left( R\left( K_{i},\hat{X%
}_{j}\right) \right) +\nu \nabla _{\hat{X}}F_{1}\left( \bar{R}\left( K_{i},%
\hat{X}_{j}\right) \right) \right) =0  \label{prd}
\end{equation}%
where the factor $\sum_{i}\delta \left( X_{i}-\hat{X}_{j}\right) $ is the
agents' density in the sector $\hat{X}_{j}$, so that the more competitors in
a sector, the slower the move.

In equation (\ref{prd}), $F_{0}\left( R\left( K_{i},\hat{X}_{j}\right)
\right) $ is a function of long-term returns representing the tendency of
investors to invest in sectors with the highest returns. The gradient $%
\nabla _{\hat{X}}F_{0}\left( R\left( K_{i},\hat{X}_{j}\right) \right) $%
\textbf{\ }induces a shift of investors towards sectors with the highest
returns, as described in equation (\ref{prd}).

The function $F_{1}\left( \bar{R}\left( K_{i},\hat{X}_{j}\right) \right) $\
measures the expected variations in the stock prices of the firm, so that $%
\nu \nabla _{\hat{X}}F_{1}\left( \bar{R}\left( K_{i},\hat{X}_{j}\right)
\right) $\ describes the investors' tendency to shift towards stocks with
the highest price-dividend ratio.

\section{Translation of the framework}

Let us now use the method summed-up above to translate the classical
framework developed in section 3. The details are given in (Gosselin Lotz
Wambst 2022).

There are two types of agents in our model, firms and investors. Each type
of agent is described by two dynamic equations, thus, there are four
minimization functions to find. These minimization functions will be
translated into four functionals of two independent fields\footnote{%
The term functional refers to a function of a function, i.e. a function
whose argument is itself a function.}, one for producers $\Psi \left(
K,X\right) $\footnote{%
Note that the field should be writen $\Psi \left( K,X,s\right) $ to include
the variable $s$. However, since $s_{i}$ is static, we can omit it when
writing the fields. It will only appear as a parameter in the formula for $%
\hat{F}_{2}$ and in the transition functions.} and one for investors $\hat{%
\Psi}\left( \hat{K},\hat{X}\right) $. The sum of these four functionals will
be the "field action functional" that describes the whole system in terms of
fields\footnote{%
Details about the probabilistic step will be given as a reminder along the
text and in appendix 1.}.

\subsubsection{The Real Economy}

We show in (Gosselin Lotz Wambst 2022) that the capital allocation dynamics (%
\ref{dnp}) and the capital accumulation dynamics (\ref{dnk}) have the
following field equivalents.

\paragraph{Field action functional for physical capital allocation}

\begin{eqnarray}
S_{1} &=&-\int \Psi ^{\dag }\left( K,X\right) \nabla _{X}\left( \frac{\sigma
_{X}^{2}}{2}\nabla _{X}-\nabla _{X}R\left( K,X\right) H\left( K\right)
\right) \Psi \left( K,X\right) dKdX  \label{SN} \\
&&+\tau \int \left\vert \Psi \left( K^{\prime },X\right) \right\vert
^{2}\left\vert \Psi \left( K,X\right) \right\vert ^{2}dK^{\prime }dKdX 
\notag
\end{eqnarray}

\paragraph{Field action functional for physical capital accululation}

\begin{equation}
S_{2}=-\int \Psi ^{\dag }\left( K,X\right) \nabla _{K}\left( \frac{\sigma
_{K}^{2}}{2}\nabla _{K}+\frac{1}{\varepsilon }\left( K-\int \hat{F}%
_{2}\left( s,R\left( K,\hat{X}\right) ,\hat{X}\right) \hat{K}\left\vert \hat{%
\Psi}\left( \hat{K},\hat{X}\right) \right\vert ^{2}d\hat{K}d\hat{X}\right)
\right) \Psi \left( K,X\right)  \label{SD}
\end{equation}%
with:%
\begin{equation}
\hat{F}_{2}\left( s,R\left( K,\right) ,\hat{X}\right) =\frac{F_{2}\left(
s,R\left( K,X\right) ,\hat{X}\right) G\left( X-\hat{X}\right) }{\int
F_{2}\left( R\left( K^{\prime },X^{\prime }\right) ,\hat{X}\right) G\left(
X^{\prime }-\hat{X}\right) \left\vert \Psi \left( K^{\prime },X^{\prime
}\right) \right\vert ^{2}d\left( K^{\prime },X^{\prime }\right) }
\label{KDR}
\end{equation}%
Here:%
\begin{equation*}
F_{2}\left( R\left( K^{\prime },X^{\prime }\right) ,\hat{X}\right)
\end{equation*}%
is the average over the parameter $s^{\prime }$\ of:%
\begin{equation*}
F_{2}\left( s^{\prime },R\left( K^{\prime },X^{\prime }\right) ,\hat{X}%
\right)
\end{equation*}

\subsubsection{Financial markets}

Financial capital dynamics and financial capital allocation are given by
equations (\ref{nfs}) and (\ref{prd}) respectively.\ Both expressions
include a time derivative and are thus of type (\ref{edr}). Their
translation is straightforward.

\paragraph{Field action functional for financial capital dynamics}

\begin{eqnarray}
S_{3} &=&-\int \hat{\Psi}^{\dag }\left( \hat{K},\hat{X}\right) \nabla _{\hat{%
K}}\left( \frac{\sigma _{\hat{K}}^{2}}{2}\nabla _{\hat{K}}-\frac{\hat{K}}{%
\varepsilon }\int \left( r\left( K,X\right) -\gamma \frac{\int K^{\prime
}\left\Vert \Psi \left( K^{\prime },X\right) \right\Vert ^{2}}{K}\right.
\right. \\
&&\left. \left. +F_{1}\left( \bar{R}\left( K,X\right) ,\Gamma \left(
K,X\right) \right) \right) \hat{F}_{2}\left( s,R\left( K,X\right) ,\hat{X}%
\right) \left\Vert \Psi \left( K,X\right) \right\Vert ^{2}d\left( K,X\right)
\right) \hat{\Psi}\left( \hat{K},\hat{X}\right)  \notag
\end{eqnarray}%
where:%
\begin{eqnarray}
\bar{R}\left( K,X\right) &=&\frac{R\left( K,X\right) }{\int R\left(
K^{\prime },X^{\prime }\right) \left\Vert \Psi \left( K^{\prime },X^{\prime
}\right) \right\Vert ^{2}d\left( K^{\prime },X^{\prime }\right) }
\label{RPN} \\
\Gamma \left( K,X\right) &=&\frac{\int \hat{F}_{2}\left( s,R\left(
K,X\right) ,\hat{X}\right) \hat{K}\left\vert \hat{\Psi}\left( \hat{K},\hat{X}%
\right) \right\vert ^{2}d\hat{K}d\hat{X}}{K}-1  \label{Gmm}
\end{eqnarray}

\paragraph{Field action for financial capital allocation}

\begin{eqnarray}
S_{4} &=&-\int \hat{\Psi}^{\dag }\left( \hat{K},\hat{X}\right)  \label{SQ} \\
&&\times \left( \nabla _{\hat{X}}\sigma _{\hat{X}}^{2}\left( \nabla _{\hat{X}%
}-\int \left( \frac{\nabla _{\hat{X}}F_{0}\left( R\left( K,\hat{X}\right)
\right) +\nu \nabla _{\hat{X}}F_{1}\left( \bar{R}\left( K,X\right) ,\Gamma
\left( K,X\right) \right) }{\int \left\Vert \Psi \left( K^{\prime },\hat{X}%
\right) \right\Vert ^{2}dK^{\prime }}\right) \left\Vert \Psi \left( K,\hat{X}%
\right) \right\Vert ^{2}dK\right) \right) \hat{\Psi}\left( \hat{K},\hat{X}%
\right)  \notag
\end{eqnarray}

\subsubsection{Gathering contributions: the action functional}

Once these translations are performed, the action functional of the system
is described by the sum of all the contributions, i.e. (\ref{SN}),(\ref{SD}%
),(\ref{ST}),(\ref{SQ}):%
\begin{equation*}
S=S_{1}+S_{2}+S_{3}+S_{4}
\end{equation*}%
We can assume to simplify that investors invest in only one sector, which
translates into the following condition:%
\begin{equation}
G\left( X-\hat{X}\right) =\delta \left( X-\hat{X}\right)  \label{smp}
\end{equation}%
This simplification does not reduce the generality of our model since an
investor acting in several sectors can be modelled as an aggregation of
several investors possibly moving from one sector to another.

A compact form for the action functional $S$ can be written: 
\begin{eqnarray}
S &=&-\int \Psi ^{\dag }\left( K,X\right) \left( \nabla _{X}\left( \frac{%
\sigma _{X}^{2}}{2}\nabla _{X}-\nabla _{X}R\left( K,X\right) H\left(
K\right) \right) -\tau \left( \int \left\vert \Psi \left( K^{\prime
},X\right) \right\vert ^{2}dK^{\prime }\right) \right.  \label{fcn} \\
&&+\left. \nabla _{K}\left( \frac{\sigma _{K}^{2}}{2}\nabla _{K}+u\left(
K,X,\Psi ,\hat{\Psi}\right) \right) \right) \Psi \left( K,X\right) dKdX 
\notag \\
&&-\int \hat{\Psi}^{\dag }\left( \hat{K},\hat{X}\right) \left( \nabla _{\hat{%
K}}\left( \frac{\sigma _{\hat{K}}^{2}}{2}\nabla _{\hat{K}}-\hat{K}f\left( 
\hat{X},\Psi ,\hat{\Psi}\right) \right) +\nabla _{\hat{X}}\left( \frac{%
\sigma _{\hat{X}}^{2}}{2}\nabla _{\hat{X}}-g\left( K,X,\Psi ,\hat{\Psi}%
\right) \right) \right) \hat{\Psi}\left( \hat{K},\hat{X}\right)  \notag
\end{eqnarray}%
where we have defined:%
\begin{eqnarray}
u\left( K,X,\Psi ,\hat{\Psi}\right) &=&\frac{1}{\varepsilon }\left( K-\int 
\hat{F}_{2}\left( s,R\left( K,X\right) \right) \hat{K}\left\vert \hat{\Psi}%
\left( \hat{K},X\right) \right\vert ^{2}d\hat{K}\right)  \label{fcs} \\
f\left( \hat{X},\Psi ,\hat{\Psi}\right) &=&\frac{1}{\varepsilon }\int \left(
r\left( K,X\right) -\frac{\gamma \int K^{\prime }\left\vert \Psi \left(
K,X\right) \right\vert ^{2}}{K}+F_{1}\left( \bar{R}\left( K,X\right) ,\Gamma
\left( K,X\right) \right) \right) \\
&&\times \hat{F}_{2}\left( R\left( K,X\right) \right) \left\vert \Psi \left(
K,\hat{X}\right) \right\vert ^{2}dK  \label{fcS} \\
g\left( K,\hat{X},\Psi ,\hat{\Psi}\right) &=&\int \frac{\nabla _{\hat{X}%
}F_{0}\left( R\left( K,\hat{X}\right) \right) +\nu \nabla _{\hat{X}%
}F_{1}\left( \bar{R}\left( K,\hat{X}\right) ,\Gamma \left( K,X\right)
\right) }{\int \left\vert \Psi \left( K^{\prime },\hat{X}\right) \right\vert
^{2}dK^{\prime }}\left\vert \Psi \left( K,\hat{X}\right) \right\vert ^{2}dK
\label{fCS}
\end{eqnarray}%
and where $\bar{R}\left( K,X\right) $ is given by (\ref{RPN}). Under
assumption (\ref{smp}), functions $\hat{F}_{2}$ and\textbf{\ }$\Gamma $
write:%
\begin{eqnarray}
\hat{F}_{2}\left( s,R\left( K,X\right) \right) &=&\frac{F_{2}\left(
s,R\left( K,X\right) \right) }{\int F_{2}\left( R\left( K^{\prime },X\right)
\right) \left\vert \Psi \left( K^{\prime },X\right) \right\vert
^{2}dK^{\prime }}  \label{FH} \\
\Gamma \left( K,X\right) &=&\frac{\int \hat{F}_{2}\left( s,R\left(
K,X\right) \right) \hat{K}\left\vert \hat{\Psi}\left( \hat{K},X\right)
\right\vert ^{2}d\hat{K}}{K}-1  \label{GM}
\end{eqnarray}%
where:%
\begin{equation*}
F_{2}\left( R\left( K^{\prime },X\right) \right)
\end{equation*}%
is the average over the parameter $s^{\prime }$\ of:%
\begin{equation*}
F_{2}\left( s^{\prime },R\left( K^{\prime },X\right) \right)
\end{equation*}

Recall that $H\left( K_{X}\right) $\ is a function that encompasses the
determinants of the firms' mobility across the sector space. We will specify
later this function as a function of expected long term-returns and capital.

Function $u$\ describes the evolution of the capital of a firm, that is
located at $X$. Its dynamics depends on the relative value of a function $%
F_{2}$ that is itself a function of the firms' expected returns $R\left(
K,X\right) $. Investors allocate their capital based on their expectations
of the firms' long-term returns.

Function $f$\ describes the returns of investors located at $\hat{X}$
investing in sector $X$\ a capital $K$. These returns depend on short-term
dividends $r\left( K,X\right) $, the field-equivalent cost of capital\textbf{%
\ }$\frac{\gamma \int K^{\prime }\left\Vert \Psi \left( K,X\right)
\right\Vert ^{2}}{K}$\textbf{,\ }and firms' expected long-term stock
valuations through the function $F_{1}$..\ These returns themselves depend
on the relative attractiveness of a firm expected long-term returns
vis-a-vis its competitors.

Function $g$\ describes investors' reallocation of capital across the
sectors' space. These reallocation are driven by the gradient of expected
long-term returns and stocks valuations.

Recall that here we depart from the general formalism. We do not introduce a
time variable in the present model. Our purpose is to find collective, or
characteristic, configurations of the system that can, as such, be
considered static. It is only when we will derive these configurations that
a macro time scale will be introduced to study how the background states
evolve over time.

\section{Computation of transition functions}

We use the results of section 5.2 to compute the agents' transition
functions. To do so we need to derive the background fields of the system
and then compute the effective action of the system by expanding the action
functional around these background fields.

\subsection{Background fields and averages}

Minimizing the field action functional $S$ yields the background field of
the system. In turn, the background fields allow to compute the average
quantities of the system in that particular state. As shown in (Gosselin
Lotz Wambst 2022) both averages and background field are interrelated, and
that the system is solved by finding a system of equations for the fields
and the averages.

We first recall the definitions of the averages and then the solutions for
the background field and the defining equation for average capital that
define the system at the collective level.

\subsubsection{Averages}

At the collective level, the squared background fields of the system $%
\left\vert \Psi \left( K,X\right) \right\vert ^{2}$\ and $\left\vert \hat{%
\Psi}\left( \hat{K},\hat{X}\right) \right\vert ^{2}$, represent the density
of firms and the density of investors per sector and for a given capital $K$%
, in the collective state defined by $\Psi \left( K,X\right) $\ and $\hat{%
\Psi}\left( \hat{K},\hat{X}\right) $. Thus, the collective state determines,
for each sector and for a given capital $K$, the density of firms and the
density of investors.

Moreover, these two functions squared allow to compute various global
quantities of the system\ in the collective state $\Psi \left( K,X\right) $\
and $\hat{\Psi}\left( \hat{K},\hat{X}\right) $.

The number of producers in sector $X$, $N\left( X\right) $\ and the number
of investors in sector $\hat{X}$, $\hat{N}\left( \hat{X}\right) $ are
computed using the formula:%
\begin{eqnarray}
N\left( X\right) &=&\int \left\vert \Psi \left( K,X\right) \right\vert ^{2}dK
\label{Nx} \\
\hat{N}\left( \hat{X}\right) &=&\int \left\vert \hat{\Psi}\left( \hat{K},%
\hat{X}\right) \right\vert ^{2}d\hat{K}  \label{Nxh}
\end{eqnarray}%
The average values of total invested capital $\hat{K}_{X}$\ for each sector $%
X$ is:%
\begin{equation*}
\hat{K}_{\hat{X}}=\int \hat{K}\left\vert \hat{\Psi}\left( \hat{K},X\right)
\right\vert ^{2}d\hat{K}
\end{equation*}%
and the average invested capital per firm for sector $X$ is:\textbf{\ }%
\begin{equation}
K_{X}=\frac{\int \hat{K}\left\vert \hat{\Psi}\left( \hat{K},X\right)
\right\vert ^{2}d\hat{K}}{N\left( X\right) }  \label{kx}
\end{equation}%
Note that this $K_{X}$ is also equal to the average physical capital per
firm for sector $X$, i.e. :%
\begin{equation}
K_{X}=\frac{\int K\left\vert \Psi \left( K,X\right) \right\vert ^{2}dK}{%
N\left( X\right) }  \label{KX}
\end{equation}%
\textbf{\ }Indeed, given our assumptions, the total physical capital is
equal to the total capital invested:%
\begin{equation*}
\int K\left\vert \Psi \left( K,X\right) \right\vert ^{2}dK=\int \hat{K}%
\left\vert \hat{\Psi}\left( \hat{K},\hat{X}\right) \right\vert ^{2}d\hat{K}
\end{equation*}%
Ultimately, a collective state determines the distribution of invested
capital and capital per firm across sectors, which are given by $\frac{%
\left\vert \hat{\Psi}\left( \hat{K},X\right) \right\vert ^{2}}{\hat{N}\left(
X\right) }$\ and $\frac{\left\vert \Psi \left( K,X\right) \right\vert ^{2}}{%
N\left( X\right) }$\ respectively.

Equations (\ref{Nx}), (\ref{Nxh}) and (\ref{kx}) show that each collective
state defined by $\Psi \left( K,X\right) $ and $\hat{\Psi}\left( \hat{K},%
\hat{X}\right) $ is determined by the collection of data that characterizes
each sector: the number of firms for each sector, the number of investors
for each sector, the average capital for each sector and the density of
distribution of capital in each sector.

The above quantities allow for the study of capital allocation among sectors
and how it depends on system parameters such as expected long-term return,
short-term return, and any other parameters involved in the model. These
quantities shape the landscape in which individual agents operate.

\subsubsection{Background fields}

In (Gosselin Lotz Wambst 2022) we derived expressions for the background
fields and the densities of agents (firms and investors). This ultimately
led us to find a defining equation for the average capital invested per
sector. \ 

\paragraph{Background field $\Psi \left( K,X\right) $}

In (Gosselin Lotz Wambst 2022), we computed the density of firms in a given
sector as a function of the average capital in that sector. This density is
given by:%
\begin{equation*}
\left\Vert \Psi \left( X\right) \right\Vert ^{2}=\mathcal{N}\exp \left(
-\left( K-\hat{F}_{2}\left( s,R\left( K,X\right) \right) K_{X}\right)
^{2}\right) \left\Vert \Psi \left( X\right) \right\Vert ^{2}
\end{equation*}%
Here, $\mathcal{N}$ is a normalization factor and $K_{X}$ represents the
average invested capital per firm in sector $X$. This quantity is given by
equations (\ref{kx}) or (\ref{KX}), and: 
\begin{equation}
\left\Vert \Psi \left( X\right) \right\Vert ^{2}=\frac{D\left( \left\Vert
\Psi \right\Vert ^{2}\right) }{2\tau }-\frac{1}{4\tau }\left( \left( \nabla
_{X}R\left( X\right) \right) ^{2}+\frac{\sigma _{X}^{2}\nabla
_{X}^{2}R\left( K_{X},X\right) }{H\left( K_{X}\right) }\right) \left( 1-%
\frac{H^{\prime }\left( \hat{K}_{X}\right) K_{X}}{H\left( \hat{K}_{X}\right) 
}\right) H^{2}\left( K_{X}\right)  \label{Frp}
\end{equation}%
The constant $D\left( \left\Vert \Psi \right\Vert ^{2}\right) $ depends on
the norm of the background field.

Formula (\ref{Frp}) will be useful below while computing the transition
functions.

\paragraph{Background fields $\hat{\Psi}\left( \hat{X},\hat{K}\right) $ and
average capital per sector}

The density of investors per sectors in the background field has the form:%
\begin{equation}
\left\vert \hat{\Psi}_{-M}\left( \hat{K},\hat{X}\right) \right\vert
^{2}\simeq C\left( \bar{p}\right) \exp \left( -\frac{\sigma _{X}^{2}\hat{K}%
^{4}\left( f^{\prime }\left( X\right) \right) ^{2}}{96\sigma _{\hat{K}%
}^{2}\left\vert f\left( \hat{X}\right) \right\vert }\right) D_{p\left( \hat{X%
}\right) }^{2}\left( \left( \frac{\left\vert f\left( \hat{X}\right)
\right\vert }{\sigma _{\hat{K}}^{2}}\right) ^{\frac{1}{2}}\left( \hat{K}+%
\frac{\sigma _{\hat{K}}^{2}F\left( \hat{X}\right) }{f^{2}\left( \hat{X}%
\right) }\right) \right)  \label{Frs}
\end{equation}%
where $D_{p}$ is the parabolic cylinder function with parameter $p\left( 
\hat{X}\right) $ and:%
\begin{equation}
p\left( \hat{X}\right) =\frac{M-A\left( \hat{X}\right) }{\sqrt{f^{2}\left( 
\hat{X}\right) }}  \label{PDF}
\end{equation}%
The constant $C\left( \bar{p}\right) $ ensures that the constraint given by
equation (\ref{Nxh}) is satisfied.\textbf{\ }

Although the background field for the financial sector does not directly
appear in the transition functions, it is central to computing the average
quantities for each sector. In fact, it directly determines $K_{\hat{X}}$,
the average capital per firm in sector $\hat{X}$, in this environment.
Indeed, the defining equation of $K_{\hat{X}}$, given in (\ref{csc}), writes:%
\begin{equation}
K_{\hat{X}}\left\Vert \Psi \left( \hat{X}\right) \right\Vert ^{2}=\int \hat{K%
}\left\Vert \hat{\Psi}\left( \hat{K},\hat{X}\right) \right\Vert ^{2}d\hat{K}
\end{equation}%
This equation is actually an equation for $K_{\hat{X}}$. Actually, we have
expressed in equation (\ref{Frp}) the squared background field $\left\Vert
\Psi \left( \hat{X}\right) \right\Vert ^{2}$ as a function of $K_{\hat{X}}$,
and the field $\hat{\Psi}\left( \hat{K},\hat{X}\right) $, and $\hat{\Psi}%
\left( \hat{K},\hat{X}\right) $ is itself a function of $K_{\hat{X}}$
through equation (\ref{Frs}).

We showed that depending on the model parameters, several possible patterns
of accumulation exist in each sector.

\subsection{Effective action expansion}

\subsubsection{Second-order expansion of effective action}

Consider the field action:%
\begin{equation*}
S=S_{1}+S_{2}+S_{3}+S_{4}
\end{equation*}%
where the $S_{i}$ are defined by equations (\ref{SN}),(\ref{SD}),(\ref{ST})
and (\ref{SQ}). Expanding the action $S$ to the second-order around the
background field will allow us to compute the transition functions of
individual agents in the background, without taking into account individual
interactions. We can rewrite the fields as follows:%
\begin{eqnarray*}
\Psi \left( K,X\right) &=&\Psi _{0}\left( K,X\right) +\Delta \Psi \left(
Z,\theta \right) \\
\hat{\Psi}\left( \hat{K},\hat{X}\right) &=&\hat{\Psi}_{0}\left( \hat{K},\hat{%
X}\right) +\Delta \hat{\Psi}\left( Z,\theta \right)
\end{eqnarray*}%
where $\Psi _{0}\left( K,X\right) ,\hat{\Psi}_{0}\left( \hat{K},\hat{X}%
\right) $ are the background fields. This yields the quadratic approximation:%
\begin{equation}
S\left( \Psi ,\hat{\Psi}\right) =S\left( \Psi _{0},\hat{\Psi}_{0}\right)
+\int \left( \Delta \Psi ^{\dag }\left( Z,\theta \right) ,\Delta \hat{\Psi}%
^{\dag }\left( Z,\theta \right) \right) \left( Z,\theta \right) O\left( \Psi
_{0}\left( Z,\theta \right) \right) \left( 
\begin{array}{c}
\Delta \Psi \left( Z,\theta \right) \\ 
\Delta \hat{\Psi}\left( Z,\theta \right)%
\end{array}%
\right)
\end{equation}%
with:%
\begin{equation}
O\left( \Psi _{0}\left( Z,\theta \right) \right) =\left( 
\begin{array}{cc}
\frac{\delta ^{2}S\left( \Psi \right) }{\delta \Psi ^{\dag }\delta \Psi } & 
\frac{\delta ^{2}S\left( \Psi \right) }{\delta \Psi ^{\dag }\left( Z,\theta
\right) \delta \hat{\Psi}} \\ 
\frac{\delta ^{2}S\left( \Psi \right) }{\delta \hat{\Psi}^{\dag }\delta \Psi 
} & \frac{\delta ^{2}S\left( \Psi \right) }{\delta \hat{\Psi}^{\dag }\delta 
\hat{\Psi}}%
\end{array}%
\right) _{\substack{ \Psi \left( Z,\theta \right) =\Psi _{0}\left( Z,\theta
\right)  \\ \hat{\Psi}\left( Z,\theta \right) =\hat{\Psi}_{0}\left( Z,\theta
\right) }}  \label{prt}
\end{equation}%
The anti-diagonal terms in equation (\ref{prt}) involve crossed derivatives
with respect to both the fields of the real economy and the financial
economy. These terms represent the interactions between the two economies.
However, as explained in (Gosselin Lotz Wambst 2022), the cross-dependency
between $\Psi \left( Z,\theta \right) $\ and $\hat{\Psi}\left( Z,\theta
\right) $\ is relatively weak, since these interactions are taken into
account by the background fields. In first approximation, the minimization
of $S\left( \Psi \right) $ can be separated between $S_{1}+S_{2}$ and $%
S_{3}+S_{4}$. Therefore, we can write:%
\begin{equation}
O\left( \Psi _{0}\left( Z,\theta \right) \right) \simeq \left( 
\begin{array}{cc}
\frac{\delta ^{2}\left( S_{1}+S_{2}\right) }{\delta \Psi ^{\dag }\delta \Psi 
} & 0 \\ 
0 & \frac{\delta ^{2}\left( S_{3}\left( \Psi \right) +S_{4}\left( \Psi
\right) \right) }{\delta \hat{\Psi}^{\dag }\delta \hat{\Psi}}%
\end{array}%
\right) _{\substack{ \Psi \left( Z,\theta \right) =\Psi _{0}\left( Z,\theta
\right)  \\ \hat{\Psi}\left( Z,\theta \right) =\hat{\Psi}_{0}\left( Z,\theta
\right) }}  \label{Dfr}
\end{equation}%
The second-order expansion then becomes:%
\begin{eqnarray}
S\left( \Psi ,\hat{\Psi}\right) &=&S\left( \Psi _{0},\hat{\Psi}_{0}\right)
+\Delta S_{2}\left( \Psi ,\hat{\Psi}\right)  \label{ftv} \\
&=&S\left( \Psi _{0},\hat{\Psi}_{0}\right) +\int \Delta \Psi ^{\dag }\left(
K,X\right) \frac{\delta ^{2}\left( S_{1}+S_{2}\right) }{\delta \Psi ^{\dag
}\left( Z,\theta \right) \delta \Psi \left( Z,\theta \right) }\Delta \Psi
\left( K,\theta \right)  \notag \\
&&+\int \Delta \hat{\Psi}^{\dag }\left( Z,\theta \right) \frac{\delta
^{2}\left( S_{3}\left( \Psi \right) +S_{4}\left( \Psi \right) \right) }{%
\delta \hat{\Psi}^{\dag }\left( Z,\theta \right) \delta \hat{\Psi}\left(
Z,\theta \right) }\Delta \hat{\Psi}\left( Z,\theta \right)  \notag
\end{eqnarray}%
Computing the second order derivatives involved in (\ref{ftv}), and using
the definition of the background fields (see appendix 1) leads to the
formulas:%
\begin{eqnarray*}
\frac{\delta ^{2}\left( S_{1}+S_{2}\right) }{\delta \Psi ^{\dag }\left(
Z,\theta \right) \delta \Psi \left( Z,\theta \right) } &=&-\frac{\sigma
_{X}^{2}}{2}\nabla _{X}^{2}-\frac{\sigma _{K}^{2}}{2}\nabla _{K}^{2}+\left(
D\left( \left\Vert \Psi \right\Vert ^{2}\right) +2\tau \frac{\left\vert \Psi
\left( X\right) \right\vert ^{2}\left( K_{X}-K\right) }{K}\right) \\
&&+\frac{1}{2\sigma _{K}^{2}}\left( K-\hat{F}_{2}\left( R\left( K,X\right)
\right) K_{X}\right) ^{2}+\frac{1-\nabla _{K}\hat{F}_{2}\left( R\left(
K,X\right) \right) K_{X}}{2}
\end{eqnarray*}%
\begin{eqnarray*}
&&\frac{\delta ^{2}\left( S_{3}\left( \Psi \right) +S_{4}\left( \Psi \right)
\right) }{\delta \hat{\Psi}^{\dag }\left( Z,\theta \right) \delta \hat{\Psi}%
\left( Z,\theta \right) } \\
&=&\left( -\frac{\sigma _{\hat{X}}^{2}}{2}\nabla _{\hat{X}}^{2}+\frac{\left(
g\left( \hat{X}\right) \right) ^{2}+\sigma _{\hat{X}}^{2}\left( f\left( \hat{%
X}\right) +\nabla _{\hat{X}}g\left( \hat{X},K_{\hat{X}}\right) -\frac{\sigma
_{\hat{K}}^{2}F^{2}\left( \hat{X},K_{\hat{X}}\right) }{2f^{2}\left( \hat{X}%
\right) }\right) }{\sigma _{\hat{X}}^{2}\sqrt{f^{2}\left( \hat{X}\right) }}%
\right. \\
&&\left. -\frac{\sigma _{\hat{K}}^{2}}{2\sqrt{f^{2}\left( \hat{X}\right) }}%
\nabla _{\hat{K}}^{2}+\left( \frac{\sqrt{f^{2}\left( \hat{X}\right) }\left( 
\hat{K}+\frac{\sigma _{\hat{K}}^{2}F\left( \hat{X},K_{\hat{X}}\right) }{%
f^{2}\left( \hat{X}\right) }\right) ^{2}}{4\sigma _{\hat{K}}^{2}}\right)
\right)
\end{eqnarray*}

\subsubsection{Fourth-order corrections}

Calculating the fourth-order corrections to the effective action is
sufficient for deriving the main aspects of the interactions in a given
background field. We show in appendix 2 that the third-order terms can be
neglected, and that the series expansion of the action to the fourth-order
writes:%
\begin{eqnarray}
S\left( \Psi ,\hat{\Psi}\right) &=&S\left( \Psi _{0},\hat{\Psi}_{0}\right) \\
&&+\int \Delta \Psi ^{\dag }\left( K,X\right) \frac{\delta ^{2}\left(
S_{1}+S_{2}\right) }{\delta \Psi ^{\dag }\left( Z,\theta \right) \delta \Psi
\left( Z,\theta \right) }\Delta \Psi \left( K,\theta \right)  \notag \\
&&+\int \Delta \hat{\Psi}^{\dag }\left( Z,\theta \right) \frac{\delta
^{2}\left( S_{3}\left( \Psi \right) +S_{4}\left( \Psi \right) \right) }{%
\delta \hat{\Psi}^{\dag }\left( Z,\theta \right) \delta \hat{\Psi}\left(
Z,\theta \right) }\Delta \hat{\Psi}\left( Z,\theta \right) +\Delta
S_{4}\left( \Psi ,\hat{\Psi}\right)  \notag
\end{eqnarray}%
with: 
\begin{eqnarray}
&&\Delta S_{4}\left( \Psi ,\hat{\Psi}\right)  \label{dtc} \\
&\simeq &2\tau \int \left\vert \Delta \Psi \left( K^{\prime },X\right)
\right\vert ^{2}dK^{\prime }\left\vert \Delta \Psi \left( K,X\right)
\right\vert ^{2}dKdX  \notag \\
&&-\Delta \Psi ^{\dag }\left( K,X\right) \Delta \Psi ^{\dag }\left(
K^{\prime },X^{\prime }\right) \nabla _{K}\frac{\delta ^{2}u\left( K,X,\Psi ,%
\hat{\Psi}\right) }{\delta \Psi \left( K^{\prime },X\right) \delta \Psi
^{\dagger }\left( K^{\prime },X\right) }\Delta \Psi \left( K^{\prime
},X^{\prime }\right) \Delta \Psi \left( K,X\right)  \notag \\
&&-\Delta \Psi ^{\dagger }\left( K,\theta \right) \Delta \hat{\Psi}^{\dagger
}\left( \hat{K},\theta \right) \nabla _{K}\frac{\delta ^{2}u\left( K,X,\Psi ,%
\hat{\Psi}\right) }{\delta \hat{\Psi}\left( \hat{K},\hat{X}\right) \delta 
\hat{\Psi}^{\dagger }\left( \hat{K},\hat{X}\right) }\Delta \hat{\Psi}\left( 
\hat{K},\theta \right) \Delta \Psi \left( K,\theta \right)  \notag \\
&&-\Delta \hat{\Psi}^{\dag }\left( \hat{K},\hat{X}\right) \Delta \Psi ^{\dag
}\left( K^{\prime },\theta \right) \left\{ \nabla _{\hat{K}}\frac{\hat{K}%
\delta ^{2}f\left( \hat{X},\Psi ,\hat{\Psi}\right) }{\delta \Psi \left(
K^{\prime },X\right) \delta \Psi ^{\dagger }\left( K^{\prime },X\right) }%
+\nabla _{\hat{X}}\frac{\delta ^{2}g\left( \hat{X},\Psi ,\hat{\Psi}\right) }{%
\delta \Psi \left( K^{\prime },X\right) \delta \Psi ^{\dagger }\left(
K^{\prime },X\right) }\right\} \Delta \Psi \left( K^{\prime },X^{\prime
}\right) \Delta \hat{\Psi}\left( \hat{K},\hat{X}\right)  \notag
\end{eqnarray}%
Computing the terms involved in (\ref{dtc}) (see appendix 2) allows us to
interpret the various terms arising in the correction to the action.

The first term in the right-hand side of (\ref{dtc}) describes the direct
repulsive interaction between firms due to competition in a given sector.

The second term describes the indirect competition between firms through
capital allocation by investors, since:%
\begin{equation}
\frac{\delta ^{2}u\left( K,X,\Psi ,\hat{\Psi}\right) }{\delta \Psi \left(
K^{\prime },X\right) \delta \Psi ^{\dagger }\left( K^{\prime },X\right) }=-%
\frac{1}{\varepsilon }\hat{F}_{2}\left( s,R\left( K,X\right) \right) \hat{F}%
_{2}\left( s,R\left( K^{\prime },X^{\prime }\right) \right) \hat{K}_{X}
\label{duf}
\end{equation}%
and this term involves the relative attractiveness of two firms with capital 
$K$ and $K^{\prime }$ respectively in sector $X$. \ 

The third term represents the firms-investors direct interactions through
investment, since: 
\begin{equation}
\frac{\delta ^{2}u\left( K,X,\Psi ,\hat{\Psi}\right) }{\delta \hat{\Psi}%
\left( \hat{K},\hat{X}\right) \delta \hat{\Psi}^{\dagger }\left( \hat{K},%
\hat{X}\right) }=\frac{1}{\varepsilon }\hat{F}_{2}\left( s,R\left(
K,X\right) \right) \hat{K}  \label{duh}
\end{equation}%
is the relative attractiveness of a firm with capital $K^{\prime }$ at
sector $X$.

The last term describes the variation of investement due to the relative
short-term and long-term return of a given firm. Specifically, we have: 
\begin{equation}
\frac{\delta ^{2}f\left( \hat{X},\Psi ,\hat{\Psi}\right) }{\delta \Psi
\left( K^{\prime },X\right) \delta \Psi ^{\dagger }\left( K^{\prime
},X\right) }\simeq \frac{1}{\varepsilon }\left( \Delta f\left( K^{\prime },%
\hat{X},\Psi ,\hat{\Psi}\right) -\gamma \frac{K^{\prime }}{K_{X}}\right)
\label{dtf}
\end{equation}%
and:%
\begin{equation}
\frac{\delta ^{2}g\left( \hat{X},\Psi ,\hat{\Psi}\right) }{\delta \Psi
\left( K^{\prime },X\right) \delta \Psi ^{\dagger }\left( K^{\prime
},X\right) }=\frac{1}{\int \left\Vert \Psi \left( K^{\prime },\hat{X}\right)
\right\Vert ^{2}dK^{\prime }}\Delta \left( g\left( K^{\prime },\hat{X},\Psi ,%
\hat{\Psi}\right) \right)  \label{dtg}
\end{equation}%
with:%
\begin{equation*}
\Delta f\left( K^{\prime },\hat{X},\Psi ,\hat{\Psi}\right) =f\left(
K^{\prime },\hat{X},\Psi ,\hat{\Psi}\right) -f\left( \hat{X},\Psi ,\hat{\Psi}%
\right)
\end{equation*}%
and:%
\begin{equation*}
\Delta g\left( K^{\prime },\hat{X},\Psi ,\hat{\Psi}\right) =g\left(
K^{\prime },\hat{X},\Psi ,\hat{\Psi}\right) -g\left( \hat{X},\Psi ,\hat{\Psi}%
\right)
\end{equation*}%
are the relative short-term return and long-term return for firm with
capital $K^{\prime }$ at sector $\hat{X}$ respectively.

\subsection{One agent transition functions}

Following section 5.2.4, we consider first the "free" transition functions
that are given by the inverse operator of:%
\begin{equation}
\left( O\left( \Psi _{0}\left( Z,\theta \right) \right) +\alpha \right) ^{-1}
\label{lpc}
\end{equation}%
Given (\ref{Dfr}), the inverse (\ref{lpc}) reduces to:%
\begin{equation*}
\left( 
\begin{array}{cc}
\left( \frac{\delta ^{2}\left( S_{1}+S_{2}\right) }{\delta \Psi ^{\dag
}\delta \Psi }+\alpha \right) ^{-1} & 0 \\ 
0 & \left( \frac{\delta ^{2}\left( S_{3}\left( \Psi \right) +S_{4}\left(
\Psi \right) \right) }{\delta \hat{\Psi}^{\dag }\delta \hat{\Psi}}+\alpha
\right) ^{-1}%
\end{array}%
\right) _{\substack{ \Psi \left( Z,\theta \right) =\Psi _{0}\left( Z,\theta
\right)  \\ \hat{\Psi}\left( Z,\theta \right) =\hat{\Psi}_{0}\left( Z,\theta
\right) }}
\end{equation*}%
This implies that the transition functions can be computed independently for
the individual firms and investors. We will write:%
\begin{equation*}
G_{1}\left( \left( K_{f},X_{f}\right) ,\left( X_{i},K_{i}\right) ,\alpha
\right)
\end{equation*}%
the transition probability for a firm between an initial state $\left(
X_{i},K_{i}\right) $ and a final state $\left( K_{f},X_{f}\right) $ during
an average timespan $\alpha ^{-1}$ and:%
\begin{equation*}
G_{2}\left( \left( \hat{K}_{f},\hat{X}_{f}\right) ,\left( \hat{X}_{i},\hat{K}%
_{i}\right) ,\alpha \right)
\end{equation*}%
the transition probability for a firm between an initial state $\left( \hat{X%
}_{i},\hat{K}_{i}\right) $ and a final state $\left( \hat{K}_{f},\hat{X}%
_{f}\right) $ average timespan $\alpha ^{-1}$. Appendix 3 computes these
transition functions. We find the following results.

\paragraph{One firm transition function}

\begin{eqnarray}
&&G_{1}\left( \left( K_{f},X_{f}\right) ,\left( X_{i},K_{i}\right) \right)
\label{Gn} \\
&=&\exp \left( D\left( \left( K_{f},X_{f}\right) ,\left( X_{i},K_{i}\right)
\right) -\alpha _{eff}\left( \Psi ,\left( K_{f},X_{f}\right) ,\left(
X_{i},K_{i}\right) \right) \sqrt{\frac{\left( X_{f}-X_{i}\right) ^{2}}{%
2\sigma _{X}^{2}}+\frac{\left( K_{f}^{\prime }-K_{i}^{\prime }\right) ^{2}}{%
2\sigma _{K}^{2}}}\right)  \notag
\end{eqnarray}%
where:%
\begin{equation}
D\left( \left( K_{f},X_{f}\right) ,\left( X_{i},K_{i}\right) \right)
=D_{1}+D_{2}+D_{3}  \label{frd}
\end{equation}%
with:%
\begin{equation}
D_{1}=\int_{X_{i}}^{X_{f}}\frac{\nabla _{X}R\left( K_{X},X\right) }{\sigma
_{X}^{2}}H\left( K_{X}\right)  \label{DTN}
\end{equation}%
\begin{equation}
D_{2}=-\int_{K_{i}}^{K_{f}}\left( K-\hat{F}_{2}\left( s,R\left( K,\bar{X}%
\right) \right) K_{\bar{X}}\right) dK  \label{DTT}
\end{equation}%
\begin{equation}
D_{3}=\int_{K_{i}}^{K_{f}}\left( \left( \frac{X_{f}-X_{i}}{2}\right) \nabla
_{X}\hat{F}_{2}\left( s,R\left( K,\bar{X}\right) \right) K_{\bar{X}}\right)
dK  \label{DTR}
\end{equation}%
\begin{eqnarray}
&&\alpha _{eff}\left( \Psi ,\left( K_{f},X_{f}\right) ,\left(
X_{i},K_{i}\right) \right)  \label{frt} \\
&=&\alpha +D\left( \left\Vert \Psi \right\Vert ^{2}\right) +\tau \left( 
\frac{\left\vert \Psi \left( X_{f}\right) \right\vert ^{2}\left(
K_{X_{f}}-K_{f}\right) }{K_{f}}-\frac{\left\vert \Psi \left( X_{i}\right)
\right\vert ^{2}\left( K_{X_{i}}-K_{i}\right) }{K_{i}}\right) +\frac{\sigma
_{K}^{2}}{2}K_{f}^{\prime }K_{i}^{\prime }  \notag
\end{eqnarray}%
and:%
\begin{eqnarray*}
K_{i}^{\prime } &=&K_{i}-\hat{F}_{2}\left( s,R\left( K_{X_{i}},X_{i}\right)
\right) K_{X_{i}} \\
K_{f}^{\prime } &=&K_{f}-\hat{F}_{2}\left( s,R\left( K_{X_{f}},X_{f}\right)
\right) K_{X_{f}}
\end{eqnarray*}

\paragraph{One investor transition function}

\begin{eqnarray}
&&G_{2}\left( \left( \hat{K}_{f},\hat{X}_{f}\right) ,\left( \hat{X}_{i},\hat{%
K}_{i}\right) \right)  \label{Gt} \\
&=&\exp \left( D^{\prime }\left( \left( \hat{K}_{f},\hat{X}_{f}\right)
,\left( \hat{X}_{i},\hat{K}_{i}\right) \right) \right)  \notag \\
&&\times \exp \left( -\alpha _{eff}^{\prime }\left( \left( \hat{K}_{f},\hat{X%
}_{f}\right) ,\left( \hat{X}_{i},\hat{K}_{i}\right) \right) \left\vert
\left( \hat{K}_{f}+\frac{\sigma _{\hat{K}}^{2}F\left( \hat{X}_{f},K_{\hat{X}%
_{f}}\right) }{f^{2}\left( \hat{X}_{f}\right) }\right) -\left( \hat{K}_{i}+%
\frac{\sigma _{\hat{K}}^{2}F\left( \hat{X}_{i},K_{\hat{X}_{i}}\right) }{%
f^{2}\left( \hat{X}_{i}\right) }\right) \right\vert \right)  \notag
\end{eqnarray}%
with:%
\begin{equation}
D^{\prime }\left( \left( \hat{K}_{f},\hat{X}_{f}\right) ,\left( \hat{X}_{i},%
\hat{K}_{i}\right) \right) =\frac{1}{\sigma _{\hat{X}}^{2}}\int_{\hat{X}%
_{i}}^{\hat{X}_{f}}g\left( \hat{X}\right) d\hat{X}+\frac{\hat{K}_{f}^{2}}{%
\sigma _{\hat{K}}^{2}}f\left( \hat{X}_{f}\right) -\frac{\hat{K}_{i}^{2}}{%
\sigma _{\hat{K}}^{2}}f\left( \hat{X}_{i}\right)  \label{Frd}
\end{equation}%
and:%
\begin{eqnarray}
&&\alpha _{eff}^{\prime }\left( \left( \hat{K}_{f},\hat{X}_{f}\right)
,\left( \hat{X}_{i},\hat{K}_{i}\right) \right)  \label{Frt} \\
&=&\left( \alpha +\frac{\sigma _{\hat{X}}^{2}}{2}\left( \hat{K}_{f}+\frac{%
\sigma _{\hat{K}}^{2}F\left( \hat{X}_{f},K_{\hat{X}_{f}}\right) }{%
f^{2}\left( \hat{X}_{f}\right) }\right) \left( \hat{K}_{i}+\frac{\sigma _{%
\hat{K}}^{2}F\left( \hat{X}_{i},K_{\hat{X}_{i}}\right) }{f^{2}\left( \hat{X}%
_{i}\right) }\right) \right) \sqrt{\frac{\left\vert f\left( \frac{\hat{X}%
_{f}+\hat{X}_{i}}{2}\right) \right\vert }{2\sigma _{\hat{X}}^{2}}}+g^{\left(
R\right) }\left( \hat{X}\right)  \notag
\end{eqnarray}%
with:%
\begin{equation*}
g^{\left( R\right) }\left( \hat{X}\right) =\int_{\hat{X}_{i}}^{\hat{X}_{f}}%
\frac{\left( g\left( \hat{X}\right) \right) ^{2}+\sigma _{\hat{X}}^{2}\left(
f\left( \hat{X}\right) +\nabla _{\hat{X}}g\left( \hat{X},K_{\hat{X}}\right) -%
\frac{\sigma _{\hat{K}}^{2}F^{2}\left( \hat{X},K_{\hat{X}}\right) }{%
2f^{2}\left( \hat{X}\right) }\right) }{\left\Vert \hat{X}_{f}-\hat{X}%
_{i}\right\Vert \sigma _{\hat{X}}^{2}\sqrt{f^{2}\left( \hat{X}\right) }}
\end{equation*}

\subsection{Two agents transition functions and Interactions between agents}

To study the agents interactions within the background field we consider the
two-agent transition functions. There are three of them. One for two firms:%
\begin{equation*}
G_{11}\left( \left[ \left( K_{f},X_{f}\right) ,\left( K_{f},X_{f}\right)
^{\prime }\right] ,\left[ \left( X_{i},K_{i}\right) ,\left(
X_{i},K_{i}\right) ^{\prime }\right] \right)
\end{equation*}%
one for one firm and one investor:%
\begin{equation*}
G_{12}\left( \left[ \left( K_{f},X_{f}\right) ,\left( \hat{K}_{f},\hat{X}%
_{f}\right) \right] ,\left[ \left( X_{i},K_{i}\right) ,\left( \hat{X}_{i},%
\hat{K}_{i}\right) \right] \right)
\end{equation*}%
and one for for two investors:%
\begin{equation*}
G_{22}\left( \left[ \left( \hat{K}_{f},\hat{X}_{f}\right) ,\left( \hat{K}%
_{f},\hat{X}_{f}\right) ^{\prime }\right] ,\left[ \left( \hat{X}_{i},\hat{K}%
_{i}\right) ,\left( \hat{X}_{i},\hat{K}_{i}\right) ^{\prime }\right] \right)
\end{equation*}%
If we neglect the terms of order greater than $2$ in the effective action,
the transition functions reduce to simple products:%
\begin{eqnarray*}
&&G_{11}\left( \left[ \left( K_{f},X_{f}\right) ,\left( K_{f},X_{f}\right)
^{\prime }\right] ,\left[ \left( X_{i},K_{i}\right) ,\left(
X_{i},K_{i}\right) ^{\prime }\right] \right) \\
&=&G_{1}\left( \left( K_{f},X_{f}\right) ,\left( X_{i},K_{i}\right) \right)
G_{1}\left( \left( K_{f},X_{f}\right) ^{\prime },\left( X_{i},K_{i}\right)
^{\prime }\right)
\end{eqnarray*}%
\begin{eqnarray*}
&&G_{12}\left( \left[ \left( K_{f},X_{f}\right) ,\left( \hat{K}_{f},\hat{X}%
_{f}\right) \right] ,\left[ \left( X_{i},K_{i}\right) ,\left( \hat{X}_{i},%
\hat{K}_{i}\right) \right] \right) \\
&=&G_{1}\left( \left( K_{f},X_{f}\right) ,\left( X_{i},K_{i}\right) \right)
G_{2}\left( \left( \hat{K}_{f},\hat{X}_{f}\right) ,\left( \hat{X}_{i},\hat{K}%
_{i}\right) \right)
\end{eqnarray*}%
\begin{eqnarray*}
&&G_{22}\left( \left[ \left( \hat{K}_{f},\hat{X}_{f}\right) ,\left( \hat{K}%
_{f},\hat{X}_{f}\right) ^{\prime }\right] ,\left[ \left( \hat{X}_{i},\hat{K}%
_{i}\right) ,\left( \hat{X}_{i},\hat{K}_{i}\right) ^{\prime }\right] \right)
\\
&=&G_{2}\left( \left( \hat{K}_{f},\hat{X}_{f}\right) ,\left( \hat{X}_{i},%
\hat{K}_{i}\right) \right) G_{2}\left( \left( \hat{K}_{f},\hat{X}_{f}\right)
^{\prime },\left( \hat{X}_{i},\hat{K}_{i}\right) ^{\prime }\right)
\end{eqnarray*}%
In first approximation, agents behave independently, solely influenced by
the given background state.

To take into account agents interactions we write the expansion:%
\begin{equation*}
\exp \left( -S\left( \Psi \right) \right) =\exp \left( -\left( S\left( \Psi
_{0},\hat{\Psi}_{0}\right) +\Delta S_{2}\left( \Psi ,\hat{\Psi}\right)
\right) \right) \left( 1+\sum_{n\geqslant 1}\frac{\left( -\Delta S_{4}\left(
\Psi ,\hat{\Psi}\right) \right) ^{n}}{n!}\right)
\end{equation*}%
as explained in section 5.2.5, the series produces corrective terms to the
transition functions. Appendix 4 presents the computations and compute the
transitions in the approximations of paths that cross each other one time at
some $X$. In this approximation, we find:

\subsubsection{Firm-firm transition function:}

\begin{eqnarray}
&&G_{11}\left( \left[ \left( K_{f},X_{f}\right) ,\left( K_{f},X_{f}\right)
^{\prime }\right] ,\left[ \left( X_{i},K_{i}\right) ,\left(
X_{i},K_{i}\right) ^{\prime }\right] \right)  \label{nng} \\
&\simeq &G_{1}\left( \left( K_{f},X_{f}\right) ,\left( X_{i},K_{i}\right)
\right) G_{1}\left( \left( K_{f},X_{f}\right) ^{\prime },\left(
X_{i},K_{i}\right) ^{\prime }\right)  \notag \\
&&-\left( 2\tau -\nabla _{K}\frac{\delta ^{2}u\left( \bar{K},\bar{X},\Psi ,%
\hat{\Psi}\right) }{\delta \Psi \left( \bar{K}^{\prime },\bar{X}\right)
\delta \Psi ^{\dagger }\left( K^{\prime },\bar{X}\right) }\right) \hat{G}%
_{1}\left( \left( K_{f},X_{f}\right) ,\left( X_{i},K_{i}\right) \right) \hat{%
G}_{1}\left( \left( K_{f},X_{f}\right) ^{\prime },\left( X_{i},K_{i}\right)
^{\prime }\right)  \notag
\end{eqnarray}

\subsubsection{Firm-investor transition function:}

\begin{eqnarray}
&&G_{12}\left( \left[ \left( K_{f},X_{f}\right) ,\left( \hat{K}_{f},\hat{X}%
_{f}\right) ^{\prime }\right] ,\left[ \left( X_{i},K_{i}\right) ,\left( \hat{%
X},\hat{K}_{i}\right) ^{\prime }\right] \right)  \label{ntg} \\
&\simeq &G_{1}\left( \left( K_{f},X_{f}\right) ,\left( X_{i},K_{i}\right)
\right) G_{2}\left( \left( \hat{K}_{f},\hat{X}_{f}\right) ^{\prime },\left( 
\hat{X},\hat{K}_{i}\right) ^{\prime }\right)  \notag \\
&&+\left( \nabla _{K}\frac{\delta ^{2}u\left( \bar{K},\bar{X},\Psi ,\hat{\Psi%
}\right) }{\delta \hat{\Psi}\left( \overline{\hat{K},\hat{X}}\right) \delta 
\hat{\Psi}^{\dagger }\left( \overline{\hat{K},\hat{X}}\right) }+\nabla _{%
\hat{K}}\frac{\hat{K}\delta ^{2}f\left( \overline{\hat{X}},\Psi ,\hat{\Psi}%
\right) }{\delta \Psi \left( \overline{K^{\prime },X}\right) \delta \Psi
^{\dagger }\left( \overline{K^{\prime },X}\right) }+\nabla _{\hat{X}}\frac{%
\delta ^{2}g\left( \overline{\hat{X}},\Psi ,\hat{\Psi}\right) }{\delta \Psi 
\overline{\left( K^{\prime },X\right) }\delta \Psi ^{\dagger }\left( 
\overline{K^{\prime },X}\right) }\right)  \notag \\
&&\times \hat{G}_{1}\left( \left( K_{f},X_{f}\right) ,\left(
X_{i},K_{i}\right) \right) \hat{G}_{2}\left( \left( \hat{K}_{f},\hat{X}%
_{f}\right) ^{\prime },\left( \hat{X},\hat{K}_{i}\right) ^{\prime }\right) 
\notag
\end{eqnarray}

\subsubsection{Investor-investor transition function:}

\begin{eqnarray}
&&G_{22}\left( \left[ \left( \hat{K}_{f},\hat{X}_{f}\right) ,\left( \hat{K}%
_{f},\hat{X}_{f}\right) ^{\prime }\right] ,\left[ \left( \hat{X},\hat{K}%
_{i}\right) ,\left( \hat{X},\hat{K}_{i}\right) ^{\prime }\right] \right)
\label{ttg} \\
&\simeq &G_{2}\left( \left( \hat{K}_{f},\hat{X}_{f}\right) ,\left( \hat{X},%
\hat{K}_{i}\right) \right) G_{2}\left( \left( \hat{K}_{f},\hat{X}_{f}\right)
^{\prime },\left( \hat{X},\hat{K}_{i}\right) ^{\prime }\right)  \notag
\end{eqnarray}%
with:%
\begin{eqnarray*}
\left( \bar{X},\bar{K}\right) &=&\frac{\left( K_{f},X_{f}\right) +\left(
X_{i},K_{i}\right) }{2} \\
\left( \bar{X},\bar{K}\right) ^{\prime } &=&\frac{\left( K_{f},X_{f}\right)
^{\prime }+\left( X_{i},K_{i}\right) ^{\prime }}{2}
\end{eqnarray*}%
The derivatives are given in (\ref{duf}), (\ref{duh}), (\ref{dtf}), (\ref%
{dtg}) and:%
\begin{equation*}
\hat{G}_{i}\left( \left( K_{f},X_{f}\right) ,\left( X,K\right) \right) \hat{G%
}_{j}\left( \left( K_{f},X_{f}\right) ^{\prime },\left( X,K\right) ^{\prime
}\right)
\end{equation*}%
is the transition function computed on paths that cross once.

\section{Results and interpretations}

\subsection{One-agent transition functions}

We present a synthesis of the results for firms and investors transition
functions. Some technical details are given in appendix 5.

\subsubsection{Firms transition function}

For a given background state, the probability of transition for a firm
between two states $K_{i},X_{i}$ and $K_{f},X_{f}$, over an average time of $%
1/\alpha $, is given by $G_{1}$ (see \ref{Gn}). This formula computes the
probability that a firm initialy endowed with a capital $K_{i}$ in sector $%
X_{i}$ will relocate to sector $X_{f}$ with capital $K_{f}$. The transition
probability is the result of competing effects, as it is composed of several
interdependent terms of similar magnitude. Firm transitions occur over the
medium to long term but at a slower time scale than transitions for
investors. Firms remain in each transitory sector long enough to resettle,
and for investors to adjust the capital allocated between firms. Thus, in
each transitory sector, firm capital evolves depending on the
characteristics of the firm, the sector, and investors expectations.

\paragraph{Attractiveness and sectors shifts}

The drift term $D$ in formula (\ref{frd}) is the average transition of a
firm between its initial and final points $\left( X_{i},K_{i}\right) $ and $%
\left( K_{f},X_{f}\right) $, respectively. This term is usually different
from zero because firms tend to shift sectors, and their capital evolves.\
This tendency for a firm to evolve depends both on the transitory sectors
and the background field, i.e., the entire landscape in which the transition
occurs. In addition, fluctuations around the drift term can alter a firm's
trajectory, contributing to the probabilistic nature of the transition.

The drift term of equation (\ref{frd}) is composed of three interacting
contributions, $D_{1}$,\textbf{\ }$D_{2}$\textbf{\ }$\ $and $D_{3}$.\textbf{%
\ }

The first component $D_{1}$\textbf{\ }shows that firms tend to relocate to
sectors with higher long-term returns, shifts which in turn modify their
present and future attractiveness to investors.

The second component $D_{2}$\ shows that the shift alters the capital of the
firm. Specifically, the amount of investment that investors are willing to
make in the firm, $\hat{F}_{2}\left( R\left( K,\bar{X}\right) \right) K_{%
\bar{X}}$ depends on three key parameters: the average capital of the new
sector,\ $K_{\bar{X}}$, the absolute average return on capital in the
sector,\ $R\left( K,\bar{X}\right) $, and the propensity of investors, $\hat{%
F}_{2}$ to invest in the firm based on its given capital compared to the
average capital of firms in the sector.

When a firm begins the process of relocating to a nearby sector, its
capitalization may differ from that of firms already present in that sector,
which in turn affects its attractiveness to investors, represented by\ $\hat{%
F}_{2}$. The shape of $\hat{F}_{2}$ reflects the propensity of investors to
invest in the firm.\ When $\hat{F}_{2}$ is concave, this propensity
marginally decreases, while a convex shape results in a marginal increase.%
\textbf{\ }

The equilibrium capital of the firm in the new sector is $\hat{F}_{2}\left(
s,R\left( K,\bar{X}\right) \right) K_{\bar{X}}$.\ When a firm relocates, its
capital may turn out to be below or above this equilibrium level. For each
of these cases, two possibilities arise depending on the shape $s$ of $\hat{F%
}_{2}$.

If $\hat{F}_{2}$ is concave, the marginal propensity of investors to invest
is decreasing: once the firm has entered the sector, its capital will
converge towards the sector's average capital.\ It will either increase or
decrease, depending on whether its initial level of capital is above or
below the equilibrium capital, respectively. If $\hat{F}_{2}\left( s,R\left(
K,\bar{X}\right) \right) $ is convex, the marginal propensity of investors
to invest is increasing: the dynamics of its capital accumulation is
unstable.\ Investors will tend to over or underinvest in the firm.

The third contribution $D_{3}$ reflects the firm's relative attractiveness
in different transition sectors. If the the firm's relative attractiveness
is reduced during the shift, such that it attracts less capital than the
average capital of the transition sector, it may become stuck in an
intermediate sector.

\paragraph{Impact of competition}

The coefficient \ $\alpha _{eff}\ $defined in equation (\ref{frt})
represents the inverse of the average mobility of a firm. This mobility
depends on the competition in transitional sectors which is captured by the
two first terms on the right-hand side of (\ref{frt}).

The first term, $D\left( \left\Vert \Psi \right\Vert ^{2}\right) $ is a
constant that characterizes the background state of the firms and is
correlated with the total number of firms in the space of sectors.\ As
competition increases, $\alpha _{eff}$ rises and firms' mobility decreases.

The second term measures the local competition that firms face as they move
through the sector space. It is determined by the density of agents in the
sector, multiplied by the variation, along the path, of the firm's excess
capital with respect to the average capital of the sector. A
well-capitalized firm facing many less-capitalized competitors will repel
them and create its own market share.\textbf{\ }Relocation will occur
towards sectors that are denser and less capitalized. Under-capitalized
firms will be forced out of their sectors and into denser, less capitalized
sectors. The relocation process may result in a capital gain or loss.
However, holding capital constant, initially under-capitalized firms will
tend to move towards sectors with lower average capital, whereas
over-capitalized firms tend to move towards sectors with higher average
capital.

\paragraph{Stabilization terms:}

The square-rooted term multiplying $\alpha _{eff}$ is written: 
\begin{equation}
\sqrt{\frac{\left( X_{f}-X_{i}\right) ^{2}}{2\sigma _{X}^{2}}+\frac{\left( 
\tilde{K}_{f}-\tilde{K}_{i}\right) ^{2}}{2\sigma _{K}^{2}}}  \label{abc}
\end{equation}%
and the last term in the right-hand side of equation (\ref{frd}) both
describe random oscillations around a path of zero marginal capital demand.
Changes in equity, investments, for instance, may modify (\ref{abc}). and
the oscillations are of magnitude $\frac{\sigma _{K}^{2}}{2}$. However,
these oscillations do not necessarily imply a return to the initial point.
The larger the deviation from the average, the more likely firms are to
deviate from the average, and possibly shift to a new trajectory. Therefore,
a capital increase above the average may induce a shift in sector, which in
turn may modify the firm's accumulation and prospects. Thus, oscillations do
not prevent trends and may even initiate them.\ However, such "random
shifts" may prove disadvantageous as they could harm the firm's position and
reduce its capital compared to the sector.

\paragraph{Possible paths}

Overall, what are the possible dynamics for a firm in terms of capital and
sector? If a firm experiences capital growth in a sector where the investor
propensity, $F_{2}$, is concave, the accumulation of its capital could cause
the firm to shift to a higher-return sector, but this may result in the firm
being perceived as less attractive by investors in this new sector.

This shift can lead to a change in the firm's attractiveness to investors, $%
F_{2}$. The growth or decline of the firm in the new sector will depend on
both its capital level and the shape of\textbf{\ }$F_{2}$.\textbf{\ }These
factors will also determine the speed\textbf{\ }of this change. If the
firm's capital level gradually declines, it may have time to react and
reposition itself. However, if the decline in capital is sudden, the firm
may not have enough resources to reposition itself.\ The new sector may turn
out to be a capital trap.

The patterns of possible trajectories are various and may be irregular, with
some transitions occurring at a constant rate, while others may involve
discontinuities and sudden increases or reductions in capital, depending on
the characteristics of the landscape such as expected returns in sectors,
density of firms, and other background factors.

\subsubsection{Investors transition functions}

\paragraph{Drift term}

Short-term and long-term returns are the two parameters that determine
investors' capital allocation. Short-term returns include the firm's
dividends and increase with the value of its shares, while long-term returns
reflect the market's expectations for the firm's future growth potential,
which in turn affect expectations for higher dividends and share price
appreciation. Both types of returns are captured in the drift term $%
D^{\prime }$, which is defined in equation (\ref{Frd}). The most likely
paths are those that maximize both short-term and long-term returns.

However, these returns are not independent, since faltering share prices in
the short-term impact long-term returns expectations, and vice versa.

Ideally, to maximize their capital, investors seek both higher short-term
and long-term returns. Therefore, capital allocation within and across
sectors will depend on firms share prices volatility and dividends.

A sector in which share prices increase tends to attract capital, since
investors can maximize both short-term and long-term returns: an increase in
share prices sustains the firm's growth expectations. Investors tend to move
towards the next local maximum of long-term returns while also maximizing
their short-term return. In this case, there is no trade-off between the two
objectives.

In a sector where stock prices fall or remain stagnant, investors are faced
with a trade-off between short- and long-term returns. When stock prices no
longer support long-term earnings expectations, capital allocation is
determined by short-term dividends. Capital reallocation will depend on the
level of capital held by investors. While investors may consider long-term
expectationsthey must also generate short-term returns to maintain their
capital. An investor who ignores dividends in a context of falling share
prices would eventually see his capital depleted, which could hinder or
impair his ability to reallocate capital in the long term.

\paragraph{Stabilization terms:}

Similarly to firms, investors have an effective inverse mobility $\alpha
_{eff}^{\prime }$, defined in equation (\ref{Frt}). This formula shows that
mobility $\frac{1}{\alpha _{eff}^{\prime }}$ decreases with the average
short-term return along the path\ : the higher the returns, the lower the
incentive to switch from one sector to another. Similarly, mobility
increases with $g^{\left( R\right) }\left( \hat{X}\right) $, which measures
the relative long-term return of the sectors along the path. The higher this
relative return, the lower the incentive to switch to another sector.

Moreover, $\frac{1}{\alpha _{eff}^{\prime }}$ decreases with the final level
of capital $\hat{K}_{f}$ increases, impairing the firm's capacity to reach
high levels of capital. Conversely, $\frac{1}{\alpha _{eff}^{\prime }}$
decreases with the initial capital $\hat{K}_{i}$ decreases, indicating that
investors with high capitalization are less likely to experience significant
capital losses. This is supported by the factor multiplying $\alpha
_{eff}^{\prime }$: 
\begin{equation*}
\left\vert \left( \hat{K}_{f}+\frac{\sigma _{\hat{K}}^{2}F\left( \hat{X}%
_{f},K_{\hat{X}_{f}}\right) }{f^{2}\left( \hat{X}_{f}\right) }\right)
-\left( \hat{K}_{i}+\frac{\sigma _{\hat{K}}^{2}F\left( \hat{X}_{i},K_{\hat{X}%
_{i}}\right) }{f^{2}\left( \hat{X}_{i}\right) }\right) \right\vert
\end{equation*}%
As a result, the probability for an investor to deviate significantly\textbf{%
\ }from its initial capital value, apart from the smoothing term which can
be neglected, is relatively low.

\subsection{Two-agent transition functions}

First, it should be noted that the transition function $G_{22}$, as defined
in equation (\ref{ttg}), does not include any interaction corrections.
Specifically, the transition probability for two investors is simply the
product of their individual transition probabilities. In our model,
investors do not directly interact with each other, but only through their
investments in various firms. Only two transition functions are affected by
these indirect interactions.

\subsubsection{Firm-firm interactions}

First the transition $G_{11}$ is modified by the term:%
\begin{equation*}
I=2\tau -\nabla _{K}\frac{\delta ^{2}u\left( \bar{K},\bar{X},\Psi ,\hat{\Psi}%
\right) }{\delta \Psi \left( \bar{K}^{\prime },\bar{X}\right) \delta \Psi
^{\dagger }\left( K^{\prime },\bar{X}\right) }
\end{equation*}%
The interaction $I$ measures the interactions between two firms in the same
sector. The first contribution to $I$ describes a direct competition between
firms in a given sector, whereas the second term describes the competition
of the firms to attract investors that share their investments between the
two firms. Given that the 2-agents transition functions are modified by (see
(\ref{nng})):%
\begin{equation*}
\left( 2\tau -\nabla _{K}\frac{\delta ^{2}u\left( \bar{K},\bar{X},\Psi ,\hat{%
\Psi}\right) }{\delta \Psi \left( \bar{K}^{\prime },\bar{X}\right) \delta
\Psi ^{\dagger }\left( K^{\prime },\bar{X}\right) }\right) \hat{G}_{1}\left(
\left( K_{f},X_{f}\right) ,\left( X_{i},K_{i}\right) \right) \hat{G}%
_{1}\left( \left( K_{f},X_{f}\right) ^{\prime },\left( X_{i},K_{i}\right)
^{\prime }\right)
\end{equation*}%
and since $I>0$, the contribution to the green function of paths crossing at
some point are underweighted. The competition between the two firms repell
them from the sector where they interact. If we consider that the competion
factor $\tau $ is capital-dependent (see (\ref{tfc})), the less capitalized
firm is relatively more repelled than the more capitalized one.

\subsubsection{Firm-investor interactions}

Second, the firm-investor transition function $G_{12}$ is modified by the
term: 
\begin{equation*}
\left( \nabla _{K}\frac{\delta ^{2}u\left( \bar{K},\bar{X},\Psi ,\hat{\Psi}%
\right) }{\delta \hat{\Psi}\left( \overline{\hat{K},\hat{X}}\right) \delta 
\hat{\Psi}^{\dagger }\left( \overline{\hat{K},\hat{X}}\right) }+\left\{
\nabla _{\hat{K}}\frac{\hat{K}\delta ^{2}f\left( \overline{\hat{X}},\Psi ,%
\hat{\Psi}\right) }{\delta \Psi \left( \overline{K^{\prime },X}\right)
\delta \Psi ^{\dagger }\left( \overline{K^{\prime },X}\right) }+\nabla _{%
\hat{X}}\frac{\delta ^{2}g\left( \overline{\hat{X}},\Psi ,\hat{\Psi}\right) 
}{\delta \Psi \overline{\left( K^{\prime },X\right) }\delta \Psi ^{\dagger
}\left( \overline{K^{\prime },X}\right) }\right\} \right)
\end{equation*}%
Given (\ref{duh}), (\ref{dth}), (\ref{dtg}), this term depends mainly on
three contributions: 
\begin{equation*}
\nabla _{K}\frac{F_{2}\left( s,R\left( K^{\prime },X\right) \right) \hat{K}}{%
\int F_{2}\left( s,R\left( K^{\prime },X\right) \right) \left\Vert \Psi
\left( K^{\prime },X\right) \right\Vert ^{2}dK^{\prime }}
\end{equation*}%
\begin{equation*}
\nabla _{\hat{K}}\Delta f\left( K^{\prime },\hat{X},\Psi ,\hat{\Psi}\right)
\end{equation*}%
\begin{equation*}
\nabla _{\hat{X}}\Delta \left( g\left( K^{\prime },\hat{X},\Psi ,\hat{\Psi}%
\right) \right)
\end{equation*}%
each of this contribution describes the relative perspectives of the firm in
his path through the sectors.

The first one represents the gradient of firm"s attractiveness with respect
to capital. The investor willl decide to invest or not depending on the
marginal gain of long term returns of the firm.

The second term represents the marginal short-term return of an investment
of the firm, and the third one measures the reltive attractiveness of the
firm with respect to his neighbours (see Gosselin Lotz Wambst 2022).

The interaction between the firm and the investor is a combination of these
three quantities.

When the combination of these term is positive, the firm has positive
perspective either in terms of short or long term returns, or relatively to
his neighbors.

In this case the associated corrections to the path crossing at some points
is positive and this paths will be overweighted: in probability, this
translates by the fact that paths in which a firm presents above average
perspectives in his capital accumaltion and shift in sectors, will be
favoured by an increase in investment. The firm will take advantage from its
interaction with the investor, except if this one experiences, for any
reason, an decrease of capital. On the contrary, a firm perceived as moving
toward lower perspective will experience in average a decrease in
investment. This decrease will be dampened if its investor has itself low
capital to invest. Some mixed situation may arise: good short term
perspective, but uncertain long term expectations may cancel or compensate
each other.

\subsection{Discussion}

The central feature of the formalism presented and applied in this paper is
to encompass macro and microeconomic elements: the macro scale keeps track
of the entire set of agents and, in turn, influences the microeconomic
scale, allowing for two-level interpretations.

The macroeconomic scale was studied in (Gosselin Lotz Wambst 2022).\ We
showed that the underlying macroeconomic state of the model reveals
disparities among sectors and instabilities in capital accumulation.
Different sectors can exhibit distinct accumulation patterns. Some sectors
may attract significant capital, while others may experience depletion.
Moreover, these accumulation patterns undergo changes in perspectives or
expectations, leading to potential large fluctuations in capital
allocations. These fluctuations can even result in a change of accumulation
pattern. Furthermore, individual dynamics heavily rely on the underlying
macroeconomic state. Some parameters governing these dynamics depend on the
average capital and the number of firms per sector, which are both
characteristics of the collective or macroeconomic state.

The present paper studies the microeconomic scale and mechanisms resulting
from macroeconomic states and fluctuations.

In the face of these fluctuations, investors may experience capital losses.
However they can always shield their capital by reallocating it to more
profitable or stable sectors. In doing so, they may amplify global capital
fluctuations for firms, which are unable to react at the same pace.
Financial risk is therefore limited in our model. Investors can always
reposition themselves and, as a result, do not bear the same risk as firms
that move to attract investors. The primary burden of risk falls on the
firms themselves, not the financial sector. Our model demonstrates that
investors do not experience the eviction phenomenon that firms do. However,
investors may face eviction from certain investment sectors if their capital
no longer allows them to invest in sectors perceived as the most promising,
based on returns and share prices.

We posit that firms have a natural inclination to switch sectors. Indeed,
firms tend to change due to the continuous evolution and transformation of
sectors and the changing economic environment. In our model, the historical
development of a sector is not depicted by a specific variable, but rather
by firms shifts between closely related sectors. In the shift, the initial
sector is the past state of the sector, and the final sector its present
state. Thus, firms transitions captures both firm reorientations and their
adaptation to an evolving environment.

Attracting investors is crucial to firms and can be achieved through
continuous expansion. However, firms face higher uncertainty and risk than
investors. Specifically, firms face two distinct risks:

First, the individual risk associated with seeking higher returns. Switching
to more attractive sectors may expose firms to higher competition and
faltering investors sentiment. For example, a firm shifting to a
high-capitalized sector will experience a stronger competition and weaker
prospects, potentially deterring any present or additional investment. When
these two phenomena combine, they may induce a substantial loss of capital,
and trap the firm in the sector, evict it towards less-capitalized and
less-attractive ones, and impact its ability to position itself for future
sectoral changes and transformations.

Second, the global risk, caused by exogenous and macro fluctuations.\ This
risk can alter sectoral growth prospects and, consequently, affect
individual dynamics. Our model captures these potential instabilities at the
individual level.\ Within a sector, sub-sectors may emerge, some presenting
more promising opportunities than others. However, the entire sector can be
impacted. Even though, on average, the collective state may exhibit some
stability, fluctuations among a set of similar firms can be substantial at
the individual dynamics level. Consequently, fluctuations in this context
magnify the uncertainty at the individual level, making it difficult to
identify and capitalize on profitable shifts while also increasing the risk
of making detrimental moves. To sum up, both collective and individual
results suggest that firms with high initial capitalization are generally
less exposed to market fluctuations. Note incidentally that these risks may
be amplified by swift financial reallocation in the face of global
uncertainties.

Therefore, firms can undergo sharp changes in dynamics due to variations in
the landscape of expected returns, reactivity of expectations, relative
attractiveness compared to neighboring sectors, or the number of competing
firms.

The present paper also advocates that field formalism, in addition to mixing
macro and micro analysis, provides some precise insights about the
structures of interactions inside the macroeconomic state. The technique of
series expansion of the effective action induces emerging interactions that
are not detected in the classical formalism, such as indirect emerging
competition among agents. More precisely, interactions between firms within
a sector reveal phenomena of specialization and eviction. Competition is at
first determined by the firms relative levels of capital. This is the direct
form of competition. The firm with the highest capital is more likely to
evict its competitors. However, field formalism reveals that competition
also revolves around attracting investor capital. This is the indirect form
of competition. A firm that successfully differentiates itself within a
sector, through specialization, has the potential to attract capital and
mitigate or reverse a possible eviction. However, specialization makes the
firm dependent on its investors. If investors suffer capital losses, the
firm is directly impacted.

Interactions between firms and their investors detail the impact of
investment at the individual level. A firm that attracts more investors will
be better positioned in the sector, as it enjoys a stronger position,
whereas its competitors will be compelled to reorient themselves. To attract
investors, a firm needs to demonstrate a high growth potential, which may
favor new entrants in a sector, provided they have the necessary capital to
position themselves, or better growth prospects.

To conclude, note that the concept of comparative advantage is not relevant
in our model. Indeed, given that changes are inevitable within sectors, any
comparative advantage is bound to be swept away, potentially even by
relatively distant and unexpected causes. Actually, exogenous fluctuations,
such as the perception of the sector and the firm within it (by investors),
as well as competition among firms to retain their position and attract
investors, create inherent instability within a specific sector.
Specializing in a single sector exposes firms to the risk of eventual
eviction, forming a trap.

\section{Conclusion}

We have studied the impact of financial capital on physical capital
allocation in a field-formalism setting. In a previous paper, we introduced
the concept of background or collective state of the system as a modeling of
a macroeconomic historical state. This paper presents the probabilistic
dynamics of agents in this environment. We have identified several types of
dynamics for producers, depending on the firms' landscape, returns, and the
firms' and sectors' relative attractiveness. A firm's dynamics depends on
these parameters as well as its initial sector and level of capital, and may
exhibit turning points. Increased competition in transitory sectors may
reduce the capital of the firm and impair its dynamics. The dependence of
the firm on investors and their expectations, as well as on modifications of
the macroeconomic state may lead to significant fluctuations in a firm's
growth trajectory.

\section*{References}


\begin{description}
\item Abergel F, Chakraborti A, Muni Toke I and Patriarca M (2011a)
Econophysics review: I. Empirical facts, Quantitative Finance, Vol. 11, No.
7, 991-1012

\item Abergel F, Chakraborti A, Muni Toke I and Patriarca M (2011b)
Econophysics review: II. Agent-based models, Quantitative Finance, Vol. 11,
No. 7, 1013-1041

\item Bernanke,B., Gertler, M. and S. Gilchrist (1999), "The financial
accelerator in a quantitative business cycle framework", Chapter 21 in
Handbook of Macroeconomics, 1999, vol. 1, Part C, pp 1341-1393

\item Bensoussan A, Frehse J, Yam P (2018) Mean Field Games and Mean Field
Type Control Theory. Springer, New York

\item B\"{o}hm, V., Kikuchi, T., Vachadze, G.: Asset pricing and
productivity growth: the role of consumption scenarios. Comput. Econ. 32,
163--181 (2008)

\item Caggese A, Orive A P, The Interaction between Household and Firm
Dynamics and the Amplification of Financial Shocks. Barcelona GSE Working
Paper Series, Working Paper n%
${{}^o}$
866, 2015

\item Campello, M., Graham, J. and Harvey, C.R. (2010). "The Real Effects of
Financial Constraints: Evidence from a Financial Crisis," Journal of
Financial Economics, vol. 97(3), 470-487.

\item Gaffard JL and Napoletano M Editors (2012): Agent-based models and
economic policy. OFCE, Paris

\item Gomes DA, Nurbekyan L, Pimentel EA (2015) Economic Models and
Mean-Field Games Theory, Publica\c{c}\~{o}es Matem\'{a}ticas do IMPA, 30o Col%
\'{o}quio Brasileiro de Matem\'{a}tica, Rio de Janeiro

\item Gosselin P, Lotz A and Wambst M (2017) A Path Integral Approach to
Interacting Economic Systems with Multiple Heterogeneous Agents. IF\_PREPUB.
2017. hal-01549586v2

\item Gosselin P, Lotz A and Wambst M (2020) A Path Integral Approach to
Business Cycle Models with Large Number of Agents. Journal of Economic
Interaction and Coordination volume 15, pages 899--942

\item Gosselin P, Lotz A and Wambst M (2021) A statistical field approach to
capital accumulation. Journal of Economic Interaction and Coordination 16,
pages 817--908 (2021)

\item Gosselin P, Lotz A and Wambst M (2022) Financial Markets and the Real
Economy: A Statistical Field Perspective on Capital Allocation and
Accumulation. https://hal.science/hal-03659624v2

\item Grassetti, F., Mammana, C. \& Michetti, E. A dynamical model for real
economy and finance. Math Finan Econ (2022).
https://doi.org/10.1007/s11579-021-00311-3

\item Grosshans, D., Zeisberger, S.: All's well that ends well? on the
importance of how returns are achieved. J. Bank. Finance 87, 397--410 (2018)

\item Holmstrom, B., and Tirole, J. (1997). Financial intermediation,
loanable funds, and the
\end{description}

real sector. Quarterly Journal of Economics, 663-691.

\begin{description}
\item Jackson M (2010) Social and Economic Networks. Princeton University
Press, Princeton

\item Jermann, U.J. and Quadrini, V., (2012). "Macroeconomic Effects of
Financial Shocks," American Economic Review, Vol. 102, No. 1.

\item Khan, A., and Thomas, J. K. (2013). "Credit Shocks and Aggregate
Fluctuations in an Economy with Production Heterogeneity," Journal of
Political Economy, 121(6), 1055-1107.

\item Kaplan G, Violante L (2018) Microeconomic Heterogeneity and
Macroeconomic Shocks, Journal of Economic Perspectives, Vol. 32, No. 3,
167-194

\item Kleinert H (1989) Gauge fields in condensed matter Vol. I , Superflow
and vortex lines, Disorder Fields, Phase Transitions, Vol. II, Stresses and
defects, Differential Geometry, Crystal Melting. World Scientific, Singapore

\item Kleinert H (2009) Path Integrals in Quantum Mechanics, Statistics,
Polymer Physics, and Financial Markets 5th edition. World Scientific,
Singapore

\item Krugman P (1991) Increasing Returns and Economic Geography. Journal of
Political Economy, 99(3), 483-499

\item Lasry JM, Lions PL, Gu\'{e}ant O (2010a) Application of Mean Field
Games to Growth Theory \newline
https://hal.archives-ouvertes.fr/hal-00348376/document

\item Lasry JM, Lions PL, Gu\'{e}ant O (2010b) Mean Field Games and
Applications. Paris-Princeton lectures on Mathematical Finance, Springer.%
\textbf{\ }https://hal.archives-ouvertes.fr/hal-01393103

\item Lux T (2008) Applications of Statistical Physics in Finance and
Economics. Kiel Institute for the World Economy (IfW), Kiel

\item Lux T (2016) Applications of Statistical Physics Methods in Economics:
Current state and perspectives. Eur. Phys. J. Spec. Top. (2016) 225: 3255.
https://doi.org/10.1140/epjst/e2016-60101-x

\item Mandel A, Jaeger C, F\"{u}rst S, Lass W, Lincke D, Meissner F,
Pablo-Marti F, Wolf S (2010). Agent-based dynamics in disaggregated growth
models. Documents de travail du Centre d'Economie de la Sorbonne. Centre
d'Economie de la Sorbonne, Paris

\item Mandel A (2012) Agent-based dynamics in the general equilibrium model.
Complexity Economics 1, 105--121

\item Monacelli, T., Quadrini, V. and A. Trigari (2011). "Financial Markets
and Unemployment," NBER Working Papers 17389, National Bureau of Economic
Research.

\item Sims C A (2006) Rational inattention: Beyond the Linear Quadratic
Case, American Economic Review, vol. 96, no. 2, 158-163

\item Yang J (2018) Information theoretic approaches to economics, Journal
of Economic Survey, Vol. 32, No. 3, 940-960

\item Cochrane, J.H. (ed.): Financial Markets and the Real Economy,
International Library of Critical Writings in Financial Economics, vol. 18.
Edward Elgar (2006)\pagebreak
\end{description}

\section*{Appendix 1 Computation of effective action at the second order}

We compute the second-order derivatives for the real and the financial
economy respectively.

\subsection*{Real economy}

In first approximation:

\begin{eqnarray}
&&\frac{\delta ^{2}\left( S_{1}+S_{2}\right) }{\delta \Psi ^{\dag }\left(
Z,\theta \right) \delta \Psi \left( Z,\theta \right) }  \label{SCD} \\
&\simeq &-\int \delta \Psi ^{\dag }\left( K,X\right) \left( \nabla
_{X}\left( \frac{\sigma _{X}^{2}}{2}\nabla _{X}-\nabla _{X}R\left(
K,X\right) H\left( K\right) \right) -4\tau \left( \left\vert \Psi _{0}\left(
X\right) \right\vert ^{2}\right) \right.  \notag \\
&&+\left. \nabla _{K}\left( \frac{\sigma _{K}^{2}}{2}\nabla _{K}+u\left(
K,X,\Psi _{0},\hat{\Psi}_{0}\right) \right) \right) \delta \Psi \left(
K,X\right) dKdX  \notag
\end{eqnarray}%
where:%
\begin{equation*}
\left\vert \Psi _{0}\left( X\right) \right\vert ^{2}=\int \left\vert \Psi
_{0}\left( K^{\prime },X\right) \right\vert ^{2}dK^{\prime }
\end{equation*}%
and:%
\begin{equation}
u\left( K,X,\Psi _{0},\hat{\Psi}_{0}\right) \rightarrow \frac{1}{\varepsilon 
}\left( K-\int \hat{F}_{2}\left( s,R\left( K,X\right) \right) \hat{K}%
\left\Vert \hat{\Psi}_{0}\left( \hat{K},X\right) \right\Vert ^{2}d\hat{K}%
\right) =\frac{1}{\varepsilon }\left( K-\hat{F}_{2}\left( s,R\left(
K,X\right) \right) K_{X}d\hat{K}\right)  \label{FR}
\end{equation}%
In equation (\ref{FR}), we used the notation:%
\begin{equation*}
\int \hat{F}_{2}\left( s,R\left( K,X\right) \right) \hat{K}\left\Vert \hat{%
\Psi}_{0}\left( \hat{K},X\right) \right\Vert ^{2}d\hat{K}=\hat{F}_{2}\left(
s,R\left( K,X\right) \right) \hat{K}_{X}
\end{equation*}%
We perform a change of variables in (\ref{SCD}):%
\begin{eqnarray}
\Delta \Psi \left( K,X\right) &=&\exp \left( \int^{X}\frac{\nabla
_{X}R\left( X\right) }{\sigma _{X}^{2}}H\left( \frac{\int \hat{K}\left\Vert 
\hat{\Psi}\left( \hat{K},X\right) \right\Vert ^{2}d\hat{K}}{\left\Vert \Psi
\left( X\right) \right\Vert ^{2}}\right) \right)  \label{chn} \\
&&\times \exp \left( \int \left( K-\frac{F_{2}\left( R\left( K,X\right)
\right) K_{X}}{F_{2}\left( R\left( K_{X},X\right) \right) }\right) dK\right)
\delta \Psi \left( K,X\right)  \notag \\
\Delta \Psi ^{\dag }\left( K,X\right) &=&\exp \left( -\int^{X}\frac{\nabla
_{X}R\left( X\right) }{\sigma _{X}^{2}}H\left( \frac{\int \hat{K}\left\Vert 
\hat{\Psi}\left( \hat{K},X\right) \right\Vert ^{2}d\hat{K}}{\left\Vert \Psi
\left( X\right) \right\Vert ^{2}}\right) \right)  \notag \\
&&\times \exp \left( -\int \left( K-\frac{F_{2}\left( R\left( K,X\right)
\right) K_{X}}{F_{2}\left( R\left( K_{X},X\right) \right) }\right) dK\right)
\delta \Psi ^{\dag }\left( K,X\right)  \notag
\end{eqnarray}%
where $K_{X}$, the average invested capital per firm in sector $X$:%
\begin{equation}
K_{X}=\frac{\int \hat{K}\left\Vert \hat{\Psi}\left( \hat{K},X\right)
\right\Vert ^{2}d\hat{K}}{\left\Vert \Psi \left( X\right) \right\Vert ^{2}}
\end{equation}%
so that the effective action (\ref{SCD}) for the real economy\textbf{\ }%
becomes:

\begin{eqnarray}
&&\Delta \Psi ^{\dag }\left( Z,\theta \right) \left( \frac{\delta ^{2}\left(
S_{1}+S_{2}\right) }{\delta \Psi ^{\dag }\left( Z,\theta \right) \delta \Psi
\left( Z,\theta \right) }\right) _{\Psi \left( Z,\theta \right) =\Psi
_{0}\left( Z,\theta \right) }\Delta \Psi \left( Z,\theta \right)  \label{SCN}
\\
&=&\int \Delta \Psi ^{\dag }\left( Z,\theta \right) \left( -\frac{\sigma
_{X}^{2}}{2}\nabla _{X}^{2}+\frac{\left( \nabla _{X}R\left( K,X\right)
H\left( K_{X}\right) \right) ^{2}}{2\sigma _{X}^{2}}+\frac{\nabla
_{X}^{2}R\left( K,X\right) }{2}H\left( K\right) +4\tau \left\vert \Psi
\left( X\right) \right\vert ^{2}\right) \Delta \Psi \left( Z,\theta \right) 
\notag \\
&&+\int \Delta \Psi ^{\dag }\left( Z,\theta \right) \left( -\frac{\sigma
_{K}^{2}}{2}\nabla _{K}^{2}+\frac{1}{2\sigma _{K}^{2}}\left( K-\hat{F}%
_{2}\left( s,R\left( K,X\right) \right) K_{X}\right) ^{2}+\frac{1-\nabla _{K}%
\hat{F}_{2}\left( s,R\left( K,X\right) \right) K_{X}}{2}\right) \Delta \Psi
\left( Z,\theta \right)  \notag
\end{eqnarray}%
As explained in section 7.2.1, the effects of competition can be refined by%
\textbf{\ }considering repulsive forces that are capital dependent\textbf{. }%
It amounts to replace in (\ref{SCN}), the term:%
\begin{equation*}
\int \Delta \Psi ^{\dag }\left( Z,\theta \right) \left( 2\tau \left\vert
\Psi \left( X\right) \right\vert ^{2}\right) \Delta \Psi \left( Z,\theta
\right)
\end{equation*}%
by the term:%
\begin{eqnarray}
&&\int \Delta \Psi ^{\dag }\left( K,X\right) \left( 2\tau \frac{\int
K^{\prime }\left\vert \Psi \left( K^{\prime },X\right) \right\vert
^{2}dK^{\prime }}{K}\right) \Delta \Psi \left( K,\theta \right)  \label{NT}
\\
&=&\int \Delta \Psi ^{\dag }\left( K,X\right) \left( 2\tau \frac{\left\vert
\Psi \left( X\right) \right\vert ^{2}K_{X}}{K}\right) \Delta \Psi \left(
K,\theta \right)  \notag
\end{eqnarray}%
with:%
\begin{eqnarray*}
\left\vert \Psi \left( X\right) \right\vert ^{2} &=&\int \left\vert \Psi
\left( K^{\prime },X\right) \right\vert ^{2}dK^{\prime } \\
K_{X} &=&\frac{\int K^{\prime }\left\vert \Psi \left( K^{\prime },X\right)
\right\vert ^{2}dK^{\prime }}{\left\vert \Psi \left( X\right) \right\vert
^{2}}
\end{eqnarray*}%
This models repulsive forces that are inversely proportional to capital and
mainly affect low-capital firms. Note that this change in the interaction
does not modify the collective state, since by setting $K=K_{X}$, we recover
the previous repulsive term. Ultimately, using:%
\begin{equation}
\left\Vert \Psi \left( X\right) \right\Vert ^{2}=\left( 2\tau \right)
^{-1}\left( D\left( \left\Vert \Psi \right\Vert ^{2}\right) -\frac{1}{%
2\sigma _{X}^{2}}\left( \left( \nabla _{X}R\left( X\right) \right) ^{2}+%
\frac{\sigma _{X}^{2}\nabla _{X}^{2}R\left( K_{X},X\right) }{H\left(
K_{X}\right) }\right) H^{2}\left( K_{X}\right) \left( 1-\frac{H^{\prime
}\left( \hat{K}_{X}\right) K_{X}}{H\left( \hat{K}_{X}\right) }\right) \right)
\label{PFR}
\end{equation}%
the interaction term (\ref{NT}) becomes:%
\begin{eqnarray*}
&&\int \Delta \Psi ^{\dag }\left( K,X\right) \frac{1}{2}\left( \left( \nabla
_{X}R\left( X\right) \right) ^{2}+\frac{\sigma _{X}^{2}\nabla
_{X}^{2}R\left( K_{X},X\right) }{H\left( K_{X}\right) }\right) \\
&&\times H^{2}\left( K_{X}\right) \left( 1-\frac{H^{\prime }\left( \hat{K}%
_{X}\right) K_{X}}{H\left( \hat{K}_{X}\right) }\right) +2\tau \frac{%
\left\vert \Psi \left( X\right) \right\vert ^{2}K_{X}}{K}\Delta \Psi \left(
K,\theta \right) \\
&=&\int \Delta \Psi ^{\dag }\left( K,X\right) \left( D\left( \left\Vert \Psi
\right\Vert ^{2}\right) +2\tau \frac{\left\vert \Psi \left( X\right)
\right\vert ^{2}\left( K_{X}-K\right) }{K}\right) \Delta \Psi \left(
K,\theta \right)
\end{eqnarray*}%
When the above expression is used to rewrite (\ref{SCN}), it yields the
formula:%
\begin{eqnarray}
\frac{\delta ^{2}\left( S_{1}+S_{2}\right) }{\delta \Psi ^{\dag }\left(
Z,\theta \right) \delta \Psi \left( Z,\theta \right) } &=&-\frac{\sigma
_{X}^{2}}{2}\nabla _{X}^{2}-\frac{\sigma _{K}^{2}}{2}\nabla _{K}^{2}+\left(
D\left( \left\Vert \Psi \right\Vert ^{2}\right) +2\tau \frac{\left\vert \Psi
\left( X\right) \right\vert ^{2}\left( K_{X}-K\right) }{K}\right)
\label{SCR} \\
&&+\frac{1}{2\sigma _{K}^{2}}\left( K-\hat{F}_{2}\left( s,R\left( K,X\right)
\right) K_{X}\right) ^{2}+\frac{1-\nabla _{K}\hat{F}_{2}\left( s,R\left(
K,X\right) \right) K_{X}}{2}  \notag
\end{eqnarray}%
as stated in the text.

\subsubsection*{Financial economy}

For the financial sector, we consider the field-action for $\hat{\Psi}^{\dag
}\left( \hat{K},\hat{X}\right) $: 
\begin{equation}
S_{3}+S_{4}=-\int \hat{\Psi}^{\dag }\left( \hat{K},\hat{X}\right) \left(
\nabla _{\hat{K}}\left( \frac{\sigma _{\hat{K}}^{2}}{2}\nabla _{\hat{K}}-%
\hat{K}f\left( \hat{X},K_{\hat{X}}\right) \right) +\nabla _{\hat{X}}\left( 
\frac{\sigma _{\hat{X}}^{2}}{2}\nabla _{\hat{X}}-g\left( \hat{X},K_{\hat{X}%
}\right) \right) \right) \hat{\Psi}\left( \hat{K},\hat{X}\right)  \label{STF}
\end{equation}%
with:

\begin{eqnarray}
f\left( \hat{X},K_{\hat{X}}\right) &=&\frac{1}{\varepsilon }\left( r\left(
K_{\hat{X}},\hat{X}\right) -\gamma \left\Vert \Psi \left( \hat{X}\right)
\right\Vert ^{2}+F_{1}\left( \frac{R\left( K_{\hat{X}},\hat{X}\right) }{\int
R\left( K_{X^{\prime }}^{\prime },X^{\prime }\right) \left\Vert \Psi \left(
X^{\prime }\right) \right\Vert ^{2}dX^{\prime }}\right) \right) \\
g\left( \hat{X},K_{\hat{X}}\right) &=&\left( \frac{\nabla _{\hat{X}%
}F_{0}\left( R\left( K_{\hat{X}},\hat{X}\right) \right) }{\left\Vert \nabla
_{\hat{X}}R\left( K_{\hat{X}},\hat{X}\right) \right\Vert }+\nu \nabla _{\hat{%
X}}F_{1}\left( \frac{R\left( K_{\hat{X}},\hat{X}\right) }{\int R\left(
K_{X^{\prime }}^{\prime },X^{\prime }\right) \left\Vert \Psi \left(
X^{\prime }\right) \right\Vert ^{2}dX^{\prime }}\right) \right)
\end{eqnarray}%
Using a change of variable (see appendix 3.1.2):%
\begin{eqnarray}
\hat{\Psi} &\rightarrow &\exp \left( \frac{1}{\sigma _{\hat{X}}^{2}}\int
g\left( \hat{X}\right) d\hat{X}+\frac{\hat{K}^{2}}{\sigma _{\hat{K}}^{2}}%
f\left( \hat{X}\right) \right) \hat{\Psi}  \label{chg} \\
\hat{\Psi}^{\dag } &\rightarrow &\exp \left( \frac{1}{\sigma _{\hat{X}}^{2}}%
\int g\left( \hat{X}\right) d\hat{X}+\frac{\hat{K}^{2}}{\sigma _{\hat{K}}^{2}%
}f\left( \hat{X}\right) \right) \hat{\Psi}^{\dag }  \notag
\end{eqnarray}%
the action (\ref{STF}) becomes:%
\begin{eqnarray}
&&S_{3}+S_{4}=-\int \hat{\Psi}^{\dag }\left( \frac{\sigma _{\hat{X}}^{2}}{2}%
\nabla _{\hat{X}}^{2}-\frac{1}{2\sigma _{\hat{X}}^{2}}\left( g\left( \hat{X}%
,K_{\hat{X}}\right) \right) ^{2}-\frac{1}{2}\nabla _{\hat{X}}g\left( \hat{X}%
,K_{\hat{X}}\right) \right) \hat{\Psi}  \label{stm} \\
&&-\int \hat{\Psi}^{\dag }\left( \nabla _{\hat{K}}\left( \frac{\sigma _{\hat{%
K}}^{2}}{2}\nabla _{\hat{K}}-\hat{K}f\left( \hat{X},K_{\hat{X}}\right)
\right) \right) \hat{\Psi}  \notag
\end{eqnarray}

To obtain the second-order expansion of the field's action, we start by the
first derivative of (\ref{stm}) arising in the minimization equation in
(Gosselin Lotz Wambst 2022):%
\begin{eqnarray}
\frac{\delta \left( S_{3}\left( \Psi \right) +S_{4}\left( \Psi \right)
\right) }{\delta \hat{\Psi}^{\dag }\left( Z,\theta \right) } &=&-\frac{%
\sigma _{\hat{X}}^{2}}{2}\nabla _{\hat{X}}^{2}\hat{\Psi}-\frac{\sigma _{\hat{%
K}}^{2}}{2}\nabla _{\hat{K}}^{2}\hat{\Psi}+\frac{1}{2\sigma _{\hat{X}}^{2}}%
\left( g\left( \hat{X},K_{\hat{X}}\right) \right) ^{2}+\frac{1}{2}\nabla _{%
\hat{X}}g\left( \hat{X},K_{\hat{X}}\right) \hat{\Psi}  \label{SCP} \\
&&+\frac{\hat{K}^{2}}{2\sigma _{\hat{K}}^{2}}f^{2}\left( \hat{X}\right) +%
\frac{1}{2}f\left( \hat{X},K_{\hat{X}}\right) \hat{\Psi}+F\left( \hat{X},K_{%
\hat{X}}\right) \hat{K}\hat{\Psi}  \notag
\end{eqnarray}%
with:%
\begin{eqnarray}
F\left( \hat{X},K_{\hat{X}}\right) &=&\nabla _{K_{\hat{X}}}\left( \frac{%
\left( g\left( \hat{X},K_{\hat{X}}\right) \right) ^{2}}{2\sigma _{\hat{X}%
}^{2}}+\frac{1}{2}\nabla _{\hat{X}}g\left( \hat{X},K_{\hat{X}}\right)
+f\left( \hat{X},K_{\hat{X}}\right) \right) \frac{\left\Vert \hat{\Psi}%
\left( \hat{X}\right) \right\Vert ^{2}}{\left\Vert \Psi \left( \hat{X}%
\right) \right\Vert ^{2}} \\
&&+\frac{\nabla _{K_{\hat{X}}}f^{2}\left( \hat{X},K_{\hat{X}}\right) }{%
\sigma _{\hat{K}}^{2}\left\Vert \Psi \left( \hat{X}\right) \right\Vert ^{2}}%
\left\langle \hat{K}^{2}\right\rangle _{\hat{X}}  \notag
\end{eqnarray}%
so that :%
\begin{eqnarray*}
\frac{\delta ^{2}\left( S_{3}\left( \Psi \right) +S_{4}\left( \Psi \right)
\right) }{\delta \hat{\Psi}^{\dag }\left( Z,\theta \right) \delta \hat{\Psi}%
\left( Z,\theta \right) } &=&-\frac{\sigma _{\hat{X}}^{2}}{2}\nabla _{\hat{X}%
}^{2}-\frac{\sigma _{\hat{K}}^{2}}{2}\nabla _{\hat{K}}^{2}+\frac{1}{2\sigma
_{\hat{X}}^{2}}\left( g\left( \hat{X},K_{\hat{X}}\right) \right) ^{2}+\frac{1%
}{2}\nabla _{\hat{X}}g\left( \hat{X},K_{\hat{X}}\right) \\
&&+\frac{\hat{K}^{2}}{2\sigma _{\hat{K}}^{2}}f^{2}\left( \hat{X}\right) +%
\frac{1}{2}f\left( \hat{X},K_{\hat{X}}\right) +F\left( \hat{X},K_{\hat{X}%
}\right) \hat{K}-\hat{\Psi}^{\dag }\frac{\delta F\left( \hat{X},K_{\hat{X}%
}\right) }{\delta \left\Vert \hat{\Psi}\left( \hat{K},\hat{X}\right)
\right\Vert ^{2}}\hat{K}\hat{\Psi}
\end{eqnarray*}%
where the last term is given by:%
\begin{eqnarray*}
\frac{\delta F\left( \hat{X},K_{\hat{X}}\right) }{\delta \hat{\Psi}\left(
Z,\theta \right) } &\simeq &\nabla _{K_{\hat{X}}}\left( \frac{\left( g\left( 
\hat{X},K_{\hat{X}}\right) \right) ^{2}}{2\sigma _{\hat{X}}^{2}}+\frac{1}{2}%
\nabla _{\hat{X}}g\left( \hat{X},K_{\hat{X}}\right) +f\left( \hat{X},K_{\hat{%
X}}\right) \right) \frac{\hat{\Psi}^{\dagger }\left( \hat{K},\hat{X}\right) 
}{\left\Vert \Psi \left( \hat{X}\right) \right\Vert ^{2}} \\
&&+\frac{\nabla _{K_{\hat{X}}}f^{2}\left( \hat{X},K_{\hat{X}}\right) }{%
\sigma _{\hat{K}}^{2}\left\Vert \Psi \left( \hat{X}\right) \right\Vert ^{2}}%
\hat{K}^{2}\hat{\Psi}^{\dagger }\left( \hat{K},\hat{X}\right)
\end{eqnarray*}%
Following (Gosselin Lotz Wambst 2022) we neglect in first approximation the
derivatives with respect to $K_{\hat{X}}$ , and define the new variable:%
\begin{equation}
y=\frac{\hat{K}+\frac{\sigma _{\hat{K}}^{2}F\left( \hat{X},K_{\hat{X}%
}\right) }{f^{2}\left( \hat{X}\right) }}{\sqrt{\sigma _{\hat{K}}^{2}}}\left(
f^{2}\left( \hat{X}\right) \right) ^{\frac{1}{4}}  \label{cvr}
\end{equation}%
\begin{equation}
\frac{\delta ^{2}\left( S_{3}\left( \Psi \right) +S_{4}\left( \Psi \right)
\right) }{\delta \hat{\Psi}^{\dag }\left( Z,\theta \right) \delta \hat{\Psi}%
\left( Z,\theta \right) }=-\frac{\sigma _{\hat{X}}^{2}}{2}\nabla _{\hat{X}%
}^{2}-\nabla _{y}^{2}+\left( \frac{y^{2}}{4}+\frac{\left( g\left( \hat{X}%
\right) \right) ^{2}+\sigma _{\hat{X}}^{2}\left( f\left( \hat{X}\right)
+\nabla _{\hat{X}}g\left( \hat{X},K_{\hat{X}}\right) -\frac{\sigma _{\hat{K}%
}^{2}F^{2}\left( \hat{X},K_{\hat{X}}\right) }{2f^{2}\left( \hat{X}\right) }%
\right) }{\sigma _{\hat{X}}^{2}\sqrt{f^{2}\left( \hat{X}\right) }}\right)
\label{SCQ}
\end{equation}%
This leads to:%
\begin{eqnarray*}
&&\Delta \hat{\Psi}^{\dag }\left( Z,\theta \right) \frac{\delta ^{2}\left(
S_{3}\left( \Psi \right) +S_{4}\left( \Psi \right) \right) }{\delta \hat{\Psi%
}^{\dag }\left( Z,\theta \right) \delta \hat{\Psi}\left( Z,\theta \right) }%
\Delta \hat{\Psi}\left( Z,\theta \right) \\
&=&\Delta \hat{\Psi}^{\dag }\left( Z,\theta \right) \left( -\frac{\sigma _{%
\hat{X}}^{2}}{2}\nabla _{\hat{X}}^{2}+\frac{\left( g\left( \hat{X}\right)
\right) ^{2}+\sigma _{\hat{X}}^{2}\left( f\left( \hat{X}\right) +\nabla _{%
\hat{X}}g\left( \hat{X},K_{\hat{X}}\right) -\frac{\sigma _{\hat{K}%
}^{2}F^{2}\left( \hat{X},K_{\hat{X}}\right) }{2f^{2}\left( \hat{X}\right) }%
\right) }{\sigma _{\hat{X}}^{2}\sqrt{f^{2}\left( \hat{X}\right) }}\right. \\
&&\left. -\frac{\sigma _{\hat{K}}^{2}}{2\sqrt{f^{2}\left( \hat{X}\right) }}%
\nabla _{\hat{K}}^{2}+\left( \frac{\sqrt{f^{2}\left( \hat{X}\right) }\left( 
\hat{K}+\frac{\sigma _{\hat{K}}^{2}F\left( \hat{X},K_{\hat{X}}\right) }{%
f^{2}\left( \hat{X}\right) }\right) ^{2}}{4\sigma _{\hat{K}}^{2}}\right)
\right) \Delta \hat{\Psi}\left( Z,\theta \right)
\end{eqnarray*}

\section*{Appendix 2 Higher order corrections to the effective action}

The higher-order corrections are obtained by expanding at higher-orders in $%
\Delta \Psi \left( Z,\theta \right) $ and $\Delta \hat{\Psi}\left( Z,\theta
\right) $. These variations around the background fields can be considered
to be orthogonals to $\Psi _{0}\left( Z,\theta \right) $ and $\hat{\Psi}%
_{0}\left( Z,\theta \right) $.

\subsection*{Third order terms}

The orthogonality condition implies that the third-order terms in the
expansion can be neglected. Actually, in first approximation the third-order
terms arising in the expansion of $S$ have the form:%
\begin{eqnarray}
&&2\tau \int \Delta \Psi \left( K^{\prime },X\right) \Psi _{0}^{\dag }\left(
K^{\prime },X^{\prime }\right) dK^{\prime }\left\vert \Delta \Psi \left(
K,X\right) \right\vert ^{2}dKdX  \label{CLC} \\
&&-\int \Delta \Psi ^{\dag }\left( K,X\right) \Psi _{0}^{\dag }\left(
K^{\prime },X^{\prime }\right) \nabla _{K}\frac{\delta u\left( K,X,\Psi ,%
\hat{\Psi}\right) }{\delta \left\vert \Psi \left( K^{\prime },X\right)
\right\vert ^{2}}\Delta \Psi \left( K^{\prime },X^{\prime }\right) \Delta
\Psi \left( K,X\right)  \notag \\
&&-\int \Delta \Psi ^{\dagger }\left( K,\theta \right) \hat{\Psi}%
_{0}^{\dagger }\left( \hat{K},\theta \right) \nabla _{K}\frac{\delta u\left(
K,X,\Psi ,\hat{\Psi}\right) }{\delta \left\vert \hat{\Psi}\left( \hat{K},%
\hat{X}\right) \right\vert ^{2}}\Delta \hat{\Psi}\left( \hat{K},\theta
\right) \Delta \Psi \left( K,\theta \right)  \notag \\
&&-\int \Delta \hat{\Psi}^{\dag }\left( \hat{K},\hat{X}\right) \Psi
_{0}^{\dag }\left( K^{\prime },\theta \right) \left\{ \nabla _{\hat{K}}\frac{%
\hat{K}\delta ^{2}f\left( \hat{X},\Psi ,\hat{\Psi}\right) }{\delta
\left\vert \Psi \left( K^{\prime },X\right) \right\vert ^{2}}+\nabla _{\hat{X%
}}\frac{\delta g\left( \hat{X},\Psi ,\hat{\Psi}\right) }{\delta \left\vert
\Psi \left( K^{\prime },X\right) \right\vert ^{2}}\right\} \Delta \Psi
\left( K^{\prime },X^{\prime }\right) \Delta \hat{\Psi}\left( \hat{K},\hat{X}%
\right)  \notag \\
&&+H.C.  \notag
\end{eqnarray}%
where the notation $H.C.$ stands for the hermitian conjugate of the
expression. Replacing the terms:%
\begin{equation*}
\nabla _{K}\frac{\delta u\left( K,X,\Psi ,\hat{\Psi}\right) }{\delta
\left\vert \Psi \left( K^{\prime },X\right) \right\vert ^{2}}\text{ , }%
\nabla _{K}\frac{\delta u\left( K,X,\Psi ,\hat{\Psi}\right) }{\delta
\left\vert \hat{\Psi}\left( \hat{K},\hat{X}\right) \right\vert ^{2}}
\end{equation*}%
and:%
\begin{equation*}
\nabla _{\hat{K}}\frac{\hat{K}\delta ^{2}f\left( \hat{X},\Psi ,\hat{\Psi}%
\right) }{\delta \left\vert \Psi \left( K^{\prime },X\right) \right\vert ^{2}%
}+\nabla _{\hat{X}}\frac{\delta g\left( \hat{X},\Psi ,\hat{\Psi}\right) }{%
\delta \left\vert \Psi \left( K^{\prime },X\right) \right\vert ^{2}}
\end{equation*}%
by their averages in (\ref{CLC}), and using the orthogonality conditions:%
\begin{equation*}
\int \hat{\Psi}_{0}^{\dagger }\left( \hat{K},\theta \right) \Delta \hat{\Psi}%
\left( \hat{K},\theta \right) =\int \Psi _{0}^{\dag }\left( K^{\prime
},X^{\prime }\right) \Delta \Psi \left( K^{\prime },X^{\prime }\right) =0
\end{equation*}%
leads to neglect the third-order terms in first approximation.

\subsection*{Fourth order terms}

\subsubsection*{\protect\bigskip General formula}

Considering the fourth-order in the action expansion yields quartic
corrections. Using that in average:

\begin{equation*}
\frac{\delta ^{2}f\left( \hat{X},\Psi ,\hat{\Psi}\right) }{\delta \hat{\Psi}%
\left( \hat{K},\hat{X}\right) \delta \hat{\Psi}^{\dagger }\left( \hat{K},%
\hat{X}\right) }\simeq 0
\end{equation*}%
\begin{equation*}
\frac{\delta ^{2}g\left( \hat{X},\Psi ,\hat{\Psi}\right) }{\delta \hat{\Psi}%
\left( \hat{K},\hat{X}\right) \delta \hat{\Psi}^{\dagger }\left( \hat{K},%
\hat{X}\right) }\simeq 0
\end{equation*}%
the fourth-order terms in the fields' action become:%
\begin{eqnarray}
&&2\tau \int \left\vert \Delta \Psi \left( K^{\prime },X\right) \right\vert
^{2}dK^{\prime }\left\vert \Delta \Psi \left( K,X\right) \right\vert ^{2}dKdX
\label{FRT} \\
&&-\Delta \Psi ^{\dag }\left( K,X\right) \Delta \Psi ^{\dag }\left(
K^{\prime },X^{\prime }\right) \nabla _{K}\frac{\delta ^{2}u\left( K,X,\Psi ,%
\hat{\Psi}\right) }{\delta \Psi \left( K^{\prime },X\right) \delta \Psi
^{\dagger }\left( K^{\prime },X\right) }\Delta \Psi \left( K^{\prime
},X^{\prime }\right) \Delta \Psi \left( K,X\right)  \notag \\
&&-\Delta \Psi ^{\dagger }\left( K,\theta \right) \Delta \hat{\Psi}^{\dagger
}\left( \hat{K},\theta \right) \nabla _{K}\frac{\delta ^{2}u\left( K,X,\Psi ,%
\hat{\Psi}\right) }{\delta \hat{\Psi}\left( \hat{K},\hat{X}\right) \delta 
\hat{\Psi}^{\dagger }\left( \hat{K},\hat{X}\right) }\Delta \hat{\Psi}\left( 
\hat{K},\theta \right) \Delta \Psi \left( K,\theta \right)  \notag \\
&&-\Delta \hat{\Psi}^{\dag }\left( \hat{K},\hat{X}\right) \Delta \Psi ^{\dag
}\left( K^{\prime },\theta \right) \left\{ \nabla _{\hat{K}}\frac{\hat{K}%
\delta ^{2}f\left( \hat{X},\Psi ,\hat{\Psi}\right) }{\delta \Psi \left(
K^{\prime },X\right) \delta \Psi ^{\dagger }\left( K^{\prime },X\right) }%
+\nabla _{\hat{X}}\frac{\delta ^{2}g\left( \hat{X},\Psi ,\hat{\Psi}\right) }{%
\delta \Psi \left( K^{\prime },X\right) \delta \Psi ^{\dagger }\left(
K^{\prime },X\right) }\right\} \Delta \Psi \left( K^{\prime },X^{\prime
}\right) \Delta \hat{\Psi}\left( \hat{K},\hat{X}\right)  \notag
\end{eqnarray}

\subsubsection*{\protect\bigskip Estimation of the various terms}

The three last terms in the rhs of (\ref{FRT}) can be evaluated. The second
term is given by:%
\begin{eqnarray*}
&&\Delta \Psi ^{\dag }\left( K,X\right) \Delta \Psi ^{\dag }\left( K^{\prime
},X^{\prime }\right) \frac{\delta ^{2}u\left( K,X,\Psi ,\hat{\Psi}\right) }{%
\delta \Psi \left( K^{\prime },X\right) \delta \Psi ^{\dagger }\left(
K^{\prime },X\right) }\Delta \Psi \left( K,X\right) \Delta \Psi \left(
K^{\prime },\theta \right) \\
&=&\Delta \Psi ^{\dag }\left( K,\theta \right) \Delta \Psi ^{\dag }\left(
K^{\prime },\theta \right) \left\{ \int \hat{F}_{2}\left( s,R\left(
K,X\right) \right) \hat{F}_{2}\left( s^{\prime },R\left( K^{\prime
},X^{\prime }\right) \right) \hat{K}\left\Vert \hat{\Psi}\left( \hat{K}%
,X\right) \right\Vert ^{2}d\hat{K}\right\} \Delta \Psi \left( K,\theta
\right) \Delta \Psi \left( K^{\prime },\theta \right) \\
&&-2\Delta \Psi ^{\dag }\left( K,\theta \right) \Delta \Psi ^{\dag }\left(
K^{\prime },\theta \right) \\
&&\times \left\{ \int \Psi _{0}^{\dag }\left( K^{\prime },X\right) \frac{%
\hat{F}_{2}\left( s,R\left( K,X\right) \right) \hat{F}_{2}\left( s^{\prime
},R\left( K^{\prime },X^{\prime }\right) \right) }{\int F_{2}\left(
s^{\prime },R\left( K^{\prime },X\right) \right) \left\Vert \Psi \left(
K^{\prime },X\right) \right\Vert ^{2}dK^{\prime }}\Psi _{0}\left( K,\hat{X}%
\right) \hat{K}\left\Vert \hat{\Psi}\left( \hat{K},X\right) \right\Vert ^{2}d%
\hat{K}\right\} \Delta \Psi \left( K,\theta \right) \Delta \Psi \left(
K^{\prime },\theta \right) \\
&\simeq &\Delta \Psi ^{\dag }\left( K,\theta \right) \Delta \Psi ^{\dag
}\left( K^{\prime },\theta \right) \left\{ \int \hat{F}_{2}\left( s,R\left(
K,X\right) \right) \hat{F}_{2}\left( s^{\prime },R\left( K^{\prime
},X^{\prime }\right) \right) \hat{K}\left\Vert \hat{\Psi}\left( \hat{K}%
,X\right) \right\Vert ^{2}d\hat{K}\right\} \Delta \Psi \left( K,\theta
\right) \Delta \Psi \left( K^{\prime },\theta \right)
\end{eqnarray*}

The second term in the rhs of (\ref{FRT}) is equal to:%
\begin{eqnarray*}
&&\Delta \Psi ^{\dagger }\left( K,\theta \right) \Delta \hat{\Psi}^{\dagger
}\left( \hat{K},\theta \right) \frac{\delta ^{2}u\left( K,X,\Psi ,\hat{\Psi}%
\right) }{\delta \hat{\Psi}\left( \hat{K},\hat{X}\right) \delta \hat{\Psi}%
^{\dagger }\left( \hat{K},\hat{X}\right) }\Delta \Psi \left( K,\theta
\right) \Delta \hat{\Psi}\left( \hat{K},\theta \right) \\
&=&-\Delta \Psi ^{\dagger }\left( K,\theta \right) \Delta \hat{\Psi}%
^{\dagger }\left( \hat{K},\theta \right) \frac{1}{\varepsilon }\hat{F}%
_{2}\left( s,R\left( K,X\right) \right) \hat{K}\Delta \Psi \left( K,\theta
\right) \Delta \hat{\Psi}\left( \hat{K},\theta \right)
\end{eqnarray*}%
Ultimately, the last term in the rhs of (\ref{FRT}):%
\begin{equation*}
\Delta \hat{\Psi}^{\dag }\left( \hat{K},\hat{X}\right) \Delta \Psi ^{\dag
}\left( K^{\prime },\theta \right) \left\{ \nabla _{\hat{K}}\frac{\hat{K}%
\delta ^{2}f\left( \hat{X},\Psi ,\hat{\Psi}\right) }{\delta \Psi \left(
K^{\prime },X\right) \delta \Psi ^{\dagger }\left( K^{\prime },X\right) }%
+\nabla _{\hat{X}}\frac{\delta ^{2}g\left( \hat{X},\Psi ,\hat{\Psi}\right) }{%
\delta \Psi \left( K^{\prime },X\right) \delta \Psi ^{\dagger }\left(
K^{\prime },X\right) }\right\} \Delta \Psi \left( K^{\prime },X^{\prime
}\right) \Delta \hat{\Psi}\left( \hat{K},\hat{X}\right)
\end{equation*}%
is obtained by using the expressions of $f\left( \hat{X},\Psi ,\hat{\Psi}%
\right) $ and $g\left( \hat{X},\Psi ,\hat{\Psi}\right) $ that compute
short-term and long-term returns, respectively:

\begin{eqnarray*}
f\left( \hat{X},\Psi ,\hat{\Psi}\right) &=&\frac{1}{\varepsilon }\int \left(
r\left( K,X\right) -\gamma \frac{\int K^{\prime }\left\Vert \Psi \left(
K^{\prime },X\right) \right\Vert ^{2}}{K}+F_{1}\left( \frac{R\left(
K,X\right) }{\int R\left( K^{\prime },X^{\prime }\right) \left\Vert \Psi
\left( K^{\prime },X^{\prime }\right) \right\Vert ^{2}d\left( K^{\prime
},X^{\prime }\right) },\Gamma \left( K,X\right) \right) \right) \\
&&\times \hat{F}_{2}\left( s,R\left( K,X\right) \right) \left\Vert \Psi
\left( K,\hat{X}\right) \right\Vert ^{2}dK \\
g\left( K,\hat{X},\Psi ,\hat{\Psi}\right) &=&\int \left( \nabla _{\hat{X}%
}F_{0}\left( R\left( K,\hat{X}\right) \right) +\nu \nabla _{\hat{X}%
}F_{1}\left( \frac{R\left( K,\hat{X}\right) }{\int R\left( K^{\prime
},X^{\prime }\right) \left\Vert \Psi \left( K^{\prime },X^{\prime }\right)
\right\Vert ^{2}d\left( K^{\prime },X^{\prime }\right) }\right) \right) \\
&&\times \frac{\left\Vert \Psi \left( K,\hat{X}\right) \right\Vert ^{2}dK}{%
\int \left\Vert \Psi \left( K^{\prime },\hat{X}\right) \right\Vert
^{2}dK^{\prime }}
\end{eqnarray*}%
We find:%
\begin{eqnarray}
&&\frac{\delta ^{2}f\left( \hat{X},\Psi ,\hat{\Psi}\right) }{\delta \Psi
\left( K^{\prime },X\right) \delta \Psi ^{\dagger }\left( K^{\prime
},X\right) }  \label{DLT} \\
&=&\frac{1}{\varepsilon }\Delta \left( r\left( K^{\prime },X\right) -\gamma 
\frac{\int K^{\prime }\left\Vert \Psi \left( K^{\prime },X\right)
\right\Vert ^{2}}{K^{\prime }}+F_{1}\left( \frac{R\left( K^{\prime
},X\right) }{\int R\left( K^{\prime },X^{\prime }\right) \left\Vert \Psi
\left( K^{\prime },X^{\prime }\right) \right\Vert ^{2}d\left( K^{\prime
},X^{\prime }\right) },\Gamma \left( K,X\right) \right) \right)  \notag \\
&&\times \frac{F_{2}\left( s^{\prime },R\left( K^{\prime },\hat{X}\right)
\right) }{\int F_{2}\left( s^{\prime }R\left( K^{\prime },\hat{X}\right)
\right) \left\Vert \Psi \left( K^{\prime },\hat{X}\right) \right\Vert
^{2}dK^{\prime }}  \notag \\
&&-\frac{1}{\varepsilon }\int \left( \gamma \frac{K^{\prime }}{K}+\frac{%
R\left( K^{\prime },X\right) R\left( K_{\hat{X}},X\right) }{\left( \int
R\left( K^{\prime },X^{\prime }\right) \left\Vert \Psi \left( K^{\prime
},X^{\prime }\right) \right\Vert ^{2}d\left( K^{\prime },X^{\prime }\right)
\right) ^{2}}F_{1}^{\prime }\left( \frac{R\left( K,X\right) }{\int R\left(
K^{\prime },X^{\prime }\right) \left\Vert \Psi \left( K^{\prime },X^{\prime
}\right) \right\Vert ^{2}d\left( K^{\prime },X^{\prime }\right) },\Gamma
\left( K,X\right) \right) \right)  \notag \\
&&\times \hat{F}_{2}\left( s,R\left( K,X\right) \right) \left\Vert \Psi
\left( K,\hat{X}\right) \right\Vert ^{2}  \notag
\end{eqnarray}%
where we define the deviation $\Delta Y$ of a quantity by the difference:%
\begin{equation}
\Delta Y=Y-\left\langle Y\right\rangle  \label{RV}
\end{equation}%
with $\left\langle Y\right\rangle $, the average of $Y$:%
\begin{equation*}
\left\langle Y\right\rangle =\int Y\left( K,X\right) dKdX
\end{equation*}%
\textbf{\ }Thus we write:%
\begin{eqnarray*}
&&\Delta \left( r\left( K^{\prime },X\right) -\gamma \frac{\int K^{\prime
}\left\Vert \Psi \left( K^{\prime },X\right) \right\Vert ^{2}}{K^{\prime }}%
+F_{1}\left( \frac{R\left( K^{\prime },X\right) }{\int R\left( K^{\prime
},X^{\prime }\right) \left\Vert \Psi \left( K^{\prime },X^{\prime }\right)
\right\Vert ^{2}d\left( K^{\prime },X^{\prime }\right) },\Gamma \left(
K,X\right) \right) \right) \\
&=&\left( r\left( K^{\prime },X\right) -\gamma \frac{\int K^{\prime
}\left\Vert \Psi \left( K^{\prime },X\right) \right\Vert ^{2}}{K^{\prime }}%
+F_{1}\left( \frac{R\left( K^{\prime },X\right) }{\int R\left( K^{\prime
},X^{\prime }\right) \left\Vert \Psi \left( K^{\prime },X^{\prime }\right)
\right\Vert ^{2}d\left( K^{\prime },X^{\prime }\right) },\Gamma \left(
K,X\right) \right) \right) \\
&&-\left\langle \left( r\left( K^{\prime },X\right) -\gamma \frac{\int
K^{\prime }\left\Vert \Psi \left( K^{\prime },X\right) \right\Vert ^{2}}{%
K^{\prime }}+F_{1}\left( \frac{R\left( K^{\prime },X\right) }{\int R\left(
K^{\prime },X^{\prime }\right) \left\Vert \Psi \left( K^{\prime },X^{\prime
}\right) \right\Vert ^{2}d\left( K^{\prime },X^{\prime }\right) },\Gamma
\left( K,X\right) \right) \right) \right\rangle
\end{eqnarray*}%
and in first approximation, (\ref{DLT}) reduces to:%
\begin{eqnarray*}
&&\frac{\delta ^{2}f\left( \hat{X},\Psi ,\hat{\Psi}\right) }{\delta \Psi
\left( K^{\prime },X\right) \delta \Psi ^{\dagger }\left( K^{\prime
},X\right) } \\
&\simeq &\frac{1}{\varepsilon }\left( \Delta \left( r\left( K^{\prime
},X\right) -\gamma \frac{K_{X}}{K^{\prime }}+F_{1}\left( \frac{R\left(
K^{\prime },X\right) }{\int R\left( K^{\prime },X^{\prime }\right)
\left\Vert \Psi \left( K^{\prime },X^{\prime }\right) \right\Vert
^{2}d\left( K^{\prime },X^{\prime }\right) },\Gamma \left( K,X\right)
\right) \right) -\gamma \frac{K^{\prime }}{K_{X}}\right) \\
&\simeq &\frac{1}{\varepsilon }\left( \Delta f\left( K^{\prime },\hat{X}%
,\Psi ,\hat{\Psi}\right) -\gamma \frac{K^{\prime }}{K_{X}}\right)
\end{eqnarray*}%
where:%
\begin{equation*}
\Delta f\left( K^{\prime },\hat{X},\Psi ,\hat{\Psi}\right) =f\left(
K^{\prime },\hat{X},\Psi ,\hat{\Psi}\right) -f\left( K_{\hat{X}},\hat{X}%
,\Psi ,\hat{\Psi}\right)
\end{equation*}%
is the relative short-term return for firm with capital $K^{\prime }$ at
sector $\hat{X}$.

Similarly, the second derivative for $g\left( \hat{X},\Psi ,\hat{\Psi}%
\right) $ is:%
\begin{eqnarray*}
&&\frac{\delta ^{2}g\left( \hat{X},\Psi ,\hat{\Psi}\right) }{\delta \Psi
\left( K^{\prime },X\right) \delta \Psi ^{\dagger }\left( K^{\prime
},X\right) } \\
&=&\frac{1}{\int \left\Vert \Psi \left( K^{\prime },\hat{X}\right)
\right\Vert ^{2}dK^{\prime }}\Delta \left( \nabla _{\hat{X}}F_{0}\left(
R\left( K^{\prime },\hat{X}\right) \right) +\nu \nabla _{\hat{X}}F_{1}\left( 
\frac{R\left( K^{\prime },\hat{X}\right) }{\int R\left( K^{\prime
},X^{\prime }\right) \left\Vert \Psi \left( K^{\prime },X^{\prime }\right)
\right\Vert ^{2}d\left( K^{\prime },X^{\prime }\right) }\right) \right) \\
&=&\frac{1}{\int \left\Vert \Psi \left( K^{\prime },\hat{X}\right)
\right\Vert ^{2}dK^{\prime }}\Delta \left( g\left( K^{\prime },\hat{X},\Psi ,%
\hat{\Psi}\right) \right)
\end{eqnarray*}%
with:%
\begin{eqnarray*}
&&\Delta \left( \nabla _{\hat{X}}F_{0}\left( R\left( K^{\prime },\hat{X}%
\right) \right) +\nu \nabla _{\hat{X}}F_{1}\left( \frac{R\left( K^{\prime },%
\hat{X}\right) }{\int R\left( K^{\prime },X^{\prime }\right) \left\Vert \Psi
\left( K^{\prime },X^{\prime }\right) \right\Vert ^{2}d\left( K^{\prime
},X^{\prime }\right) }\right) \right) \\
&=&\left( \nabla _{\hat{X}}F_{0}\left( R\left( K^{\prime },\hat{X}\right)
\right) +\nu \nabla _{\hat{X}}F_{1}\left( \frac{R\left( K^{\prime },\hat{X}%
\right) }{\int R\left( K^{\prime },X^{\prime }\right) \left\Vert \Psi \left(
K^{\prime },X^{\prime }\right) \right\Vert ^{2}d\left( K^{\prime },X^{\prime
}\right) }\right) \right) \\
&&-\left\langle \nabla _{\hat{X}}F_{0}\left( R\left( K^{\prime },\hat{X}%
\right) \right) +\nu \nabla _{\hat{X}}F_{1}\left( \frac{R\left( K^{\prime },%
\hat{X}\right) }{\int R\left( K^{\prime },X^{\prime }\right) \left\Vert \Psi
\left( K^{\prime },X^{\prime }\right) \right\Vert ^{2}d\left( K^{\prime
},X^{\prime }\right) }\right) \right\rangle
\end{eqnarray*}%
in other words:%
\begin{equation*}
\Delta g\left( K^{\prime },\hat{X},\Psi ,\hat{\Psi}\right) =g\left(
K^{\prime },\hat{X},\Psi ,\hat{\Psi}\right) -g\left( \hat{X},\Psi ,\hat{\Psi}%
\right)
\end{equation*}%
is the relative long-term return for firm with capital $K^{\prime }$ at
sector $\hat{X}$.

\section*{Appendix 3: "free" transition functions}

Given the second-order operator arising in the expansion for the fields'
action:%
\begin{equation}
O\left( \Psi _{0}\left( Z,\theta \right) \right) \simeq \left( 
\begin{array}{cc}
\frac{\delta ^{2}\left( S_{1}+S_{2}\right) }{\delta \Psi ^{\dag }\left(
Z,\theta \right) \delta \Psi \left( Z,\theta \right) } & 0 \\ 
0 & \frac{\delta ^{2}\left( S_{3}\left( \Psi \right) +S_{4}\left( \Psi
\right) \right) }{\delta \hat{\Psi}^{\dag }\left( Z,\theta \right) \delta 
\hat{\Psi}\left( Z,\theta \right) }%
\end{array}%
\right) _{\substack{ \Psi \left( Z,\theta \right) =\Psi _{0}\left( Z,\theta
\right)  \\ \hat{\Psi}\left( Z,\theta \right) =\hat{\Psi}_{0}\left( Z,\theta
\right) }}
\end{equation}%
The transition functions for the individual firms:

\begin{equation*}
G_{1}\left( \left( K_{f},X_{f}\right) ,\left( X_{i},K_{i}\right) ,\alpha
\right)
\end{equation*}%
and investors:%
\begin{equation*}
G_{2}\left( \left( \hat{K}_{f},\hat{X}_{f}\right) ,\left( \hat{X}_{i},\hat{K}%
_{i}\right) ,\alpha \right)
\end{equation*}%
satisfy:%
\begin{eqnarray*}
\left( \frac{\delta ^{2}\left( S_{1}+S_{2}\right) }{\delta \Psi ^{\dag
}\left( Z,\theta \right) \delta \Psi \left( Z,\theta \right) }+\alpha
\right) G_{1}\left( \left( K_{f},X_{f}\right) ,\left( X_{i},K_{i}\right)
,\alpha \right) &=&\delta \left( \left( K_{f},X_{f}\right) -\left(
X_{i},K_{i}\right) \right) \\
\left( \frac{\delta ^{2}\left( S_{3}\left( \Psi \right) +S_{4}\left( \Psi
\right) \right) }{\delta \hat{\Psi}^{\dag }\left( Z,\theta \right) \delta 
\hat{\Psi}\left( Z,\theta \right) }+\alpha \right) G_{2}\left( \left( \hat{K}%
_{f},\hat{X}_{f}\right) ,\left( \hat{X}_{i},\hat{K}_{i}\right) ,\alpha
\right) &=&\delta \left( \left( \hat{K}_{f},\hat{X}_{f}\right) -\left( \hat{X%
}_{i},\hat{K}_{i}\right) \right)
\end{eqnarray*}%
The functions $G_{1}\left( \left( K_{f},X_{f}\right) ,\left(
X_{i},K_{i}\right) ,\alpha \right) $ and $G_{2}\left( \left( \hat{K}_{f},%
\hat{X}_{f}\right) ,\left( \hat{X}_{i},\hat{K}_{i}\right) ,\alpha \right) $\
are the Laplace transforms of the following transition functions:%
\begin{equation*}
T_{1}\left( \left( K_{f},X_{f}\right) ,\left( X_{i},K_{i}\right) ,t\right)
\end{equation*}%
\begin{equation*}
T_{2}\left( \left( \hat{K}_{f},\hat{X}_{f}\right) ,\left( \hat{X}_{i},\hat{K}%
_{i}\right) ,t\right)
\end{equation*}%
satisfying:%
\begin{equation}
-\frac{\partial }{\partial t}T_{1}\left( \left( K_{f},X_{f}\right) ,\left(
X_{i},K_{i}\right) ,t\right) =\left( \frac{\delta ^{2}\left(
S_{1}+S_{2}\right) }{\delta \Psi ^{\dag }\left( Z,\theta \right) \delta \Psi
\left( Z,\theta \right) }\right) T_{1}\left( \left( K_{f},X_{f}\right)
,\left( X_{i},K_{i}\right) ,t\right)  \label{TNQ}
\end{equation}%
\begin{equation}
-\frac{\partial }{\partial t}T_{2}\left( \left( \hat{K}_{f},\hat{X}%
_{f}\right) ,\left( \hat{X}_{i},\hat{K}_{i}\right) ,t\right) =\left( \frac{%
\delta ^{2}\left( S_{3}\left( \Psi \right) +S_{4}\left( \Psi \right) \right) 
}{\delta \hat{\Psi}^{\dag }\left( Z,\theta \right) \delta \hat{\Psi}\left(
Z,\theta \right) }\right) T_{2}\left( \left( \hat{K}_{f},\hat{X}_{f}\right)
,\left( \hat{X}_{i},\hat{K}_{i}\right) ,t\right)  \label{TSQ}
\end{equation}

\subsection*{Approximations to and (\protect\ref{TNQ}) \ and (\protect\ref%
{TSQ})}

We consider some approximations to find the solutions of equations (\ref{SCR}%
) and (\ref{SCQ}). We first assume that:%
\begin{equation*}
\frac{\nabla _{K}\frac{F_{2}\left( R\left( K,X\right) \right) }{\left\langle
F_{2}\left( R\left( K,X\right) \right) \right\rangle _{K}}K_{X}}{2}<<1
\end{equation*}%
so that:%
\begin{eqnarray*}
K-\hat{F}_{2}\left( s,R\left( K,X\right) \right) K_{X} &\simeq &K-\hat{F}%
_{2}\left( R\left( K_{X},X\right) \right) K_{X}-\nabla _{K_{X}}\hat{F}%
_{2}\left( R\left( K_{X},X\right) \right) \left( K-K_{X}\right) \\
&\simeq &K-\hat{F}_{2}\left( R\left( K_{X},X\right) \right) K_{X}
\end{eqnarray*}%
Equation (\ref{SCR}) then simplifies as:%
\begin{eqnarray}
\frac{\delta ^{2}\left( S_{1}+S_{2}\right) }{\delta \Psi ^{\dag }\left(
Z,\theta \right) \delta \Psi \left( Z,\theta \right) } &=&-\frac{\sigma
_{X}^{2}}{2}\nabla _{X}^{2}-\frac{\sigma _{K}^{2}}{2}\nabla _{K}^{2}+\left(
D\left( \left\Vert \Psi \right\Vert ^{2}\right) +2\tau \frac{\left\vert \Psi
\left( X\right) \right\vert ^{2}\left( K_{X}-K\right) }{K}\right) \\
&&+\frac{1}{2\sigma _{K}^{2}}\left( K-\hat{F}_{2}\left( s,R\left(
K_{X},X\right) \right) K_{X}\right) ^{2}  \notag
\end{eqnarray}%
and equation (\ref{TNQ}) becomes:%
\begin{eqnarray}
&&-\frac{\partial }{\partial t}T_{1}\left( \left( K_{f},X_{f}\right) ,\left(
X_{i},K_{i}\right) ,t\right)  \label{TNV} \\
&=&\left( -\frac{\sigma _{X}^{2}}{2}\nabla _{X}^{2}+D\left( \left\Vert \Psi
\right\Vert ^{2}\right) +2\tau \frac{K_{X}-K}{K}\left\Vert \Psi \left(
X\right) \right\Vert ^{2}\right) T_{1}\left( \left( K_{f},X_{f}\right)
,\left( X_{i},K_{i}\right) ,t\right)  \notag \\
&&+\left( -\frac{\sigma _{K}^{2}}{2}\nabla _{K}^{2}+\frac{1}{2\sigma _{K}^{2}%
}\left( K-\hat{F}_{2}\left( R\left( K_{X},X\right) \right) K_{X}\right)
^{2}\right) T_{1}\left( \left( K_{f},X_{f}\right) ,\left( X_{i},K_{i}\right)
,t\right)  \notag
\end{eqnarray}%
Second, we assumed from the beginning that the motion of firms in the
sectors space is at slower pace than capital fluctuations. Moreover, we may
assume that in average $\left\vert \frac{K_{X}-K}{K}\right\vert <<1$. As a
consequence, along the path from the initial point $\left(
X_{i},K_{i}\right) $ to the final point $\left( K_{f},X_{f}\right) $, we can
consider that:%
\begin{equation*}
\frac{K_{X}-K}{K}\left\Vert \Psi \left( X\right) \right\Vert ^{2}
\end{equation*}%
is slowly varying and can be replaced by its average.

The equation for $T_{1}$ thus rewrites:

\begin{eqnarray}
&&-\frac{\partial }{\partial t}T_{1}\left( \left( K_{f},X_{f}\right) ,\left(
X_{i},K_{i}\right) \right)  \label{trn} \\
&=&\left( -\frac{\sigma _{X}^{2}}{2}\nabla _{X}^{2}+D\left( \left\Vert \Psi
\right\Vert ^{2}\right) +\tau \left( \frac{\left\vert \Psi \left(
X_{f}\right) \right\vert ^{2}\left( K_{X_{f}}-K_{f}\right) }{K_{f}}+\frac{%
\left\vert \Psi \left( X_{i}\right) \right\vert ^{2}\left(
K_{X_{i}}-K_{i}\right) }{K_{i}}\right) \right) T_{1}\left( \left(
K_{f},X_{f}\right) ,\left( X_{i},K_{i}\right) \right)  \notag \\
&&+\left( -\frac{\sigma _{K}^{2}}{2}\nabla _{K}^{2}+\frac{1}{2\sigma _{K}^{2}%
}\left( K-\hat{F}_{2}\left( R\left( K_{X},X\right) \right) K_{X}\right)
^{2}\right) T_{1}\left( \left( K_{f},X_{f}\right) ,\left( X_{i},K_{i}\right)
\right)  \notag
\end{eqnarray}%
On the other hand, the derivation of the equation for $T_{2}$ yields
directly: 
\begin{eqnarray}
&&-\frac{\partial }{\partial t}T_{2}\left( \left( \hat{K}_{f},\hat{X}%
_{f}\right) ,\left( \hat{X}_{i},\hat{K}_{i}\right) \right)  \label{TRV} \\
&=&\left( -\frac{\sigma _{\hat{X}}^{2}}{2}\nabla _{\hat{X}}^{2}-\nabla
_{y}^{2}\right) T_{2}\left( \left( \hat{K}_{f},\hat{X}_{f}\right) ,\left( 
\hat{X}_{i},\hat{K}_{i}\right) \right)  \notag \\
&&+\left( \frac{y^{2}}{4}+\frac{\left( g\left( \hat{X}\right) \right)
^{2}+\sigma _{\hat{X}}^{2}\left( f\left( \hat{X}\right) +\nabla _{\hat{X}%
}g\left( \hat{X},K_{\hat{X}}\right) -\frac{\sigma _{\hat{K}}^{2}F^{2}\left( 
\hat{X},K_{\hat{X}}\right) }{2f^{2}\left( \hat{X}\right) }\right) }{\sigma _{%
\hat{X}}^{2}\sqrt{f^{2}\left( \hat{X}\right) }}\right) T_{2}\left( \left( 
\hat{K}_{f},\hat{X}_{f}\right) ,\left( \hat{X}_{i},\hat{K}_{i}\right) \right)
\notag
\end{eqnarray}

\subsection*{Computation of $T_{1}$}

\subsubsection{Solution of (\protect\ref{trn})}

We first rewrite the competition term in (\ref{trn}) as:%
\begin{eqnarray*}
&&\frac{\left\vert \Psi \left( X_{f}\right) \right\vert ^{2}\left(
K_{X_{f}}-K_{f}\right) }{K_{f}}+\frac{\left\vert \Psi \left( X_{i}\right)
\right\vert ^{2}\left( K_{X_{i}}-K_{i}\right) }{K_{i}} \\
&=&\left( \frac{\left\vert \Psi \left( X_{f}\right) \right\vert ^{2}\left(
K_{X_{f}}-K_{f}\right) }{K_{f}}-\frac{\left\vert \Psi \left( X_{i}\right)
\right\vert ^{2}\left( K_{X_{i}}-K_{i}\right) }{K_{i}}\right) +\frac{%
\left\vert \Psi \left( X_{i}\right) \right\vert ^{2}\left(
K_{X_{i}}-K_{i}\right) }{K_{i}}
\end{eqnarray*}%
Then, we normalize the transition functions by factoring the solution of (%
\ref{trn}):%
\begin{eqnarray}
&&T_{1}\left( \left( K_{f},X_{f}\right) ,\left( X_{i},K_{i}\right) \right)
\label{PNL} \\
&=&\exp \left( -t\left( D\left( \left\Vert \Psi \right\Vert ^{2}\right)
+\tau \left( \frac{\left\vert \Psi \left( X_{f}\right) \right\vert
^{2}\left( K_{X_{f}}-K_{f}\right) }{K_{f}}+\frac{\left\vert \Psi \left(
X_{i}\right) \right\vert ^{2}\left( K_{X_{i}}-K_{i}\right) }{K_{i}}\right)
\right) \right) \hat{T}_{1}\left( \left( K_{f},X_{f}\right) ,\left(
X_{i},K_{i}\right) \right)  \notag
\end{eqnarray}%
so that the transition equation writes:%
\begin{eqnarray}
&&-\frac{\partial }{\partial t}\hat{T}_{1}\left( \left( K_{f},X_{f}\right)
,\left( X_{i},K_{i}\right) \right)  \label{TRN} \\
&=&\left( -\frac{\sigma _{X}^{2}}{2}\nabla _{X}^{2}-\frac{\sigma _{K}^{2}}{2}%
\nabla _{K}^{2}+\frac{1}{2\sigma _{K}^{2}}\left( K-\hat{F}_{2}\left( R\left(
K,X\right) \right) K_{X}\right) ^{2}\right) \hat{T}_{1}\left( \left(
K_{f},X_{f}\right) ,\left( X_{i},K_{i}\right) \right)  \notag
\end{eqnarray}%
Note that, given the exponential factor, if $K_{i}<<K_{X_{i}}$, $\frac{%
\left\vert \Psi \left( X_{i}\right) \right\vert ^{2}\left(
K_{X_{i}}-K_{i}\right) }{K_{i}}<0$ and the probability to move away from $%
X_{i}$ is very low. The same applies for $\frac{\left\vert \Psi \left(
X_{f}\right) \right\vert ^{2}\left( K_{X_{f}}-K_{f}\right) }{K_{f}}>0$.

The transition function $\hat{T}_{1}\left( \left( K_{f},X_{f}\right) ,\left(
X_{i},K_{i}\right) \right) $ can be found by using our assumption that
shifts in sectors space are slower than the fluctuations in capital. In (\ref%
{TRN}) we can thus consider in first approximation that the term:%
\begin{equation*}
K-\hat{F}_{2}\left( R\left( K,X\right) \right) K_{X}
\end{equation*}%
shifts the inital and final values of capital:%
\begin{eqnarray*}
K_{i} &\rightarrow &K_{i}-\hat{F}_{2}\left( s,R\left( K_{X_{i}},X_{i}\right)
\right) K_{X_{i}}=K_{i}^{\prime } \\
K_{f} &\rightarrow &K_{f}-\hat{F}_{2}\left( s,R\left( K_{X_{f}},X_{f}\right)
\right) K_{X_{f}}=K_{f}^{\prime }
\end{eqnarray*}%
So that we have:%
\begin{equation}
\hat{T}_{1}\left( \left( K_{f},X_{f}\right) ,\left( X_{i},K_{i}\right)
\right) \simeq \tilde{T}_{1}\left( \left( K_{f}-\hat{F}_{2}\left( s,R\left(
K_{X_{f}},X_{f}\right) \right) K_{X_{f}},X_{f}\right) ,\left( K_{i}-\hat{F}%
_{2}\left( R\left( s,K_{X_{i}},X_{i}\right) \right) K_{X_{i}},K_{i}\right)
\right)  \label{NRM}
\end{equation}%
where $\tilde{T}_{1}$ satisfies:%
\begin{equation}
-\frac{\partial }{\partial t}\tilde{T}_{1}=\left( -\frac{\sigma _{X}^{2}}{2}%
\nabla _{X}^{2}-\frac{\sigma _{K}^{2}}{2}\nabla _{K}^{2}+\frac{1}{2\sigma
_{K}^{2}}K^{2}\right) \tilde{T}_{1}  \label{NRP}
\end{equation}%
Up to a normalization factor, the solution of (\ref{NRP}) is:%
\begin{equation*}
\tilde{T}_{1}\left( \left( K_{f}^{\prime },X_{f}\right) ,\left(
X_{i},K_{i}^{\prime }\right) \right) =\exp \left( -\left( \frac{\left(
X_{f}-X_{i}\right) ^{2}}{2t\sigma _{X}^{2}}+\frac{\left( K_{f}^{\prime
}-K_{i}^{\prime }\right) ^{2}}{2t\sigma _{K}^{2}}-\frac{t\sigma _{K}^{2}}{2}%
K_{f}^{\prime }K_{f}^{\prime }\right) \right)
\end{equation*}%
Using (\ref{NRM}) and (\ref{PNL}), we find the solution of (\ref{trm}):%
\begin{eqnarray}
&&\exp \left( -t\left( D\left( \left\Vert \Psi \right\Vert ^{2}\right) +\tau
\left( \frac{\left\vert \Psi \left( X_{f}\right) \right\vert ^{2}\left(
K_{X_{f}}-K_{f}\right) }{K_{f}}+\frac{\left\vert \Psi \left( X_{i}\right)
\right\vert ^{2}\left( K_{X_{i}}-K_{i}\right) }{K_{i}}\right) \right) \right)
\label{SPR} \\
&&\times \exp \left( -\left( \frac{\left( X_{f}-X_{i}\right) ^{2}}{2t\sigma
_{X}^{2}}+\frac{\left( K_{f}^{\prime }-K_{i}^{\prime }\right) ^{2}}{2t\sigma
_{K}^{2}}-\frac{t\sigma _{K}^{2}}{2}K_{f}^{\prime }K_{f}^{\prime }\right)
\right)  \notag
\end{eqnarray}

\subsubsection{Full transition function}

\bigskip

To obtain the full transition function, recall that (\ref{SPR}) has been
obtained by a change of variable (\ref{chn}). To come back to the initial
variables we have to introduce an other exponential factor to account for
the trend of the transition, and we find:%
\begin{eqnarray*}
&&T_{1}\left( \left( K_{f},X_{f}\right) ,\left( X_{i},K_{i}\right) \right) \\
&\simeq &\exp \left( \int_{X_{i}}^{X_{f}}\frac{\nabla _{X}R\left(
K_{X},X\right) }{\sigma _{X}^{2}}H\left( K_{X}\right) -t\left( D\left(
\left\Vert \Psi \right\Vert ^{2}\right) +\tau \left( \frac{\left\vert \Psi
\left( X_{f}\right) \right\vert ^{2}\left( K_{X_{f}}-K_{f}\right) }{K_{f}}+%
\frac{\left\vert \Psi \left( X_{i}\right) \right\vert ^{2}\left(
K_{X_{i}}-K_{i}\right) }{K_{i}}\right) \right) \right) \\
&&\times \exp \left( -\int^{K_{f}}\left( K-\hat{F}_{2}\left( R\left(
s,K,X_{f}\right) \right) K_{X_{f}}\right) dK+\int^{K_{i}}\left( K-\hat{F}%
_{2}\left( s,R\left( K,X_{i}\right) \right) K_{X_{i}}\right) dK\right) \\
&&\times \exp \left( -\left( \frac{\left( X_{f}-X_{i}\right) ^{2}}{2t\sigma
_{X}^{2}}+\frac{\left( \tilde{K}_{f}-\tilde{K}_{i}\right) ^{2}}{2t\sigma
_{K}^{2}}\right) \right) \\
&&\times \exp \left( -\frac{t\sigma _{K}^{2}}{2}\left( K_{f}-\hat{F}%
_{2}\left( s,R\left( K_{f},\bar{X}\right) \right) K_{\bar{X}}\right) \left(
K_{i}-\hat{F}_{2}\left( s,R\left( K_{i},\bar{X}\right) \right) K_{\bar{X}%
}\right) \right)
\end{eqnarray*}%
\begin{eqnarray}
&&T_{1}\left( \left( K_{f},X_{f}\right) ,\left( X_{i},K_{i}\right) \right) \\
&\simeq &\exp \left( \int_{X_{i}}^{X_{f}}\frac{\nabla _{X}R\left(
K_{X},X\right) }{\sigma _{X}^{2}}H\left( K_{X}\right) -t\left( D\left(
\left\Vert \Psi \right\Vert ^{2}\right) +\tau \left( \frac{\left\vert \Psi
\left( X_{f}\right) \right\vert ^{2}\left( K_{X_{f}}-K_{f}\right) }{K_{f}}+%
\frac{\left\vert \Psi \left( X_{i}\right) \right\vert ^{2}\left(
K_{X_{i}}-K_{i}\right) }{K_{i}}\right) \right) \right)  \notag \\
&&\times \exp \left( -\int^{K_{f}}\left( K-\hat{F}_{2}\left( s,R\left(
K,X_{f}\right) \right) K_{X_{f}}\right) dK+\int^{K_{i}}\left( K-\hat{F}%
_{2}\left( s,R\left( K,X_{i}\right) \right) K_{X_{i}}\right) dK\right) 
\notag \\
&&\times \exp \left( -\left( \frac{\left( X_{f}-X_{i}\right) ^{2}}{2t\sigma
_{X}^{2}}+\frac{\left( K_{f}^{\prime }-K_{i}^{\prime }\right) ^{2}}{2t\sigma
_{K}^{2}}-\frac{t\sigma _{K}^{2}}{2}K_{f}^{\prime }K_{f}^{\prime }\right)
\right)  \notag
\end{eqnarray}%
with:%
\begin{eqnarray*}
K_{i}^{\prime } &=&K_{i}-\hat{F}_{2}\left( R\left( K_{X_{i}},X_{i}\right)
\right) K_{X_{i}} \\
K_{f}^{\prime } &=&K_{f}-\hat{F}_{2}\left( R\left( K_{X_{f}},X_{f}\right)
\right) K_{X_{f}}
\end{eqnarray*}%
The Laplace transform of this function is the transition function given in
the text.

\subsection*{Computation of $T_{2}$}

\subsubsection{Solution ot (\protect\ref{TRV})}

Solving (\ref{TRV}) is straightforward, and similar to the derivation $T_{1} 
$.

We first introduce a change of variable:%
\begin{eqnarray*}
&&T_{2}\left( \left( \hat{K}_{f},\hat{X}_{f}\right) ,\left( \hat{X}_{i},\hat{%
K}_{i}\right) \right) \\
&=&\exp \left( -t\int_{\hat{X}_{i}}^{\hat{X}_{f}}\frac{\left( g\left( \hat{X}%
\right) \right) ^{2}+\sigma _{\hat{X}}^{2}\left( f\left( \hat{X}\right)
+\nabla _{\hat{X}}g\left( \hat{X},K_{\hat{X}}\right) -\frac{\sigma _{\hat{K}%
}^{2}F^{2}\left( \hat{X},K_{\hat{X}}\right) }{2f^{2}\left( \hat{X}\right) }%
\right) }{\left\Vert \hat{X}_{f}-\hat{X}_{i}\right\Vert \sigma _{\hat{X}}^{2}%
\sqrt{f^{2}\left( \hat{X}\right) }}\right) \hat{T}_{2}\left( \left( \hat{K}%
_{f},\hat{X}_{f}\right) ,\left( \hat{X}_{i},\hat{K}_{i}\right) \right)
\end{eqnarray*}%
The term in the exponential is the average of the relative return:%
\begin{equation*}
\frac{\left( g\left( \hat{X}\right) \right) ^{2}+\sigma _{\hat{X}}^{2}\left(
f\left( \hat{X}\right) +\nabla _{\hat{X}}g\left( \hat{X},K_{\hat{X}}\right) -%
\frac{\sigma _{\hat{K}}^{2}F^{2}\left( \hat{X},K_{\hat{X}}\right) }{%
2f^{2}\left( \hat{X}\right) }\right) }{\sigma _{\hat{X}}^{2}\sqrt{%
f^{2}\left( \hat{X}\right) }}
\end{equation*}%
along the average path, considered as a straight line, from $\hat{X}_{i}$ to 
$\hat{X}_{f}$. We have assumed slow shifts in the sectors space, so that $%
\hat{T}_{2}$ satisfies the following equation in first approximation: 
\begin{eqnarray}
&&-\frac{\partial }{\partial t}\hat{T}_{2}\left( \left( \hat{K}_{f},\hat{X}%
_{f}\right) ,\left( \hat{X}_{i},\hat{K}_{i}\right) \right)  \label{TRZ} \\
&\simeq &\left( -\frac{\sigma _{\hat{X}}^{2}}{2}\nabla _{\hat{X}}^{2}-\nabla
_{y}^{2}\right) \hat{T}_{2}\left( \left( \hat{K}_{f},\hat{X}_{f}\right)
,\left( \hat{X}_{i},\hat{K}_{i}\right) \right) +\frac{y^{2}}{4}\hat{T}%
_{2}\left( \left( \hat{K}_{f},\hat{X}_{f}\right) ,\left( \hat{X}_{i},\hat{K}%
_{i}\right) \right)  \notag
\end{eqnarray}%
Given (\ref{cvr}), we can assume that $y$ is independent from $\hat{X}$ in
first approximation. Thus, solving (\ref{TRZ}) yields:

\begin{eqnarray}
&&\hat{T}_{2}\left( \left( \hat{K}_{f},\hat{X}_{f}\right) ,\left( \hat{X}%
_{i},\hat{K}_{i}\right) \right)  \label{PRF} \\
&\simeq &\exp \left( -\left( \frac{\sigma _{\hat{X}}^{2}}{2}t\left( \hat{K}%
_{f}+\frac{\sigma _{\hat{K}}^{2}F\left( \hat{X}_{f},K_{\hat{X}_{f}}\right) }{%
f^{2}\left( \hat{X}_{f}\right) }\right) \left( \hat{K}_{i}+\frac{\sigma _{%
\hat{K}}^{2}F\left( \hat{X}_{i},K_{\hat{X}_{i}}\right) }{f^{2}\left( \hat{X}%
_{i}\right) }\right) \right) \right)  \notag \\
&&\times \exp \left( -\frac{\sqrt{f^{2}\left( \frac{\hat{X}_{f}+\hat{X}_{i}}{%
2}\right) }}{2t\sigma _{\hat{X}}^{2}}\left( \left( \hat{K}_{f}+\frac{\sigma
_{\hat{K}}^{2}F\left( \hat{X}_{f},K_{\hat{X}_{f}}\right) }{f^{2}\left( \hat{X%
}_{f}\right) }\right) -\left( \hat{K}_{i}+\frac{\sigma _{\hat{K}}^{2}F\left( 
\hat{X}_{i},K_{\hat{X}_{i}}\right) }{f^{2}\left( \hat{X}_{i}\right) }\right)
\right) ^{2}\right)  \notag
\end{eqnarray}

\subsubsection{Full transition function}

Reintroducing the change of variables (\ref{chg}) amounts to introduce a
factor:%
\begin{equation*}
\exp \left( \frac{1}{\sigma _{\hat{X}}^{2}}\int_{\hat{X}_{i}}^{\hat{X}%
_{f}}g\left( \hat{X}\right) d\hat{X}+\frac{\hat{K}_{f}^{2}}{\sigma _{\hat{K}%
}^{2}}f\left( \hat{X}_{f}\right) -\frac{\hat{K}_{i}^{2}}{\sigma _{\hat{K}%
}^{2}}f\left( \hat{X}_{i}\right) \right)
\end{equation*}%
in the formula for $T_{2}$\ and this leads to the full formula for the
transition function:%
\begin{eqnarray}
&&T_{2}\left( \left( \hat{K}_{f},\hat{X}_{f}\right) ,\left( \hat{X}_{i},\hat{%
K}_{i}\right) \right)  \label{TRC} \\
&\simeq &\exp \left( -t\int_{\hat{X}_{i}}^{\hat{X}_{f}}\frac{\left( g\left( 
\hat{X}\right) \right) ^{2}+\sigma _{\hat{X}}^{2}\left( f\left( \hat{X}%
\right) +\nabla _{\hat{X}}g\left( \hat{X},K_{\hat{X}}\right) -\frac{\sigma _{%
\hat{K}}^{2}F^{2}\left( \hat{X},K_{\hat{X}}\right) }{2f^{2}\left( \hat{X}%
\right) }\right) }{\left\Vert \hat{X}_{f}-\hat{X}_{i}\right\Vert \sigma _{%
\hat{X}}^{2}\sqrt{f^{2}\left( \hat{X}\right) }}\right)  \notag \\
&&\times \exp \left( \frac{1}{\sigma _{\hat{X}}^{2}}\int_{\hat{X}_{i}}^{\hat{%
X}_{f}}g\left( \hat{X}\right) d\hat{X}+\frac{\hat{K}_{f}^{2}}{\sigma _{\hat{K%
}}^{2}}f\left( \hat{X}_{f}\right) -\frac{\hat{K}_{i}^{2}}{\sigma _{\hat{K}%
}^{2}}f\left( \hat{X}_{i}\right) \right) \hat{T}_{2}\left( \left( \hat{K}%
_{f},\hat{X}_{f}\right) ,\left( \hat{X}_{i},\hat{K}_{i}\right) \right) 
\notag
\end{eqnarray}%
with:%
\begin{equation*}
\hat{T}_{2}\left( \left( \hat{K}_{f},\hat{X}_{f}\right) ,\left( \hat{X}_{i},%
\hat{K}_{i}\right) \right)
\end{equation*}%
given by (\ref{PRF}).

The Laplace transform of (\ref{TRC}) is the formula presented in the text.

\section*{Appendix 4}

We write the series expansion in $\Delta S_{\text{fourth order}}$ of $\exp
\left( -S\left( \Psi \right) \right) $: 
\begin{eqnarray*}
\exp \left( -S\left( \Psi \right) \right) &=&\exp \left( -\left( S\left(
\Psi _{0},\hat{\Psi}_{0}\right) +\int \left( \Delta \Psi ^{\dag }\left(
Z,\theta \right) ,\Delta \hat{\Psi}^{\dag }\left( Z,\theta \right) \right)
\left( Z,\theta \right) O\left( \Psi _{0}\left( Z,\theta \right) \right)
\left( 
\begin{array}{c}
\Delta \Psi \left( Z,\theta \right) \\ 
\Delta \hat{\Psi}\left( Z,\theta \right)%
\end{array}%
\right) \right) \right) \\
&&\left( 1+\sum_{n\geqslant 1}\frac{\left( -\Delta S_{\text{fourth order}%
}\left( \Psi ,\hat{\Psi}\right) \right) ^{n}}{n!}\right)
\end{eqnarray*}%
where $O\left( \Psi _{0}\left( Z,\theta \right) \right) $ is defined in (\ref%
{Dfr}).

Then, we decompose $\Delta S_{\text{fourth order}}\left( \Psi ,\hat{\Psi}%
\right) $ as a sum of two combinations:%
\begin{eqnarray*}
\Delta S_{\text{fourth order}}\left( \Psi ,\hat{\Psi}\right) &=&\int \Delta
\Psi ^{\dag }\left( K,X\right) \Delta \Psi ^{\dag }\left( K^{\prime
},X^{\prime }\right) \Delta S_{11}\Delta \Psi \left( K^{\prime },X^{\prime
}\right) \Delta \Psi \left( K,X\right) \\
&&+\Delta \Psi ^{\dagger }\left( K^{\prime },X^{\prime }\right) \Delta \hat{%
\Psi}^{\dagger }\left( \hat{K},\hat{X}\right) \Delta S_{12}\Delta \Psi
\left( K^{\prime },X^{\prime }\right) \Delta \hat{\Psi}\left( \hat{K},\hat{X}%
\right)
\end{eqnarray*}%
with:%
\begin{equation}
\Delta S_{11}=\left( 2\tau -\nabla _{K}\frac{\delta ^{2}u\left( K,X,\Psi ,%
\hat{\Psi}\right) }{\delta \Psi \left( K^{\prime },X\right) \delta \Psi
^{\dagger }\left( K^{\prime },X\right) }\right) \delta \left( X-X^{\prime
}\right)  \notag
\end{equation}%
and:%
\begin{equation}
\Delta S_{12}=-\left( \nabla _{K}\frac{\delta ^{2}u\left( K,X,\Psi ,\hat{\Psi%
}\right) }{\delta \hat{\Psi}\left( \hat{K},\hat{X}\right) \delta \hat{\Psi}%
^{\dagger }\left( \hat{K},\hat{X}\right) }+\left\{ \nabla _{\hat{K}}\frac{%
\hat{K}\delta ^{2}f\left( \hat{X},\Psi ,\hat{\Psi}\right) }{\delta \Psi
\left( K^{\prime },X\right) \delta \Psi ^{\dagger }\left( K^{\prime
},X\right) }+\nabla _{\hat{X}}\frac{\delta ^{2}g\left( \hat{X},\Psi ,\hat{%
\Psi}\right) }{\delta \Psi \left( K^{\prime },X\right) \delta \Psi ^{\dagger
}\left( K^{\prime },X\right) }\right\} \right) \delta \left( X-X^{\prime
}\right)  \notag
\end{equation}%
Application of (\ref{trg}) leads to the following form of the transition
functions:%
\begin{eqnarray}
&&G_{ij}\left( \left[ \left( K_{f},X_{f}\right) ,\left( K_{f},X_{f}\right)
^{\prime }\right] ,\left[ \left( X_{i},K_{i}\right) ,\left(
X_{i},K_{i}\right) ^{\prime }\right] \right)  \label{frg} \\
&=&G_{i}\left( \left( K_{f},X_{f}\right) ,\left( X_{i},K_{i}\right) \right)
G_{j}\left( \left( K_{f},X_{f}\right) ^{\prime },\left( X_{i},K_{i}\right)
^{\prime }\right)  \notag \\
&&+\sum_{p\geqslant 1}\frac{\left( -1\right) ^{p}}{p!}\int G_{i}\left(
\left( K_{f},X_{f}\right) ,\left( X_{p},K_{p}\right) \right) G_{j}\left(
\left( K_{f},X_{f}\right) ^{\prime },\left( X_{p},K_{p}\right) ^{\prime
}\right) \Delta S_{ij}\left( \left( X_{p},K_{p}\right) ,\left(
X_{p},K_{p}\right) ^{\prime }\right)  \notag \\
&&\times G_{i}\left( \left( X_{p},K_{p}\right) ,\left(
X_{p-1},K_{p-1}\right) \right) G_{j}\left( \left( X_{p},K_{p}\right)
^{\prime },\left( X_{p-1},K_{p-1}\right) ^{\prime }\right) \Delta
S_{ij}\left( \left( X_{p-1},K_{p-1}\right) ,\left( X_{p-1},K_{p-1}\right)
^{\prime }\right)  \notag \\
&&...\times \Delta S_{ij}\left( \left( K_{1},X_{1}\right) ,\left(
K_{1},X_{1}\right) ^{\prime }\right) G_{1}\left( \left( K_{1},X_{1}\right)
,\left( X_{i},K_{i}\right) \right) G_{1}\left( \left( K_{1},X_{1}\right)
^{\prime },\left( X_{i},K_{i}\right) ^{\prime }\right)
\prod\limits_{k\leqslant p}d\left( \left( X_{k},K_{k}\right) ,\left(
X_{k},K_{k}\right) ^{\prime }\right)  \notag
\end{eqnarray}

These corrections modify the $n$ agents Green functions and can be computed
using graphs expansion. In the sequel we will focus only on the first order
corrections to the four agents Green functions. This is sufficient to stress
the impact of interactions of agents in the background state.

The term $\Delta S_{11}$ measures the interaction between firms, and $\Delta
S_{12}$ the firms-investors interactions. There is no term $\Delta S_{22}$
of investors-investors interaction. In our model all interactions depend on
firms.

To estimate the impact of interactions, we can assume the paths from $\left(
\left( X_{i},K_{i}\right) ,\left( K_{f},X_{f}\right) \right) $ to $\left(
\left( X_{i},K_{i}\right) ,\left( K_{f},X_{f}\right) ^{\prime }\right) $
cross each other one time at some $X$ and approximate the terms $\Delta
S_{ij}$ by their average value estimated on the average paths from $%
K_{i},K_{i}^{\prime }$ to $K_{f},K_{f}^{\prime }$,

In this approximation, we find:%
\begin{eqnarray*}
&&G_{ij}\left( \left[ \left( K_{f},X_{f}\right) ,\left( K_{f},X_{f}\right)
^{\prime }\right] ,\left[ \left( X_{i},K_{i}\right) ,\left(
X_{i},K_{i}\right) ^{\prime }\right] \right) \\
&\simeq &G_{i}\left( \left( K_{f},X_{f}\right) ,\left( X_{i},K_{i}\right)
\right) G_{j}\left( \left( K_{f},X_{f}\right) ^{\prime },\left(
X_{i},K_{i}\right) ^{\prime }\right) \\
&&-G_{i}\left( \left( K_{f},X_{f}\right) ,\left( X,K\right) \right)
G_{j}\left( \left( K_{f},X_{f}\right) ^{\prime },\left( X,K\right) ^{\prime
}\right) \\
&&\times \Delta S_{ij}\left( \left( X,\bar{K}\right) ,\left( X,\bar{K}%
\right) ^{\prime }\right) G_{1}\left( \left( X,K\right) ,\left(
X_{i},K_{i}\right) \right) G_{1}\left( \left( X,K\right) ^{\prime },\left(
X_{i},K_{i}\right) ^{\prime }\right) \\
&\simeq &G_{i}\left( \left( K_{f},X_{f}\right) ,\left( X_{i},K_{i}\right)
\right) G_{j}\left( \left( K_{f},X_{f}\right) ^{\prime },\left(
X_{i},K_{i}\right) ^{\prime }\right) \\
&&-\Delta S_{ij}\left( \left( \bar{X},\bar{K}\right) ,\left( \bar{X},\bar{K}%
\right) ^{\prime }\right) \hat{G}_{i}\left( \left( K_{f},X_{f}\right)
,\left( X,K\right) \right) \hat{G}_{j}\left( \left( K_{f},X_{f}\right)
^{\prime },\left( X,K\right) ^{\prime }\right)
\end{eqnarray*}%
with:%
\begin{eqnarray*}
\left( \bar{X},\bar{K}\right) &=&\frac{\left( K_{f},X_{f}\right) +\left(
X_{i},K_{i}\right) }{2} \\
\left( \bar{X},\bar{K}\right) ^{\prime } &=&\frac{\left( K_{f},X_{f}\right)
^{\prime }+\left( X_{i},K_{i}\right) ^{\prime }}{2}
\end{eqnarray*}%
and:%
\begin{equation*}
\hat{G}_{i}\left( \left( K_{f},X_{f}\right) ,\left( X,K\right) \right) \hat{G%
}_{j}\left( \left( K_{f},X_{f}\right) ^{\prime },\left( X,K\right) ^{\prime
}\right)
\end{equation*}%
is the transition function computed on path that cross once. Applied to the
three transition functions for two agents yields the results of the text.

\section*{Appendix 5}

\section*{One agent transition functions}

\subsection*{Firms transition function}

We interpret the various term involved in (\ref{Gn}) and their influence on
firms individual dynamics.

\subsubsection*{Drift term}

\paragraph*{The three contributions}

The first term in (\ref{Gn}):%
\begin{equation*}
D\left( \left( K_{f},X_{f}\right) ,\left( X_{i},K_{i}\right) \right)
\end{equation*}%
is a drift term between $\left( X_{i},K_{i}\right) $ and $\left(
K_{f},X_{f}\right) $. It is composed of three contributions (see (\ref{frd}%
)):

The first term of (\ref{frd}):

\begin{equation*}
\int_{X_{i}}^{X_{f}}\frac{\nabla _{X}R\left( K_{X},X\right) }{\sigma _{X}^{2}%
}H\left( K_{X}\right)
\end{equation*}%
models the shift of producers towards sectors that have the highest
long-term returns.

To interpret the second contribution to $D\left( \left( K_{f},X_{f}\right)
,\left( X_{i},K_{i}\right) \right) $:

\begin{equation}
\int_{K_{i}}^{K_{f}}\left( K-\hat{F}_{2}\left( s,R\left( K,\bar{X}\right)
\right) K_{\bar{X}}\right) dK  \label{scn}
\end{equation}%
, recall that $\frac{F_{2}\left( R\left( K,\bar{X}\right) \right) }{%
F_{2}\left( R\left( K_{\bar{X}},\bar{X}\right) \right) }$ models the
relative expectations of returns of the firm along its path from $\left(
X_{i},K_{i}\right) $ to $\left( K_{f},X_{f}\right) $ based on their returns'
expectation $R\left( K,\bar{X}\right) $ and that $\frac{F_{2}\left( R\left(
K,\bar{X}\right) \right) K_{\bar{X}}}{F_{2}\left( R\left( K_{\bar{X}},\bar{X}%
\right) \right) }$ represents the capital investors are ready to invest in
the firm along this path. Along the path from $\left( X_{i},K_{i}\right) $
to $\left( K_{f},X_{f}\right) $, the capital invested in the firm will
increase as long as the investors expect growth and as long as additional
investment is likely to increase the firm's returns. Once the level of
capital reaches their expectations, that is 
\begin{equation}
K_{T}\left( \bar{X}\right) -\hat{F}_{2}\left( R\left( K_{T},\bar{X}\right)
\right) K_{\bar{X}}=0  \label{hl}
\end{equation}%
i.e., when the firm has reached the capital threshold, investment stops.

However, this condition is not always fulfilled. The shape of $F_{2}$ is
critical. If $F_{2}$ is above the line $Y=K$, then for $K<K_{T}$, the
threshold $K_{T}$ will be reached gradually. In this case, $K_{T}$ is an
equilibrium point. If, on the contrary, $F_{2}$ is below the line $Y=K$,
then for $K<K_{T}$, the threshold $K_{T}$ will never be reached, and $%
K\rightarrow 0$. If $K>K_{T}$, $K$ can increase indefinitely. This
corresponds to firms whose profitability is perceived as boundless as long
as more capital is invested in.

The third term in (\ref{frd}):

\begin{equation*}
\int_{K_{i}}^{K_{f}}\left( \left( \frac{X_{f}-X_{i}}{2}\right) \nabla _{X}%
\hat{F}_{2}\left( s,R\left( K,\bar{X}\right) \right) K_{\bar{X}}\right) dK
\end{equation*}%
induces firms to move towards more appropriate sectors, according to
investors and given the capital of the firm. The firm does not solely move
according to the new investors it could attract but must also take into
account its current investors. If it moves, it risks losing the investors it
has already attracted.

\paragraph*{Trade-off between terms}

There is a trade-off between the first and the third terms: firms want to
move towards sectors with higher returns, but differences in average capital
between sectors could make a firm unattractive in a new sector. The loss of
investors \textbf{incured }during a shift of sector must be compared with
the number of investors possibly attracted in the new sector: the level of
attractiveness may decrease for a given amount of capital.

The second contribution to $D$ is an indicator of the firm's growth
potential in a given sector.\ It depends on the firm's level of capital
compared to the threshold capital requirement and its dynamics in this
sector.

A move along sectors due to the terms 1 and 3 modifies the firm's relative
capital, which is sector-dependant: $F_{2}$, measuring the firm
attractiveness in the sector and indirectly the threshold of capital $K_{T}$
defined in (\ref{hl}) are modified by the shift from one sector to another.
Therefore, a firm could be below the value of $K_{T}$ in one sector, then
above in the next sector, which will reverse its capital dynamics. The
firm's capital dynamics remains the same as long as its relative
attractiveness in a sector does not change significantly.

\subsubsection*{Effective time of transition}

The following term depends on the competition in a sector:%
\begin{eqnarray}
\alpha _{eff}\left( \Psi ,\left( K_{f},X_{f}\right) ,\left(
X_{i},K_{i}\right) \right) &=&\alpha +D\left( \left\Vert \Psi \right\Vert
^{2}\right) +\frac{\tau }{2}\left( \frac{\left\vert \Psi \left( X_{f}\right)
\right\vert ^{2}\left( K_{X_{f}}-K_{f}\right) }{K_{f}}-\frac{\left\vert \Psi
\left( X_{i}\right) \right\vert ^{2}\left( K_{X_{i}}-K_{i}\right) }{K_{i}}%
\right)  \label{fct} \\
&&+\frac{\sigma _{K}^{2}}{2}\left( K_{f}-\hat{F}_{2}\left( s,R\left( K_{f},%
\bar{X}\right) \right) K_{\bar{X}}\right) \left( K_{i}-\hat{F}_{2}\left(
s,R\left( K_{i},\bar{X}\right) \right) K_{\bar{X}}\right)  \notag
\end{eqnarray}

Recall that the constant $\alpha $ is the inverse of the average lifetime of
the agents. The larger it is, the lower the probability of transition. In
the transition functions, $\alpha $ is shifted by two path-dependent terms
and replaced by $\alpha _{eff}$. Thus, $\alpha _{eff}$ is the inverse
mobility of the firm during its transition from one point to another.

Therefore, the likelihood of shifts in capital and sectors depends not only
on the average lifespan of the firm, but also on terms that are directly
related to the collective state.

The first correction to $\alpha $ is $D\left( \left\Vert \Psi \right\Vert
^{2}\right) $ that is related to competition. We have shown in (Gosselin
Lotz Wambst 2022):

\begin{equation*}
D\left( \left\Vert \Psi \right\Vert ^{2}\right) \simeq 2\tau \frac{N}{V-V_{0}%
}+\frac{1}{2\sigma _{X}^{2}}\left\langle \left( \nabla _{X}R\left( X\right)
\right) ^{2}\right\rangle _{V/V_{0}}H^{2}\left( \frac{\left\langle \hat{K}%
\right\rangle }{N}\right) \left( 1-\frac{H^{\prime }\left( \frac{%
\left\langle \hat{K}\right\rangle }{N}\right) }{H\left( \frac{\left\langle 
\hat{K}\right\rangle }{N}\right) }\frac{\left\langle \hat{K}\right\rangle }{N%
}\right)
\end{equation*}%
where $V$ is the volume of the sectors space and $V_{0}$ is the locus where $%
\left\Vert \Psi \left( X\right) \right\Vert ^{2}=0$. As a consequence, the
stronger the competition, i.e., the larger $\tau $, the greater $D\left(
\left\Vert \Psi \right\Vert ^{2}\right) $, and the less possibilities of
shifting from a sector to another.

The third term in (\ref{fct}):%
\begin{equation}
\frac{\tau }{2}\left( \frac{\left\vert \Psi \left( X_{f}\right) \right\vert
^{2}\left( K_{X_{f}}-K_{f}\right) }{K_{f}}-\frac{\left\vert \Psi \left(
X_{i}\right) \right\vert ^{2}\left( K_{X_{i}}-K_{i}\right) }{K_{i}}\right)
\label{stv}
\end{equation}%
is also linked to the competition, but depends on the level of capital of
the agent, and the number of d agents in the sectors crossed. This term
measures the strength of agent's shift from one sector to another.

It is negative when $K_{f}>K_{X_{f}}$, and when $K_{i}<K_{X_{i}}$. In other
words, when a firm has less capital than the average in its initial sector,
and when, it ends up in a sector in which it has more capital than the
average, the probability of transition from $K_{i}$ to $K_{f}$ is high. In
other words, under-average capital favors the exit from a sector.
Above-average capital promotes entry into the sector. Shifts from high
average capital sectors to lower-average-capital sectors are favoured.

This phenomenon is amplified by the number of agents. The greater the
competition in a sector, i.e., the more firms in the sector, the greater the
probability for a lower-than average capitalized firm to be ousted from the
sector by higher-than average capitalized firms that enter the sector.
Thus,\ the density of producers $\left\vert \Psi \left( X\right) \right\vert
^{2}$ along the movement enhances competition and favours high capitalized
firm to move towards higher capitalized sectors, and drives low capitalized
firms toward low capitalized sectors.

The last term in (\ref{fct}):%
\begin{equation*}
\frac{\sigma _{K}^{2}}{2}\left( K_{f}-\hat{F}_{2}\left( s,R\left( K_{f},\bar{%
X}\right) \right) K_{\bar{X}}\right) \left( K_{i}-\hat{F}_{2}\left(
s,R\left( K_{i},\bar{X}\right) \right) K_{\bar{X}}\right)
\end{equation*}%
shows that in average, shifts from the initial point to the final point is
done respecting $K=\hat{F}_{2}\left( s,R\left( K,\bar{X}\right) \right) K_{%
\bar{X}}$, the capital investors allocate to the firm. Actually, if $K_{i}-%
\hat{F}_{2}\left( s,R\left( K_{i},\bar{X}\right) \right) K_{\bar{X}}<0$,
there is a higher probability to reach $K_{f}-\hat{F}_{2}\left( s,R\left(
K_{f},\bar{X}\right) \right) K_{\bar{X}}>0$. If $K_{i}-\hat{F}_{2}\left(
s,R\left( K_{i},\bar{X}\right) \right) K_{\bar{X}}>0$, there is a higher
probability to reach $K_{f}-\hat{F}_{2}\left( s,R\left( K_{f},\bar{X}\right)
\right) K_{\bar{X}}<0$.

If a firm starts with a capital lower than $\hat{F}_{2}\left( s,R\left( K,%
\bar{X}\right) \right) K_{\bar{X}}$, it is more likely to end up with
capital above the new $\hat{F}_{2}\left( s,R\left( K,\bar{X}\right) \right)
K_{\bar{X}}$, in another sector. The movement induces a change in the $F_{2}$%
, but firms tend to fluctuate around the $F_{2}$ of transition.

\subsubsection*{Fluctuation terms}

The term:%
\begin{equation*}
\sqrt{\frac{\left( X_{f}-X_{i}\right) ^{2}}{2\sigma _{X}^{2}}+\frac{\left( 
\tilde{K}_{f}-\tilde{K}_{i}\right) ^{2}}{2\sigma _{K}^{2}}}
\end{equation*}%
describes oscillations around:%
\begin{equation*}
K_{f}-\hat{F}_{2}\left( s,R\left( K_{f},\bar{X}\right) \right) K_{\bar{X}%
}+K_{i}-\hat{F}_{2}\left( s,R\left( K_{i},\bar{X}\right) \right) K_{\bar{X}%
}=0
\end{equation*}%
On average, during a transition from the initial point to the final point,
the firm's capital is governed by the equation $K=K_{T}=\hat{F}_{2}\left(
R\left( K_{i},\bar{X}\right) \right) K_{\bar{X}}$. If a firm starts with
less capital than the threshold set by $F_{2}$, it is more likely to end up
with a capital above the new $F_{2}$, in another sector. The transition from
one sector to another involves oscillations around the sector-dependent
threshold $F_{2}$. However, these oscillations may affect the final
destination of the transition. Starting with a capital level above $K_{T}$,
i.e. $K>K_{T}$, the firm may shift towards sectors with higher perspectives,
i.e. with a higher threshold $K_{T}$. This favors in turn accumulation and
faster transition to other sectors. Finally, if $\nabla _{K}\hat{F}%
_{2}\left( s,R\left( \frac{K_{f}+K_{i}}{2},\bar{X}\right) \right) >0$, $%
F_{2} $ is highly responsive to changes in capital, and larger moves are
favored.

\subsection*{Investors transition functions}

We interpret the different contributions in (\ref{Gt}).

\subsubsection*{Drift term}

The term $D^{\prime }\left( \left( \hat{K}_{f},\hat{X}_{f}\right) ,\left( 
\hat{X}_{i},\hat{K}_{i}\right) \right) $ is composed of two contributions.
The first one:%
\begin{equation*}
\frac{1}{\sigma _{\hat{X}}^{2}}\int_{\hat{X}_{i}}^{\hat{X}_{f}}g\left( \hat{X%
}\right) d\hat{X}+\frac{\hat{K}_{f}^{2}}{\sigma _{\hat{K}}^{2}}f\left( \hat{X%
}_{f}\right) -\frac{\hat{K}_{i}^{2}}{\sigma _{\hat{K}}^{2}}f\left( \hat{X}%
_{i}\right)
\end{equation*}%
is composed of two elements.

The first element is $\frac{1}{\sigma _{\hat{X}}^{2}}\int_{\hat{X}_{i}}^{%
\hat{X}_{f}}g\left( \hat{X}\right) d\hat{X}$. Since the function $g\left( 
\hat{X}\right) $ is the anticipation of higher returns and rising stock
prices, investors move towards sectors were they anticipate the highest
returns and stock prices increase.

The second element $\frac{\hat{K}_{f}^{2}}{\sigma _{\hat{K}}^{2}}f\left( 
\hat{X}_{f}\right) -\frac{\hat{K}_{i}^{2}}{\sigma _{\hat{K}}^{2}}f\left( 
\hat{X}_{i}\right) $: this element can be rewritten as two bits:

\begin{equation*}
\frac{\hat{K}_{f}^{2}-\hat{K}_{i}^{2}}{\sigma _{\hat{K}}^{2}}f\left( \frac{%
\hat{X}_{f}+\hat{X}_{i}}{2}\right) +\frac{\left( \hat{K}_{f}^{2}+\hat{K}%
_{i}^{2}\right) }{2\sigma _{\hat{K}}^{2}}\left( \hat{X}_{f}-\hat{X}%
_{i}\right) \nabla _{X}f\left( \frac{\hat{X}_{f}+\hat{X}_{i}}{2}\right)
\end{equation*}%
The first bit, $\frac{\hat{K}_{f}^{2}-\hat{K}_{i}^{2}}{\sigma _{\hat{K}}^{2}}%
f\left( \frac{\hat{X}_{f}+\hat{X}_{i}}{2}\right) $ shows that the highest
the short-term return, the more probable is the increase in capital. The
second bit, which is equal to $\frac{\left( \hat{K}_{f}^{2}+\hat{K}%
_{i}^{2}\right) }{2\sigma _{\hat{K}}^{2}}\left( \hat{X}_{f}-\hat{X}%
_{i}\right) \nabla _{X}f\left( \frac{\hat{X}_{f}+\hat{X}_{i}}{2}\right) $\
indicates that investors move towards sectors with highest returns.\ The
more capital they have, the fastest the shift.

The second term arising in $D^{\prime }\left( \left( \hat{K}_{f},\hat{X}%
_{f}\right) ,\left( \hat{X}_{i},\hat{K}_{i}\right) \right) $:

\begin{equation*}
-\frac{1}{\sigma _{\hat{X}}^{2}}\int_{\hat{X}_{i}}^{\hat{X}_{f}}\frac{\left(
g\left( \hat{X}\right) \right) ^{2}+\sigma _{\hat{X}}^{2}\left( f\left( \hat{%
X}\right) +\nabla _{\hat{X}}g\left( \hat{X},K_{\hat{X}}\right) -\frac{\sigma
_{\hat{K}}^{2}F^{2}\left( \hat{X},K_{\hat{X}}\right) }{2f^{2}\left( \hat{X}%
\right) }\right) }{\sigma _{\hat{X}}^{2}\sqrt{f^{2}\left( \hat{X}\right) }}d%
\hat{X}
\end{equation*}%
is similar to the determinant of capital accumulation in a collective state
and has the same interpretation. There is a tradeoff between long-term and
short-term returns. It further shows the importance of relative long-term
return. Investors move towards relative long-term returns. Mathematically,
we can measure the dependence of agents' capital accumulation in neighboring
sectors using the integrand: 
\begin{equation}
p=\frac{-\left( \frac{\left( g\left( \hat{X},K_{\hat{X}_{M}}\right) \right)
^{2}}{\sigma _{\hat{X}}^{2}}+\nabla _{\hat{X}}g\left( \hat{X},K_{\hat{X}%
_{M}}\right) -\frac{\sigma _{\hat{K}}^{2}F^{2}\left( \hat{X},K_{\hat{X}%
}\right) }{2f^{2}\left( \hat{X}\right) }\right) }{f\left( \hat{X}\right) }
\label{nPR}
\end{equation}%
The function $p$ represents the relative attractivity of a sector vis-a-vis
its neighbours and depends on the gradients of long-term returns $R\left( 
\hat{X}\right) $ through the function $g\left( \hat{X}\right) $, the capital
mobility at sector $\hat{X}$.\ \ This function $g\left( \hat{X}\right) $,
which depicts investors' propensity to seek higher returns across
sectors,and is indeed proportional to $\nabla _{\hat{X}}R\left( \hat{X}%
\right) $. The gradient of $g$, $\nabla _{\hat{X}}g$, is proportional to $%
\nabla _{\hat{X}}^{2}R\left( \hat{X}\right) $: it measures the position of
the sector relative to its neighbours.\ At a local maximum, the second
derivative of $R\left( \hat{X}\right) $ is negative: $\nabla _{\hat{X}%
}^{2}R\left( \hat{X}\right) <0$. At a minimum, it is positive.

The last term, $\frac{\sigma _{\hat{K}}^{2}F^{2}\left( \hat{X},K_{\hat{X}%
}\right) }{2f^{2}\left( \hat{X}\right) }$, involved in the definition of $Y(%
\hat{X})$ and $p$ is a smoothing factor between neighbouring sectors. It
reduces differences between sectors: it increases when the relative
attractivity with respect to $K_{\hat{X}}$ decreases. The number of
investors and capital will increase in sectors\ that are in the
neighbourhood of significantly more attractive sectors, i.e. with higher
average capital and number of investors. It slows down the transitions.

\subsubsection*{Fluctuation terms}

The last term involved in (\ref{Gt}):

\begin{equation*}
\alpha _{eff}^{\prime }\left( \left( \hat{K}_{f},\hat{X}_{f}\right) ,\left( 
\hat{X}_{i},\hat{K}_{i}\right) \right) \sqrt{\frac{\left\vert f\left( \frac{%
\hat{X}_{f}+\hat{X}_{i}}{2}\right) \right\vert }{2\sigma _{\hat{X}}^{2}}}%
\left\vert \left( \hat{K}_{f}+\frac{\sigma _{\hat{K}}^{2}F\left( \hat{X}%
_{f},K_{\hat{X}_{f}}\right) }{f^{2}\left( \hat{X}_{f}\right) }\right)
-\left( \hat{K}_{i}+\frac{\sigma _{\hat{K}}^{2}F\left( \hat{X}_{i},K_{\hat{X}%
_{i}}\right) }{f^{2}\left( \hat{X}_{i}\right) }\right) \right\vert
\end{equation*}%
depicts the possible oscillations around averages (that) occur (within)in a
definite timespan, so(such) that the transition probability decrases with $%
\hat{K}_{f}-\hat{K}_{i}$ and $X_{f}-X_{i}$. However, this probability
decreases with short-term returns: the higher the returns, the lower the
incentive to switch from one sector to another.

\end{document}